\documentclass[a4paper,fleqn]{cas-sc}

\usepackage{setspace} 
\usepackage{lineno}

\usepackage[authoryear]{natbib}

\usepackage{natbib}
\usepackage{subcaption}
\usepackage{longtable}
\usepackage{multirow}
\usepackage{todonotes}
\usepackage{booktabs}
\usepackage{gensymb}

\begin{document}
\let\WriteBookmarks\relax
\def\floatpagepagefraction{1}
\def\textpagefraction{.001}

\shorttitle{Impact of Storm Surge and Power Peaking on Tidal-Fluvial Dynamics in Microtidal Neretva River Estuary}

\shortauthors{Krvavica et al.}

\title [mode = title]{Impact of Storm Surge and Power Peaking on Tidal-Fluvial Dynamics in Microtidal Neretva River Estuary}                      


\author[1, 2]{Nino Krvavica}
[   type=editor,
    auid=1,
    bioid=1,
    orcid=0000-0001-5014-5476
]

\ead{nino.krvavica@uniri.hr}

\credit{Conceptualization, Methodology, Validation, Formal Analysis, Investigation, Resources, Data curation, Writing -- Original Draft Preparation, Writing -- Review \& Editing, Visualization, Supervision, Project Administration, Funding acquisition}

\affiliation[1]{organization={University of Rijeka, Faculty of Civil Engineering, Radmile Matejcic 3, Rijeka},
country={Croatia}}

\affiliation[2]{organization={University of Rijeka, Center for Artificial Intelligence and Cybersecurity, Radmile Matejcic 2, Rijeka},
country={Croatia}}


\author[1]{Marta Marija Gržić}
[   type=editor,
    auid=2,
    bioid=2,
    orcid=0009-0000-7175-2398
]

\ead{mmgrzic@gradri.uniri.hr}

\credit{Validation, Formal Analysis, Writing -- Review \& Editing, Visualization}

\author[3]{Silvia Innocenti}
[   type=editor,
    auid=3,
    bioid=3,
    orcid=0000-0002-3095-533X
]

\ead{Silvia.Innocenti@ec.gc.ca}

\credit{Conceptualization, Methodology, Software, Validation, Resources, Formal Analysis, Writing -- Original Draft Preparation, Writing -- Review \& Editing, Visualization}

\affiliation[3]{organization={Environment and Climate Change Canada, Meteorological Research Division, Quebec City, Quebec, G1J 0C3}, 
country={Canada}}

\author[3]{Pascal Matte}
[   type=editor,
    auid=4,
    bioid=4,
    orcid=0000-0003-0968-507X
]

\ead{Pascal.Matte@ec.gc.ca}

\credit{Conceptualization, Methodology, Software, Validation, Resources, Writing -- Original Draft Preparation, Writing -- Review \& Editing, Supervision}


\begin{abstract}
This study investigates the interactions between tides, storm surge, river flow, and power peaking in the microtidal Neretva River estuary, Croatia. Based on the existing NS\_Tide tool, the study proposes a new non-stationary harmonic model adapted for microtidal conditions, which incorporates linear storm surge, as well as linear and quadratic river discharge terms. This model enhances the NS\_Tide's ability to accurately predict water levels from tide-dominated sections downstream to discharge-dominated areas upstream.
River discharge was identified as the dominant factor for predicting stage levels at most stations, while the influence of storm surge, though consistent, decreased upstream. Strong tide-river interactions were observed throughout the study domain, with the stationary tidal component consistently contributing to water level fluctuations at all locations, and minimal influence from the tide-surge interaction component. Simulations using the STREAM numerical model were also used to isolate the variability in water levels caused by power peaking. These simulations demonstrated that high-frequency discharge fluctuations due to hydropower plant operations amplify the $S_1$ constituent in upstream river sections and modulate the amplitudes of other tidal constituents in the estuarine and tidal river sections.
The proposed method proved highly effective in the microtidal context of the Neretva River and shows potential for adaptation to mesotidal and macrotidal systems.
\end{abstract}


\maketitle

\doublespacing

\section{Introduction}

Estuaries and tidal rivers are transitional coastal regions characterized by the interaction between riverine and marine processes \citep{geyer2014estuarine}. These systems are influenced by various natural forces, including tidal oscillations, storm surges, and river flow. Additionally, estuaries and coastal rivers are often affected by human activities, such as water management, hydropower operations, flood control, and navigation, all of which are vital for local economies and the sustainability of coastal communities \citep{Talke2020a}.
The complexity of water level dynamics in these environments impacts sediment transport, salinity gradients, nutrient distribution, and ecological habitats \citep{Hoitink2016}. Understanding these hydrodynamic processes is, therefore, essential for predicting extreme water level events and for effective water management.

Microtidal estuaries, with tidal ranges generally less than 2 meters, are particularly sensitive to external forcings. For example, storm surges can temporarily elevate water levels, while dam operations can lead to abrupt changes in river discharge, known as power peaking \citep{Jay2015a}. In such conditions, predicting water levels is challenging, as even small variations in external forcings can alter water levels and disrupt the hydrodynamic balance, increasing the risk of flooding, salt-wedge intrusion, or changes in flow regimes \citep{geyer2011}.

Traditional harmonic analysis (HA), which assumes stationary tidal conditions, is unable to capture the temporal variability of water levels in these dynamic systems \citep{Hoitink2016}. Such analyses are not designed to account for the modulation of tidal parameters by non-astronomical external forcings.
Some early extensions of stationary HA applied short-term analysis using moving windows ranging from a few days to a month. While this approach may be useful, it lacks a clearly defined frequency response \citep{Jay1997a,Jay1999}. \citet{Guo2015} proposed a binned HA method that segments data based on observed river discharge ranges, enabling the resolution of a greater number of tidal constituents. However, this method is most effective in systems where river flow remains relatively stable and proves less reliable in environments with rapid flow variability \citep{Hoitink2016}. 

To overcome these constraints, non-stationary tidal analysis approaches have been proposed, such as complex demodulation \citep{Gasquet1997,Jalon-Rojas2018}, empirical mode decomposition \citep{Pan2018}, and variational mode decomposition \citep{Gan2021}. Other advanced methods include S\_Tide, which utilizes independent point schemes and cubic spline interpolation \citep{Pan2018a}, and continuous wavelet transforms \citep{Jay1997a,Flinchem2000}. These approaches provide different advantages, with some offering higher frequency resolution within the tidal bands \citep[e.g.,][]{Pan2018a,Lobo2024}, and others only resolving the tidal species. While these methods have proven useful in many estuarine studies, their predictive capabilities are limited compared to approaches that can directly determine the response of tides to external influences.

In contrast, NS\_Tide \citep{matte2013,matte2014temporal} has extended the HA model by integrating the non-stationary forcing directly into the solution basis functions. As a result, non-tidal variables describing the influence of river discharge or storm surge can serve as predictors for the time-varying tidal amplitudes and phases, thus enabling predictions when forecasts for the input variables are available. The original model implemented in NS\_Tide originated from \citet{Jay1991}, \citet{kukulka2003impacts, Kukulka2003b}, and \citet{Jay2011} and expresses both mean water levels (or stage) and tides as a function of river flow and ocean tidal range. Several modifications to the original NS\_Tide model have been proposed recently, for example, to include coastal influences \citep{Pan2018a} or to simplify the basis functions \citep{Wu2022a,Cai2023}.

Despite these advancements in the field of non-stationary tidal HA, significant research gaps remain, particularly in understanding the interactions between tides, storm surges, and river flow in microtidal environments. Previous studies mainly focused on mesotidal or macrotidal systems where tides dominate the hydrodynamic processes \citep[e.g.][]{Spicer2019,Spicer2021,Xiao2021}, and comparatively less attention has been paid to microtidal estuaries.
Notable exceptions include studies like \citet{McGrath2023}, who applied wavelet analysis and both stationary and non-stationary HA (NS\_Tide) to decompose water levels in the microtidal Swan River Estuary (Western Australia). This approach separated the contributions from tides, mean sea level, barometric pressure, river flows, and river-tide interactions. Similarly, \citet{Dykstra2022} utilized continuous wavelet transforms to examine the impact of river discharge on tidal interactions within the diurnal microtidal Tombigbee River-Mobile Bay system. Additionally, \citet{Du2024} explored tide-river-surge interactions and their compound flooding implications in the microtidal Pearl River Delta using a combination of hydrodynamic modeling and wavelet analysis.

In microtidal estuaries, storm surges can have a stronger influence on water levels than tides, complicating the tidal-fluvial dynamics and accurate prediction of water levels. Therefore, a robust approach is required to capture the complexity of these physical processes. 
This study addresses this gap by developing a modified non-stationary tidal HA model to investigate the combined effects of storm surges, river flow, and power peaking on tidal-fluvial dynamics in microtidal estuaries. The goals of this study are twofold: (a) to propose a non-stationary tidal HA model specifically designed for microtidal estuaries, and (b) to improve the understanding of how tides interact with storm surges, river flow and power peaking in microtidal environments. We apply and validate this method on the Neretva River estuary in Croatia. 

To support these analyses, a new non-stationary formulation for the NS\_Tide model \citep{matte2013} is proposed, which allows a versatile, site-specific definition of the model's basis functions with user-defined external forcing variables affecting the stage levels and tidal constituents. The software also incorporates recent advances in uncertainty estimation that are well suited for signals with temporally correlated noise \citep{Innocenti2022}.
In addition, we used the two-layer shallow water model STREAM \citep{krvavica2017salt}, specifically developed for the simulation of physical processes in microtidal estuaries, to assess the impact of power peaking and support our conclusions.
With this framework, our application provides insights into tidal-fluvial dynamics potentially involved in extreme events in under-researched microtidal estuaries. 

The remainder of this paper is organized as follows: Section 2 introduces the study area, focusing on the Adriatic Sea and Neretva River estuary. Section 3 outlines the data and methodology, including the non-stationary HA model for tidal-fluvial interactions, error metrics, and the numerical model for simulating water levels. Section 4 presents the spectral analysis, model comparisons, and non-stationary tidal HA, including the water level reconstructions at various stations, the contributions from different forcing mechanisms, and spatial variations of tidal constituents. Section 5 discusses the results supported by additional analysis. Section 6 summarizes findings and suggests future research directions.

\section{Study Area}

The Neretva River is the largest river in the eastern part of the Adriatic Basin. It stretches for 215 km, with most of its course flowing through Bosnia and Herzegovina (BiH). The final 22 km is located in Croatia before the river flows into the Adriatic Sea. The total catchment area of the Neretva River is around 10,240 km$^2$, of which only 280 km$^2$ lies within Croatia \citep{krvavica2021salt}. In its upper and middle reaches, the Neretva is a typical mountain river, but in its lower reaches, it forms a wide alluvial delta covering about 12,000 hectares. This study focuses on the estuarine and tidal river sections of the Neretva River, which span approximately 30 km from the mouth of the river.

\subsection{Adriatic Sea}

The Adriatic Sea, a semi-enclosed basin within the Mediterranean Sea, is characterized by its unique tidal and hydrodynamic properties due to its shape, orientation, and bathymetry. With a length of about 800 km and narrowing from south to north, the geographical configuration of the Adriatic Sea plays a crucial role in modulating tidal amplitudes and storm surges. Tides are mixed, with semi-diurnal to diurnal amplitude ratios of 1 to 2, and amplitude ranges from 30 cm in the south to 1 m in the north \citep{vilibic_adriatic_2017}. The tidal waves entering the southern Adriatic are funneled northward, resulting in amplification due to the narrowing and shallowing of the basin \citep{janekovic_numerical_2005}.The tidal dynamics in the Adriatic are also influenced by the resonance properties of the sea, with the fundamental oscillation period close to 21.5 hours \citep{medvedev_tidal_2020}. This resonance interacts with the tidal components, leading to modulation of the tidal amplitudes. In combination with meteorological conditions, these natural resonances can lead to higher tidal extremes during storm events.

Storm surges, driven by winds and low pressure, especially affect the shallow northern Adriatic, posing flood risks \citep{medugorac_adriatic_2018,sepic_climatology_2022}.
The Scirocco wind and atmospheric pressure can generate storm surges up to 1.5 m \citep{medugorac_two_2016}. Recent studies indicate that tidal dynamics in the Adriatic are becoming increasingly non-stationary due to the combined effects of changes in atmospheric circulation, sea level rise and human intervention in the coastal zone \citep{orlic_response_1994,medugorac_two_2016,medugorac_adriatic_2018}.
Furthermore, sea level extremes, higher in the north, are driven by tides, atmospheric conditions, and local events like seiches and meteotsunamis \citep{sepic_climatology_2022}.

\subsection{Neretva River estuary}

\begin{figure}[ht]
 \centering
     \begin{subfigure}{0.95\textwidth}
        \centering
        \caption{Map}
        \includegraphics[width=\textwidth]{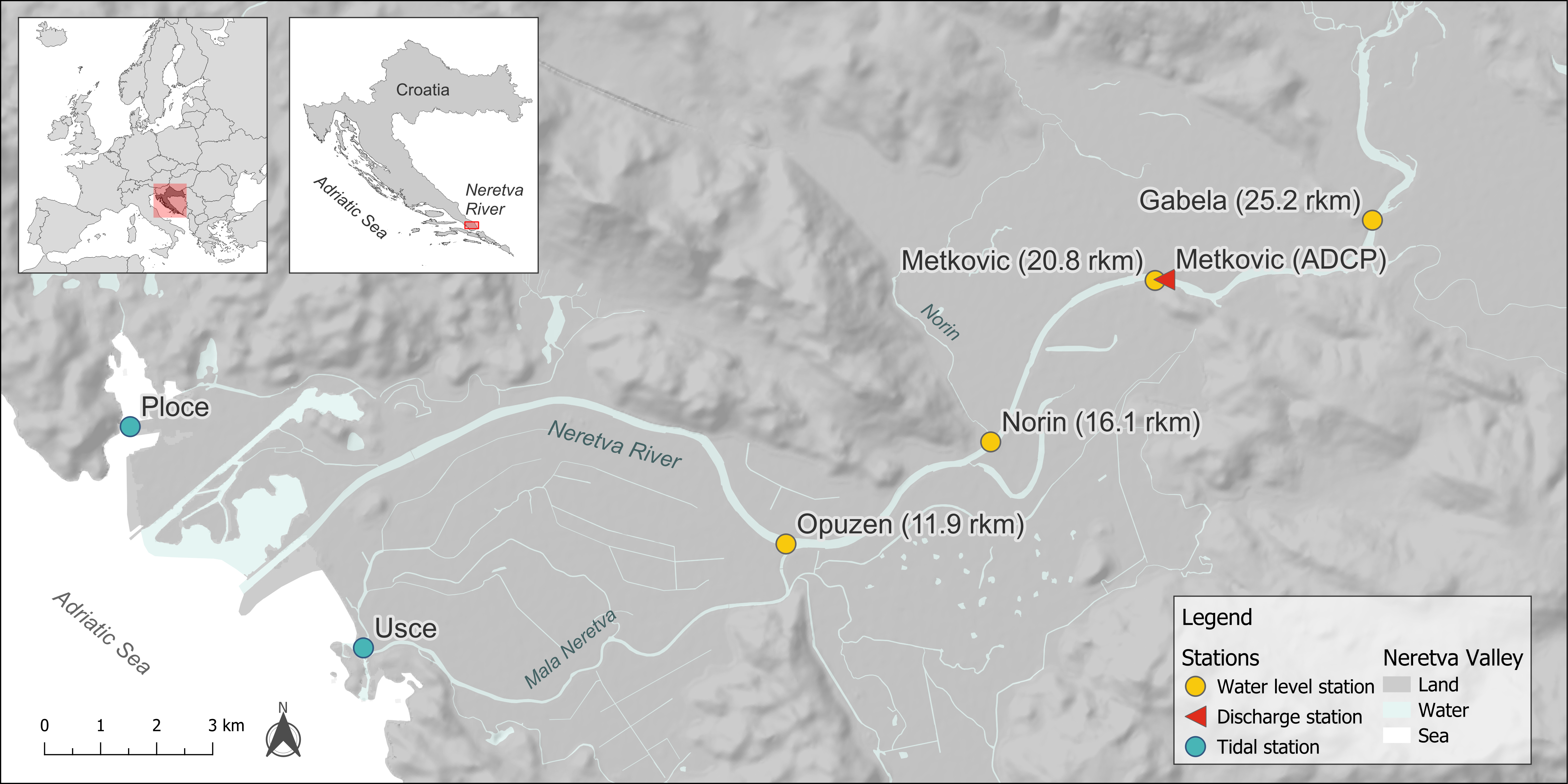}
        \label{fig:map}
    \end{subfigure}
    \begin{subfigure}{0.95\textwidth}
        \centering
        \caption{Longitudinal profile}
        \includegraphics[width=\textwidth]{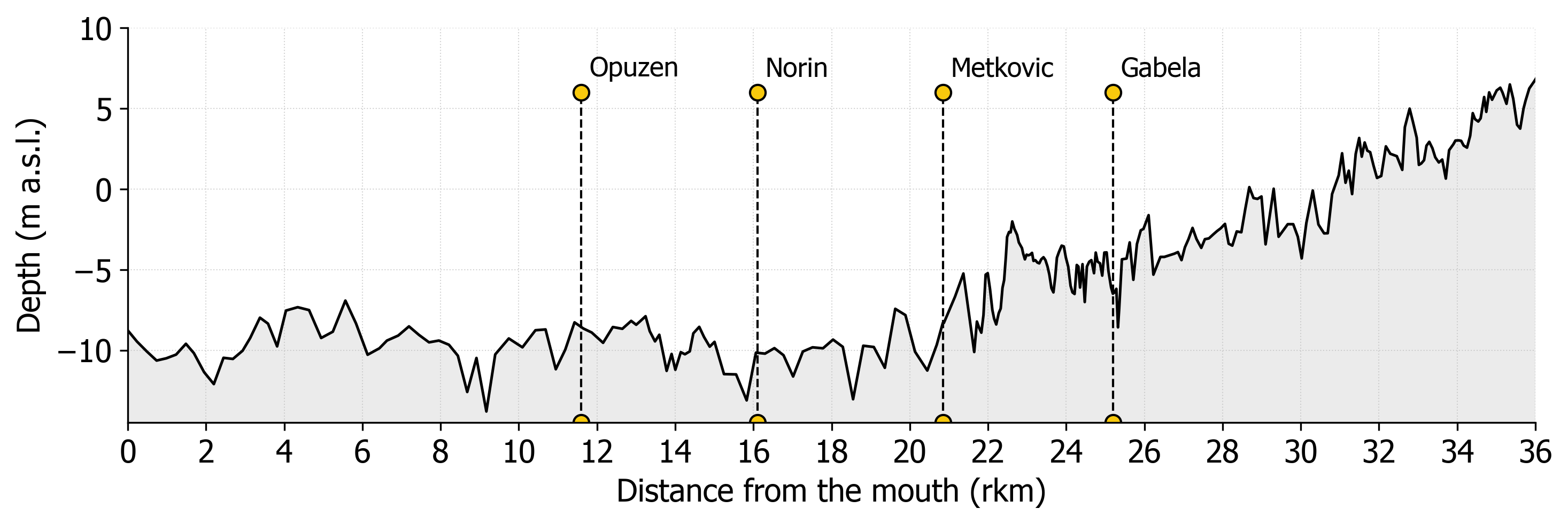}
        \label{fig:profile}
    \end{subfigure}
 \caption{The Neretva River estuary map with tidal, water level and discharge stations and their distance from the river mouth (rkm) (a) and longitudinal channel bottom profile (b).}
 \label{fig:map_profile}
\end{figure}

The Neretva River estuary is shown in Fig.~\ref{fig:map_profile}. A smaller tributary flows into the Neretva near Norin, while the Mala Neretva tributary has been separated from the main river channel and the Adriatic Sea by dams. Today, the Mala Neretva River primarily functions as a retention basin for irrigation purposes. The dynamics of water levels in the estuary of the Neretva River are governed by the non-linear interaction between sea levels of the Adriatic Sea and upstream river inflows, which are influenced by the operation and management of a system of reservoirs and dams located in BiH.

The estuarine zone, characterized by a relatively uniform bottom depth, extends over the first 22 kilometers of the river (see Fig.~\ref{fig:profile}). This section is influenced by both tides and freshwater flows. Upstream of Metkovic, the bottom of the channel becomes steeper and shallower, marking the beginning of the transitional tidal river zone. This zone extends to about 33 kilometers upstream, where the influence of tides gradually decreases. Beyond this point, the river transitions into a typical riverine regime predominantly governed by fluvial processes.
Due to the microtidal environment and weak tidal forces, the estuary of the Neretva River remains highly stratified throughout the year. This estuary is characterized by a salt wedge that intrudes beyond Metkovic in summer and is completely flushed out during periods of high river flow \citep{krvavica2021salt}.

The study area is well covered by a network of tidal and hydrological monitoring stations (Fig.~\ref{fig:map_profile}). Two tidal stations are operated near the mouth of the river: Ploce, managed by the Croatian Hydrographic Institute (HHI), and Usce, managed by the Croatian Hydrological and Meteorological Service (DHMZ). Both stations are unaffected by the river flow. Further upstream, the stations at Opuzen (11.9 rkm), Norin (16.1 rkm) and Metkovic (20.8 rkm), which are all managed by the DHMZ, measure water levels that are influenced by both sea level and river flow. The most upstream station Gabela (25.2 rkm), managed by the Agency for Watershed of the Adriatic Sea in BiH, measures the water level, which is still influenced by the sea level, but with the river flow becoming more dominant. The Capljina station (32 rkm) upstream of Gabela is weakly affected by the tides only during low flow. Due to issues with the measuring device, which caused inconsistencies in the reference level, the data from this hydrological station were not included in our analysis.

Although water level measurements have been recorded in the estuary since 1957, systematic discharge measurements were only introduced in 2015 after horizontal Acoustic Doppler Current Profiler (ADCP) devices were installed at the Metkovic station. The nearest upstream discharge station is located in Zitomislici, about 48 km upstream of the mouth, and outside the tidal influence. However, this station is not representative of the flow within the estuary due to its distance from Metkovic and tributary inflows.

The discharge of the Neretva has a pronounced seasonality typical of the Mediterranean climate, with a high-flow season from October to April and a low-flow period from May to September (see Fig.~\ref{fig:waterlevels}). In summer, rainfall and river inflows are at their lowest, allowing saltwater to regularly intrude more than 20 km upstream \citep{krvavica2021salt}. Recent short-term data (2015-2021) from the Metkovic station show that the mean annual flow rate in the Neretva River estuary is 325 m$^3$/s, with a maximum flow of 1375 m$^3$/s measured in February 2021. The minimum flow rate is regulated by an international agreement between Croatia and BiH, according to which the Mostar hydropower plant must discharge at least 50 m$^3$/s in summer.

\section{Data and methodology}

\subsection{Hydrological and tidal data}

\begin{figure}[ht]
 \centering
 \includegraphics[width=0.95\textwidth]{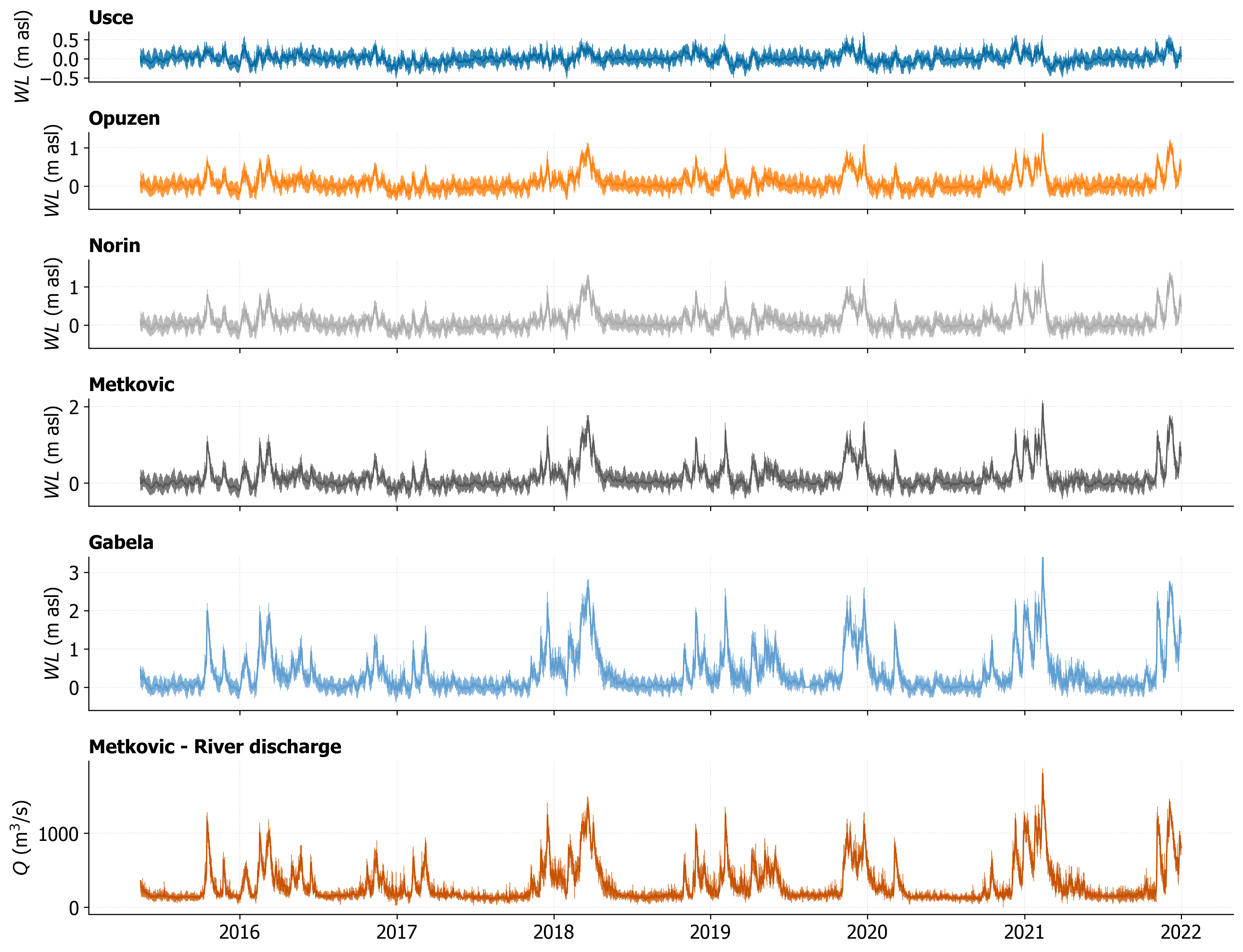}
 \caption{Water level and discharge time series at different stations measured in the Neretva River (2015-2021).}
 \label{fig:waterlevels}
\end{figure}

The dataset consists of hourly water level series measured at five stations along the Neretva River — Usce, Opuzen, Norin, Metkovic, and Gabela — and one discharge station, Metkovic, covering the period from June 2015 to December 2021. To compare the performance of the models, each time series was divided into a training period (January 2017 to December 2021) and a validation period (June 2015 to December 2016). Fig.~\ref{fig:waterlevels} shows the available time series at different stations. We observe that the water level (\(WL\)) peaks during high flow, which typically occurs in winter, and the water surface elevation increases in the upstream direction. For instance, the peak values of $WL$ at Gabela exceed 3.0 m a.s.l., while the maximum $WL$ at Opuzen barely exceed 1.0 m a.s.l. The hourly oscillations appear relatively similar across all stations at this scale.

A detail of the one-year time series of daily water levels (\(WL\)) and sub-daily residuals (\(WL_{res}\)) is presented in Fig.~A.1 in the Supplementary Materials for each station. The residuals were computed by subtracting a 24-hour moving average from the hourly time series to estimate short-term fluctuations. 
During low flow, uniform high-frequency patterns in residual water levels are observed at all stations. In contrast, high flow, especially in winter, changes both the water level and the residual amplitude at the upstream stations, where the residual signal becomes irregular and less periodic.
Changes in residual patterns during high flow emphasize the impact of the river discharge, which disrupts the regular tidal pattern and amplifies the short-term oscillations. In contrast, stations closer to the river mouth, such as Usce and Opuzen, show a more subtle response with less pronounced residual variability.
In Fig.~A.2 in the Supplementary Materials, we also show a detailed time series of the measured river discharges (\(Q\)) and the residual discharges (\(Q_{res}\)) in Metkovic for the same year. During high flow, a significant increase in the amplitude of residual discharge oscillations can be observed, with amplitudes reaching around 300 m\(^3\)/s.

\subsection{Non-stationary analysis and model definition}
\label{sec:model_definition}

In classical HA, total water levels $\eta(t)$ are typically modeled as a sum of sine waves, as follows \citep{Godin1972,pawlowicz2002classical}:
\begin{equation}
    \eta(t) = \eta_0 + \sum_{k=1}^{n} \left[ c_{k} \cos(\omega_k t)  + s_{k} \sin(\omega_k t) \right] + \epsilon(t),
    \label{eq:HA}
\end{equation}
where $\eta_0$, $c_k$, and $s_k$ are regression parameters, $\omega_k$ is the a priori known frequency of the $k$-th tidal constituent, $t$ is time, $n$ is the number of tidal constituents, and $\epsilon(t)$ is the error term.

In a non-stationary context, \cite{matte2013} expressed the harmonic model parameters as functions of time-varying processes, defining in a general form $\eta(t)$ at a given station, as follows:
\begin{equation}
    \eta(t) = S(t) + F(t) = \underbrace{S(t)}_{\text{Stage term}} + \underbrace{\sum_{k=1}^{n} \left[ c_k(t) \cos(\omega_k t) + s_k(t) \sin(\omega_k t) \right] }_{\text{Tidal-fluvial term}} + \epsilon(t),
    \label{eq:main}
\end{equation}
where $S(t)$ is the stage term that accounts for the subtidal fluctuations at frequencies lower than 0.03 cph, 
and $F(t)$ is the tidal-fluvial term that reproduces the water level variability at higher tidal frequencies (>0.03 cph) by considering temporally varying coefficients $c_k(t)$ and $s_k(t)$ for the harmonic regression cosine and sine functions. 

In the original NS\_Tide formulation, \citet{matte2013} defined the stage and tidal-fluvial terms of Eq.~\ref{eq:main} based on the analysis of \citet{kukulka2003impacts,Kukulka2003b}. Specifically, $S(t)$, $c_k(t)$, and $s_k(t)$ were expressed as functions of the river discharge $Q$ and diurnal tidal range $R$ at a station unaffected by fluvial influence, as follows:
\begin{equation}
    S(t) = a_0 + a_1 Q^{2/3}(t - \tau_Q) + a_2 \frac{R^{2}(t - \tau_R)}{Q^{4/3}(t - \tau_Q)}, \\
\label{eq:stage_model}
\end{equation}
and
\begin{equation}
   \begin{aligned}
    c_k(t) = c_{k,0} + c_{k,1} Q(t - \tau_Q) + c_{k,2} \frac{R^{2}(t - \tau_R)}{Q^{1/2}(t - \tau_Q)}, \\
    s_k(t) = s_{k,0} + s_{k,1} Q(t - \tau_Q) + s_{k,2} \frac{R^{2}(t - \tau_R)}{Q^{1/2}(t - \tau_Q)},
    \end{aligned}
\label{eq:tidal_model}
\end{equation}
where $a_i$ are regression parameters for the stage term, $c_{k,i}$ and $s_{k,i}$ are regression parameters for the tidal-fluvial term, and $\tau_Q$ and $\tau_R$ are time lags applied to the $Q(t)$ and $R(t)$ time series, respectively. The regression parameters $a_i$, $c_{k,i}$, and $s_{k,i}$, $i=0,1,2$, can be estimated through Iteratively Reweighted Least-Squares (IRLS, \citet{leffler_enhancing_2009}) as described in \citet{matte2014temporal}. 

The $a_0$ parameter in Eq.~\ref{eq:stage_model} primarily represents the impact of channel geometry at a specific location; the second and third terms represent the stage response to river discharge and coastal tidal range. Similarly, the second and third terms of Eq.~\ref{eq:tidal_model} reproduce the nonlinear response of the harmonic regression parameters to the lagged river flow and coastal tidal range, the latter representing 
the effects of frictional interaction due to neap-spring variations in mean water levels and tides. However, the third term of Eqs.~\ref{eq:stage_model} and \ref{eq:tidal_model} is expected to be inadequate in microtidal estuaries, where the amplitude of storm surge events dominates the tidal range. Therefore, our hypothesis is that replacing the tidal range term by a storm surge time series $SS(t)$ is more suitable in microtidal environments.

In Eqs. \ref{eq:stage_model}-\ref{eq:tidal_model}, the discharge and range exponents were chosen following \citet{kukulka2003impacts,Kukulka2003b} based on a critical convergence regime \citep{Jay1991}, where tidal and fluvial flows have similar magnitude and channel convergence is moderate. 
In situations where these conditions are not fully met, it has been shown that a more pragmatic approach is to tune the exponents empirically at each station along an estuary \citep{Jay2011}. 
However, this typically involves nonlinear optimization on observed data  \citep{matte2013,Cai2023}, which can yield large uncertainty in the estimated exponent values. 

Using a more efficient approach, in our application, we defined the $S(t)$ and $F(t)$ functional forms based on the signal reconstructions provided by Generalized Additive Models (GAM; \citet{hastie_generalized_1986,wood_generalized_2017}) at each specific station. 
These semi-parametric reconstructions confirmed that the theoretical exponent 2/3 derived by \cite{kukulka2003impacts} may be inadequate in Eq.~ \ref{eq:stage_model} for the Neretva River estuary. A quadratic relationship better describes the discharge influence on the stage term at the considered stations (Fig.~B.1 in the Supplementary Material). 
Similarly, applying the GAM approach to the range of higher frequency water levels suggested that linear relationships are appropriate in the $F(t)$ term across a range of $Q$ and $SS$ values encompassing most of their distribution (Fig.~A.3 and B.2 in the Supplementary Material). 
Hence, considering our preliminary analyses and the mentioned theoretical considerations on microtidal systems, three new NS\_Tide formulations were tested in the present study, each of them involving linear and quadratic storm surge $SS(t)$ and discharge $Q(t)$ predictors.  

First, to account for the nonlinear relationship between the river stage and discharge, we define the qNS\_Tide model that uses a quadratic function of $Q(t)$ in Eq.~\ref{eq:stage_model}:
\begin{equation}
    S(t) = a_0 + a_1 Q(t - \tau_Q) +  a_2 Q^{2}(t - \tau_Q) + a_3 \frac{R^{2}(t - \tau_R)}{Q^{4/3}(t - \tau_Q)}. \\
\label{eq:stage_mode3l}
\end{equation}

Second, we define the sNS\_Tide model that replaces the tidal range in Eqs.~\ref{eq:stage_model} and \ref{eq:tidal_model} by a lagged storm surge term $SS(t - \tau_{SS})$. To this end, $SS(t)$ is computed as a low-passed residual of the stationary harmonic tidal analysis at a coastal tidal station unaffected by fluvial influences:
\begin{equation}
    S(t) = a_0 + a_1 Q^{2/3}(t - \tau_Q) + a_2 SS(t - \tau_{SS}), 
\label{eq:stage_model2}
\end{equation}
and
\begin{equation}
   \begin{aligned}
    c_k(t) = c_{k,0} + c_{k,1} Q(t - \tau_Q) + c_{k,2} SS(t - \tau_{SS}), \\
    s_k(t) = s_{k,0} + s_{k,1} Q(t - \tau_Q) + s_{k,2} SS(t - \tau_{SS}).
   \end{aligned}
\label{eq:tidal_model2}
\end{equation}

Finally, we define the $\mu$NS\_Tide (microtidal NS\_Tide) 
model that considers both modifications in Eqs.~\ref{eq:stage_model} and \ref{eq:tidal_model}:
\begin{equation}
    S(t) = a_0 + a_1 Q(t - \tau_Q) +  a_2 Q^{2}(t - \tau_Q) +  a_3 SS(t - \tau_{SS}), 
\label{eq:stage_model4}
\end{equation}
and
\begin{equation}
   \begin{aligned}
     c_k(t) = c_{k,0} + c_{k,1} Q(t - \tau_Q) + c_{k,2} SS(t - \tau_{SS}), \\
     s_k(t) = s_{k,0} + s_{k,1} Q(t - \tau_Q) + s_{k,2} SS(t - \tau_{SS}).
   \end{aligned}
\label{eq:tidal_model4}
\end{equation} 

In this way, we limited the number of parameters in each model and retained the linear formulations of $c_k$ and $s_k$, consistent with the original \citet{kukulka2003impacts}'s formulation. Tab.~\ref{tab:modelsummary} summarizes the characteristics of these four models used in our analysis.

First, $R$, $Q$, and $SS$ were filtered using a Butterworth low-pass filter \citep{roberts1978use} with a cutoff frequency of 0.03 cph to remove the higher frequency components.
Then, the optimal covariate lags were estimated for each model by maximizing the cross-correlation of lagged $Q$, $R$, and $SS$ with the observed water level.
Next, the $n_{\beta}$ unknown non-stationary harmonic regression parameters (see Tab.~\ref{tab:modelsummary}) were estimated through the IRLS with the default Cauchy weight function and tuning constant. 
A total of 62 tidal constituents with frequencies higher than 0.003 cph were initially considered in each tidal-fluvial model term by excluding nodal corrections from the astronomical arguments. 
The constituents with a non-stationary amplitude Signal-to-Noise Ratio (SNR; ratio between a constituent amplitude and its standard error) greater than 2 for at least 50\% of the reconstructed time steps were retained for the analysis.
To this end, the regression coefficient variances were computed at each time step following the \textit{'correlated' NS\_Tide} procedure initially proposed by \citet{matte2014temporal} and analytically revised as the solution of a Generalized Least Squares (GLS) model 
in \citet{Innocenti2022}. For the selected tidal constituents, the NS\_Tide package also computes the tidal amplitude standard errors and phase circular variance by following the procedure detailed in \citet{Innocenti2022} and references therein.

\begin{table}[H!]
\caption{Model summary: Covariates and corresponding number of regression coefficients estimated for each model when considering 62 tidal constituents with frequency higher than 0.003 cph}
\begin{tabular}{rcccc}
\toprule
&  Stage         & Tidal-Fluvial &   No. regression \\
& $S(t)$         &  $F(t)$       &   coefficients $n_{\beta}$  \\
 \cline{2-4}\\
  NS\_Tide &  $Q^{2/3}$, $R^{2}/Q^{4/3}$ & $Q$, $R^{2}/Q^{1/2}$  & 3 + 6$\cdot$62 = 375 \hspace{0.1cm}\\
 qNS\_Tide &  $Q$, $Q^{2}$, $R^{2}/Q^{4/3}$ & $Q$, $R^{2}/Q^{1/2}$ &   4 + 6$\cdot$62 = 376  \hspace{0.1cm}\\
 sNS\_Tide &  $Q^{2/3}$, $SS$  & $Q$, $SS$   &  3 + 6$\cdot$62 = 375  \hspace{0.1cm}\\
 $\mu$NS\_Tide &  $Q$, $Q^{2}$, $SS$  &  $Q$, $SS$  & 4 + 6$\cdot$62 = 376 \hspace{0.1cm}\\
 \bottomrule \hspace{0.1cm}\\
 \end{tabular}
 \label{tab:modelsummary}
\end{table}

\subsection{Error metrics}
Each model presents a different approach for including tides, storm surges, and river flows. The model parameters and coefficients were determined using the training set (January 2016 to December 2021). We then evaluated the performance of each model using the reconstructed time series for the validation period (June 2015 to December 2016). For each reconstruction, we computed the proportion of residual variance
(ratio between residual and observed variance) and the root mean square error (RMSE) for each reconstruction to compare the fit of the model to observed data, and the Akaike Information Criterion (AIC) to weight the prediction error with the number of model parameters and allow comparison of the different NS\_Tide formulations.

\subsection{Contribution of different influences}

To assess the relative importance of each model term, we considered the partial water level reconstruction obtained separately for the stage, $S(t)$, and the tidal-fluvial terms, $F(t)$, of the $\mu$NS\_Tide model, as well as the residual. 
For these partial reconstructions, we computed the proportions of explained variance as follows:
\begin{equation}
    v_{S} = \frac{\text{var} \left[ S(t) \right]}{\text{var} \left[ \eta_{obs}(t) \right]}, \quad v_{F} = \frac{\text{var} \left[ F(t) \right]}{\text{var} \left[ \eta_{obs}(t) \right]}, \quad v_{res} = \frac{\text{var} \left[ \eta_{res}(t) \right]}{\text{var} \left[ \eta_{obs}(t) \right]},
    \label{eq:var_explained}
\end{equation}
where $\eta_{obs}(t)$ are the observed water levels, and $\eta_{res}(t) = \eta_{obs}(t) - S(t) - F(t)$ are the residuals.

Furthermore, to assess the individual contribution of each forcing (river, storm surge, tides, tide-river interaction and tide-surge interaction), we computed the corresponding partial reconstructions of $S(t)$:
\begin{equation}
    S_Q(t) = a_1 Q(t - \tau_Q) + a_2 Q^2(t - \tau_Q), \quad S_{SS}(t) = a_3 SS(t - \tau_S),
\label{eq:St_reconstructions}
\end{equation}
and $F(t)$:
\begin{equation}
   \begin{aligned}
    & F_T(t) = \sum_{k=1}^{n} \left[ c_{k,0} \cos(\omega_k t) + s_{k,0} \sin(\omega_k t) \right], \\
    & F_Q(t) = \sum_{k=1}^{n} \left[ c_{k,1} Q(t - \tau_Q) \cos(\omega_k t) + s_{k,1} Q(t - \tau_Q) \sin(\omega_k t) \right], \\
    & F_{SS}(t) = \sum_{k=1}^{n} \left[ c_{k,2} SS(t - \tau_{SS}) \cos(\omega_k t) + s_{k,2} SS(t - \tau_{SS}) \sin(\omega_k t) \right].     
   \end{aligned}
\label{eq:Ft_reconstructions}
\end{equation}

\subsection{Numerical model}

In addition to non-stationary HA, we also use a numerical model to support several arguments, which are presented in Section \ref{sec:discussion}. Discussion. The STRatified EstuArine Model (STREAM) is used to simulate hydraulic variables, water levels and discharges, along the Neretva River estuary. STREAM is a one-dimensional, time-dependent numerical model developed specifically for microtidal estuaries \citep{krvavica2017salt}.
This model has already demonstrated good performance in modelling the two-layer flow dynamics in the Rjecina and Neretva rivers \citep{krvavica2017salt, krvavica2021salt, krvavica2020assessment}.
Although relatively simple, it effectively captures the dominant hydraulic processes that occur in two-layer flow configurations in salt-wedge estuaries.
Specific details about the calibration, setup, and efficiency of the model can be found in the earlier study by \cite{krvavica2021salt}. After calibration, the simulated data set reproduces well the measured water levels at all four stations considered, as shown in \cite{krvavica2021salt}.

\section{Results}

\subsection{Spectral analysis of water levels}

\begin{figure}[ht]
 \centering
 \includegraphics[width=0.95\textwidth]{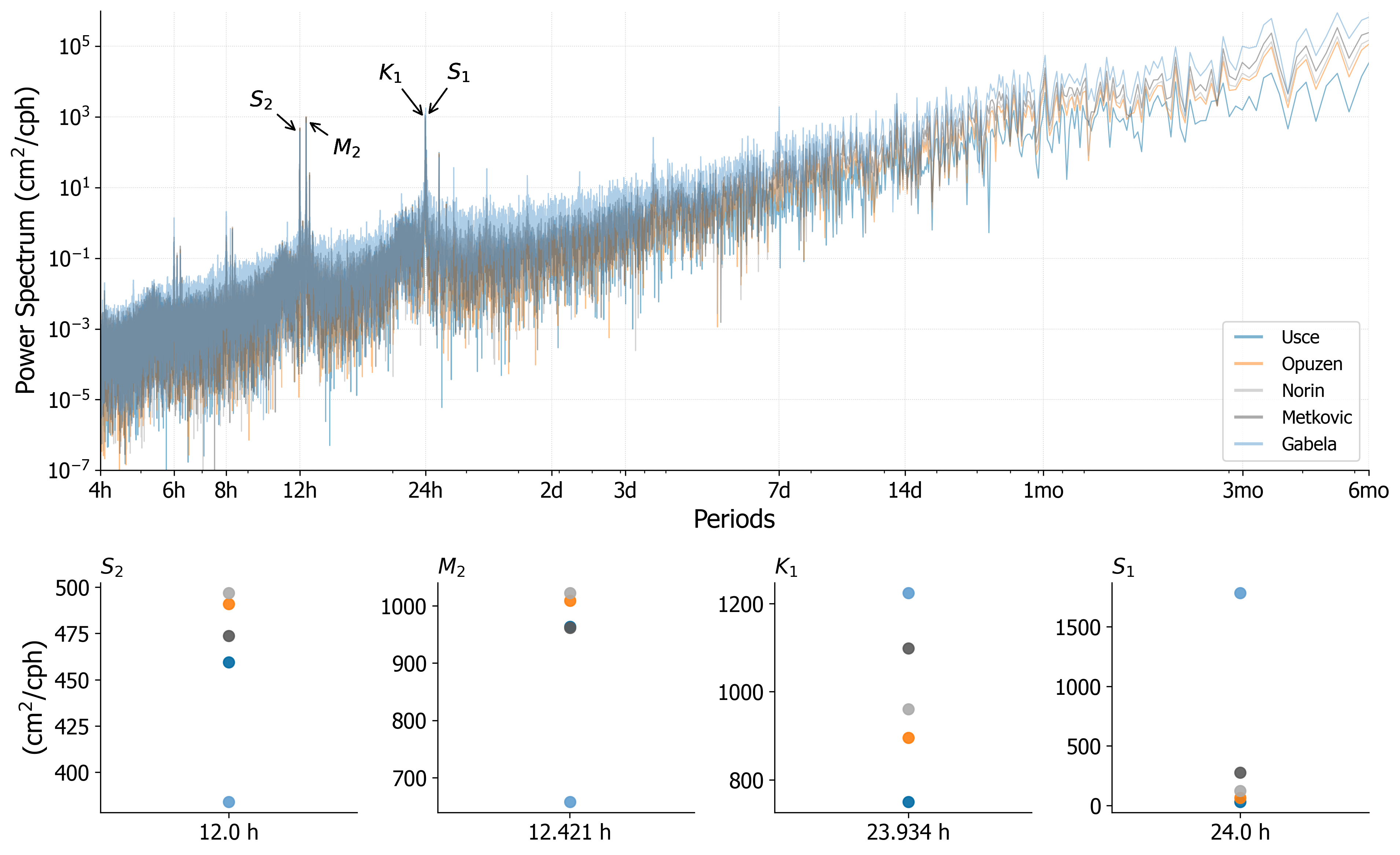}
 \caption{Power spectrum of water levels at different stations with details of the power spectrum for the main diurnal and semi-diurnal tidal harmonics.}
 \label{fig:spectral}
\end{figure}

Fig.~\ref{fig:spectral} shows the spectral properties of the water levels at the five stations — Usce, Opuzen, Norin, Metkovic and Gabela — during the observed period (June 2015 to December 2021). In the upper panel, the spectral power density is plotted against periods from 4 hours to 6 months. The spectrum is dominated by low frequencies, which is due to the strong seasonal and inter-annual variability. Spectral power at low frequencies (e.g. seasonal) is lower near the river mouth and increases at upstream stations. Spectral peaks occur at higher frequencies, especially in the diurnal and semi-diurnal frequency bands. The most prominent peaks of these higher frequencies occur at 12.0 hours (corresponding to \(S_2\)), 12.42 hours (corresponding to \(M_2\)), 23.93 hours (corresponding to \(K_1\)), and 24.0 hours (corresponding to \(S_1\)). In the lower section of Fig.~\ref{fig:spectral}, the detailed plots focus on these important tidal frequencies.

\(S_2\) and \(M_2\) semi-diurnal constituents show similar behavior, with uniform peaks at all stations, except at the upstream station Gabela, where they are slightly attenuated. This indicates a consistent semi-diurnal influence along the river up to the Gabela station, with decreased strength beyond Metkovic. On the other hand, the power of the diurnal constituents is much more variable between the stations. The \(K_1\) shows a gradual and consistent amplification from downstream to upstream stations. The \(S_1\) also shows a gradual increase in power upstream, with a particularly strong peak at Gabela.

We argue that this amplification of the \(S_1\) constituent is more likely due to variations in upstream discharge (due to power peaking) rather than the tides. This is supported by the spectral analysis of river discharges, not only at the Metkovic station but also at the Zitomislici station, located about 48 km upstream and completely outside the tidal influence, where the 24-hour period dominates. The analysis shows that the river discharges at both stations exhibit spectral peaks in 12- and 24-hour periods, with the diurnal power level being much higher than the semi-diurnal one (see Fig.~C.1 in the Supplementary Materials). Further non-stationary tidal HA will investigate how these diurnal and semi-diurnal constituents vary in time and space.

\subsection{Performance of the proposed models}

\begin{figure}[ht]
 \centering
 \includegraphics[width=0.95\textwidth]{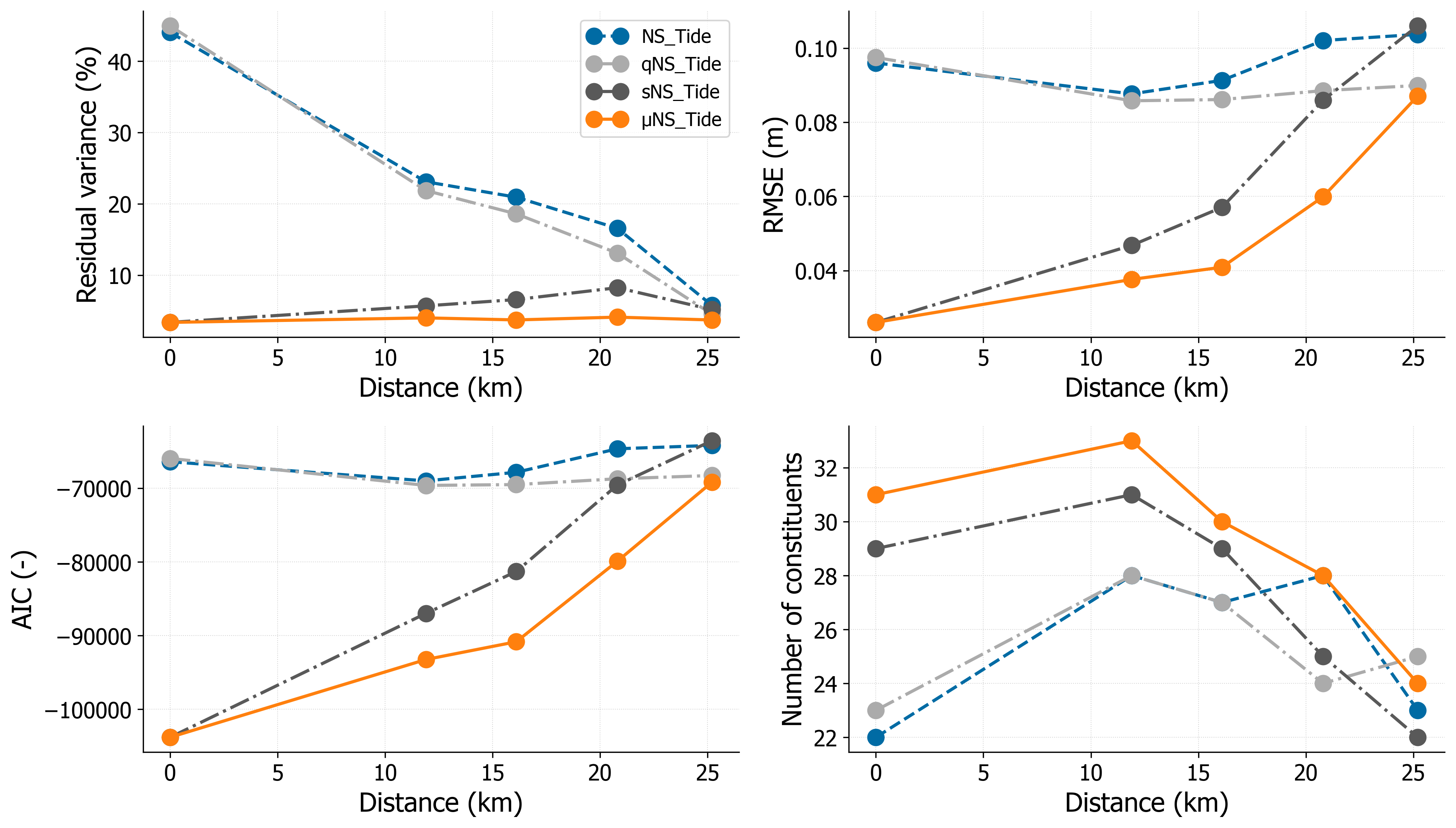}
 \caption{Statistics on water level reconstructions using the four definitions of non-stationary model for the validation period (June 2015 to December 2016): residual variance, RMSE, AIC, and number of selected tidal constituents. Station locations: Usce (0 rkm), Opuzen (11.9 rkm), Norin (16.1 rkm), Metković (20.8 rkm), and Gabela (25.2 rkm).}
 \label{fig:model_comparison}
\end{figure}

Fig.~\ref{fig:model_comparison} compares the four non-stationary models — NS\_Tide, sNS\_Tide, qNS\_Tide, and $\mu$NS\_Tide described in Section \ref{sec:model_definition} — across the five stations based on the residual variance, RMSE, and AIC of the reconstructions provided over the validation period (June 2015 to December 2016) and the number of tidal constituents selected after applying the SNR criterion.

The NS\_Tide model, which represents the original definition of stage and tidal-fluvial terms, has the highest errors. The residual variance remains consistently high, starting at above 40\% for the downstream stations and gradually decreasing upstream to around 5\%. Similarly, NS\_Tide has the largest RMSE, between 8 and 11 cm at most stations. The AIC confirms this result, although NS\_Tide has the lowest number of selected constituents at all stations except Metkovic. These results suggest that NS\_Tide has difficulty effectively capturing the non-stationary aspects of tidal and river discharge influences in the presence of a strong storm surge component.

qNS\_Tide, which introduces a quadratic term of river discharge instead of the power function in the stage term of NS\_Tide, shows slightly improved performance over NS\_Tide at upstream stations. 
RMSE and AIC remain similar to NS\_Tide at downstream stations, with slight improvements upstream. This suggests that the quadratic term of qNS\_Tide better captures the nonlinear behavior of river discharge at stations where this variable has a greater influence. This also shows that a fixed exponent for the discharge term in the stage model is not valid at all stations. Instead of opting for a station-by-station optimization of the exponent \citep[e.g.][]{Jay2011}, the definition of a quadratic function of the discharge allows comparisons between stations or spatial interpolations \citep[e.g.][]{matte2014temporal}. Interestingly, qNS\_Tide selects almost the same number of constituents as NS\_Tide and even less for Metkovic.

The sNS\_Tide model, which replaces the tidal range $R$ with storm surge $SS$, performs significantly better than the original NS\_Tide and qNS\_Tide and shows lower errors, especially at the downstream stations. At these stations sNS\_Tide also selects more tidal constituents than NS\_Tide, indicating that $SS$ improves the accuracy of predictions.  

e model $\mu$NS\_Tide, which combines both the storm surge term and the quadratic river discharge term, achieves the best overall performance. The residual variance is significantly reduced across all stations, ranging between 2\% and 3\%. $\mu$NS\_Tide also shows the lowest RMSE and AIC across all stations, consistently outperforming the other models. The number of selected tidal constituents in $\mu$NS\_Tide decreases progressively upstream, suggesting that $\mu$NS\_Tide efficiently simplifies the representation of tides when river discharge dominates. In addition, $\mu$NS\_Tide selects the most tidal constituents of all models. These results suggest that $\mu$NS\_Tide can capture both tidal and river dynamics more effectively than any other model through the combined inclusion of storm surge and quadratic discharge terms. Overall, $\mu$NS\_Tide proves to be the most suitable model for this microtidal environment.
Therefore, we retained this model for further analysis.

\subsection{Reconstructed time series}

\begin{figure}[htbp]
 \centering
 \includegraphics[width=0.85\textwidth]{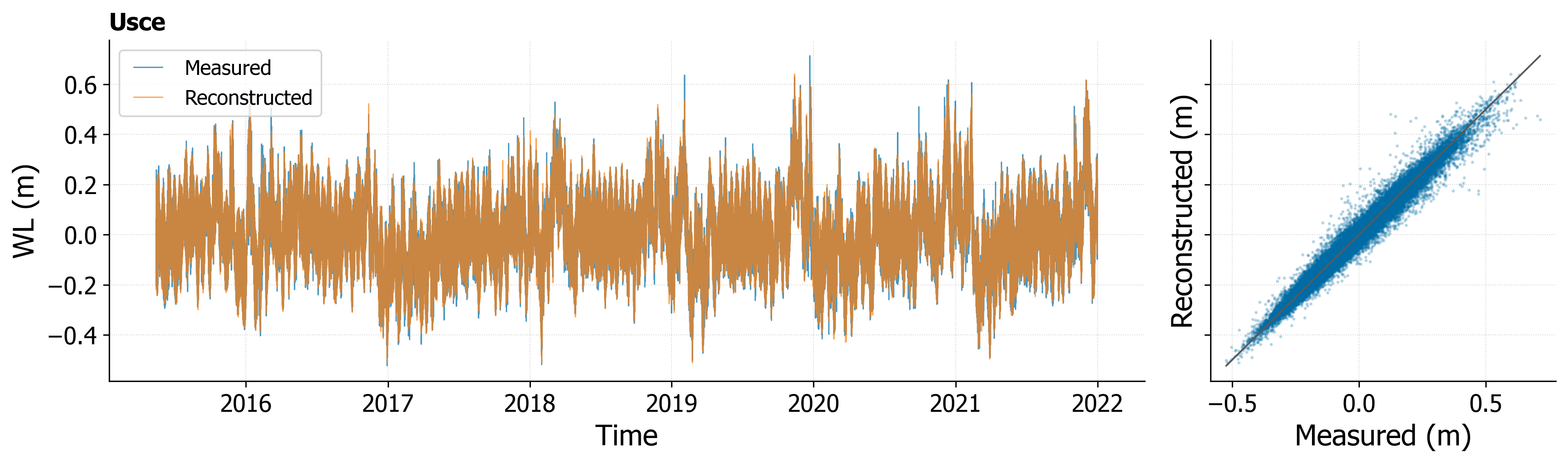}
 \includegraphics[width=0.85\textwidth]{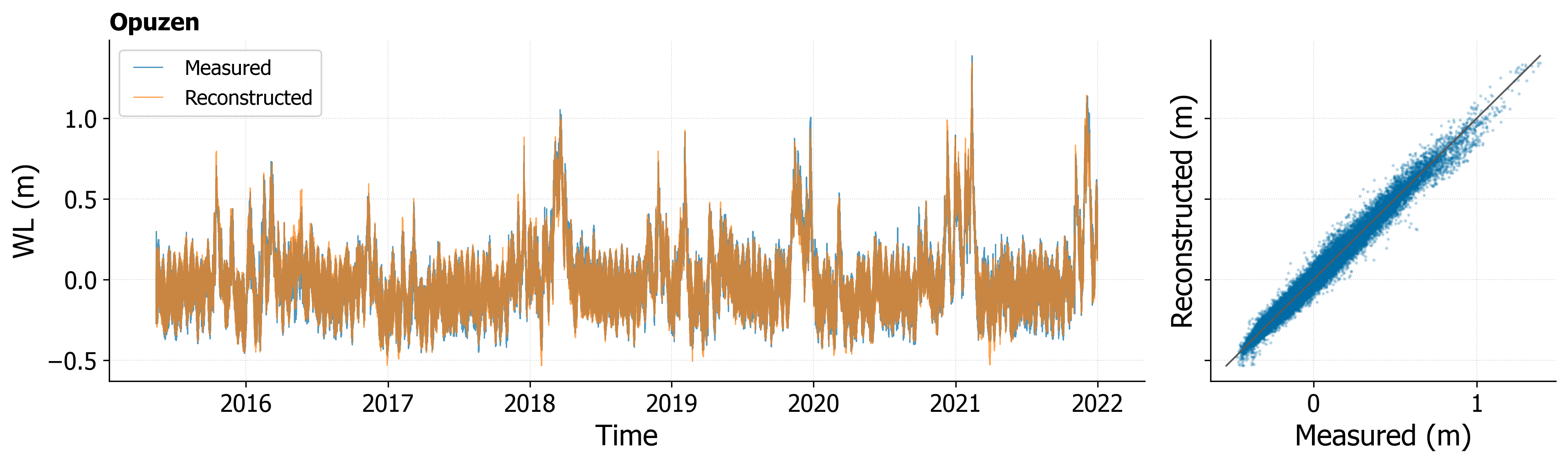}
 \includegraphics[width=0.85\textwidth]{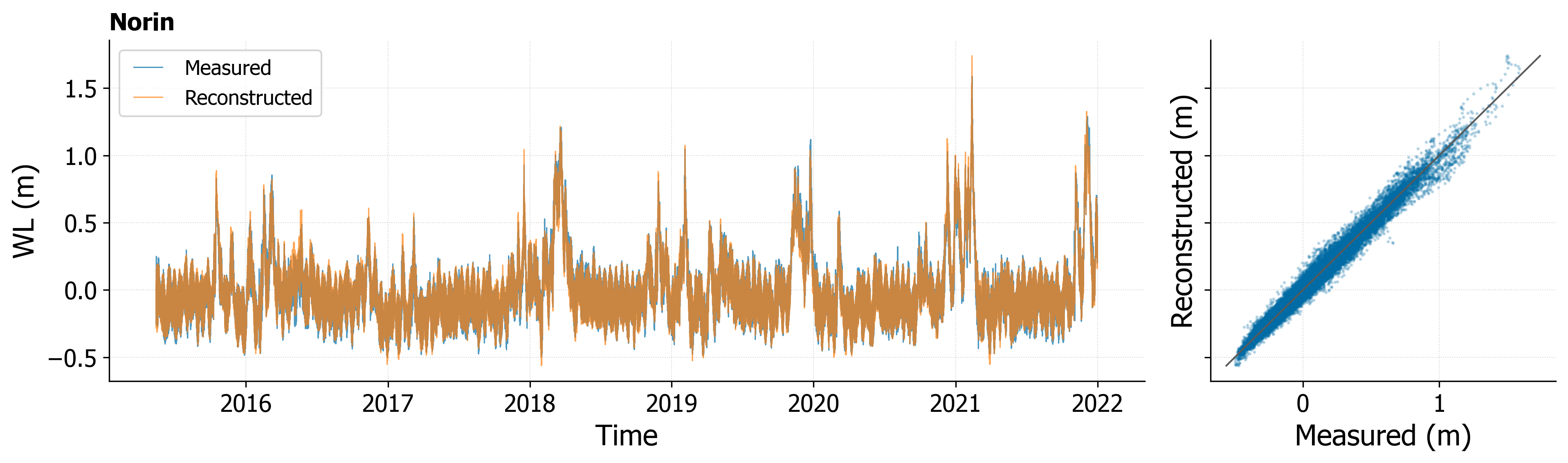}
 \includegraphics[width=0.85\textwidth]{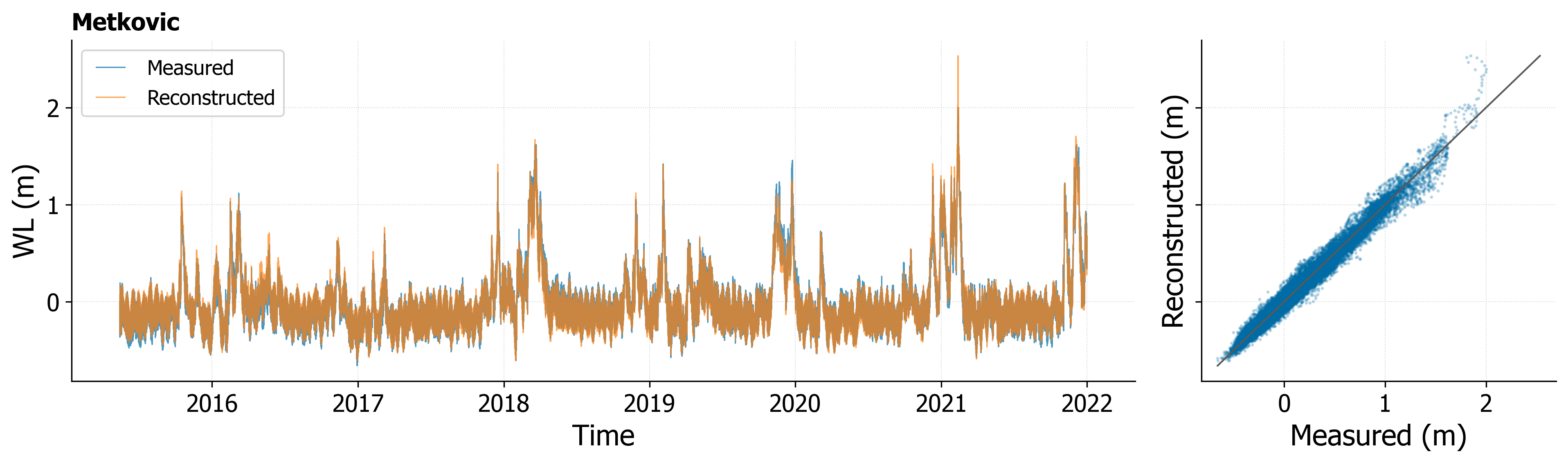}
 \includegraphics[width=0.85\textwidth]{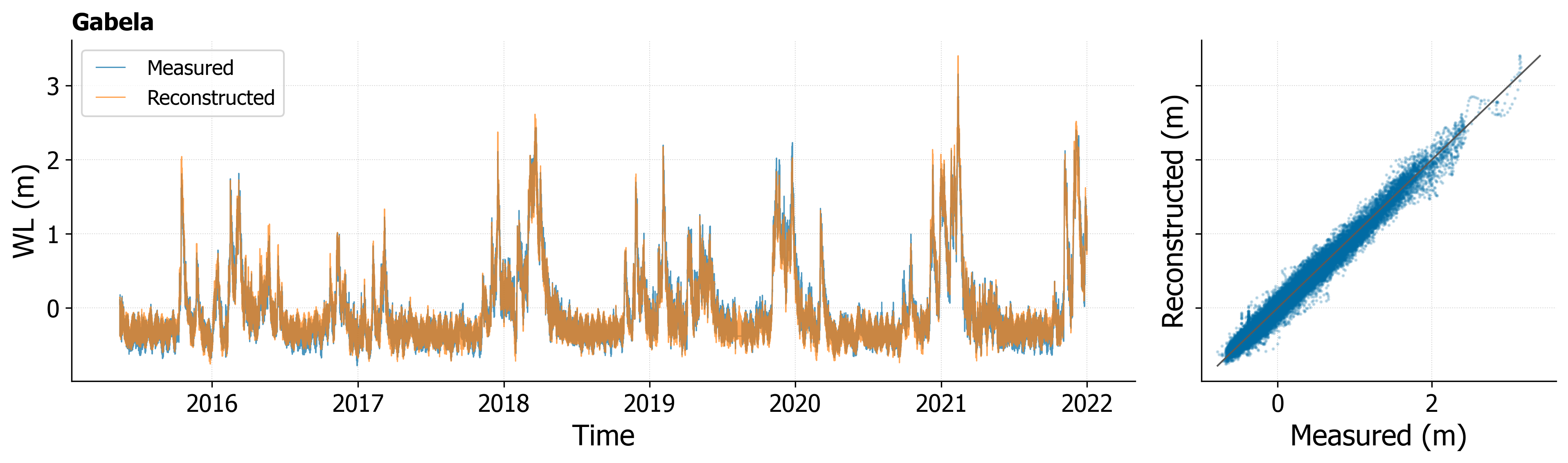}
 \caption{Reconstruction of the water levels using $\mu$NS\_Tide with scatter plots at different stations in the Neretva River for the period June 2015 to December 2021.}
 \label{fig:water_level_reconstruction}
\end{figure}

Fig.~\ref{fig:water_level_reconstruction} presents the water level reconstructions at five stations over the entire period (June 2015 to December 2021), using the $\mu$NS\_Tide model. The left panel in each subplot shows the time series comparison between the measured and reconstructed water levels. The right panel provides scatter plots of the measured versus reconstructed values at each station.

At all stations, the $\mu$NS\_Tide model demonstrates a strong ability to predict the measured water levels. The reconstructed series closely follow the observed data, with notable accuracy in capturing both the high and low water levels. The model performs well at both downstream and upstream stations such as Usce, where the tidal influence is strongest, and Gabela, where river discharge dominates.
The right-side scatter plots confirm the overall good fit between the measured and reconstructed water levels, as the points lie closely around the 1:1 line. 
esides some larger residuals visible for the highest water levels at Metkovic and Gabela 
and related to some isolated high-flow events,  the $\mu$NS\_Tide model reconstructions are accurate at the five stations.

\subsection{Variance and contributions at different stations}

Fig.~\ref{fig:contributions}a shows the percentage of variance in water levels explained by the stage and the tidal-fluvial terms across five stations. The water stage term, which mainly captures the effects of river flow and storm surge, explains most of the variance at all stations. The highest explained variance is observed at Gabela, where the stage accounts for 95\% of the total variance, while the lowest explained variance is found at Usce with $v_S = 55\%$.

\begin{figure}[ht]
    \centering
        \includegraphics[width=0.8\textwidth]{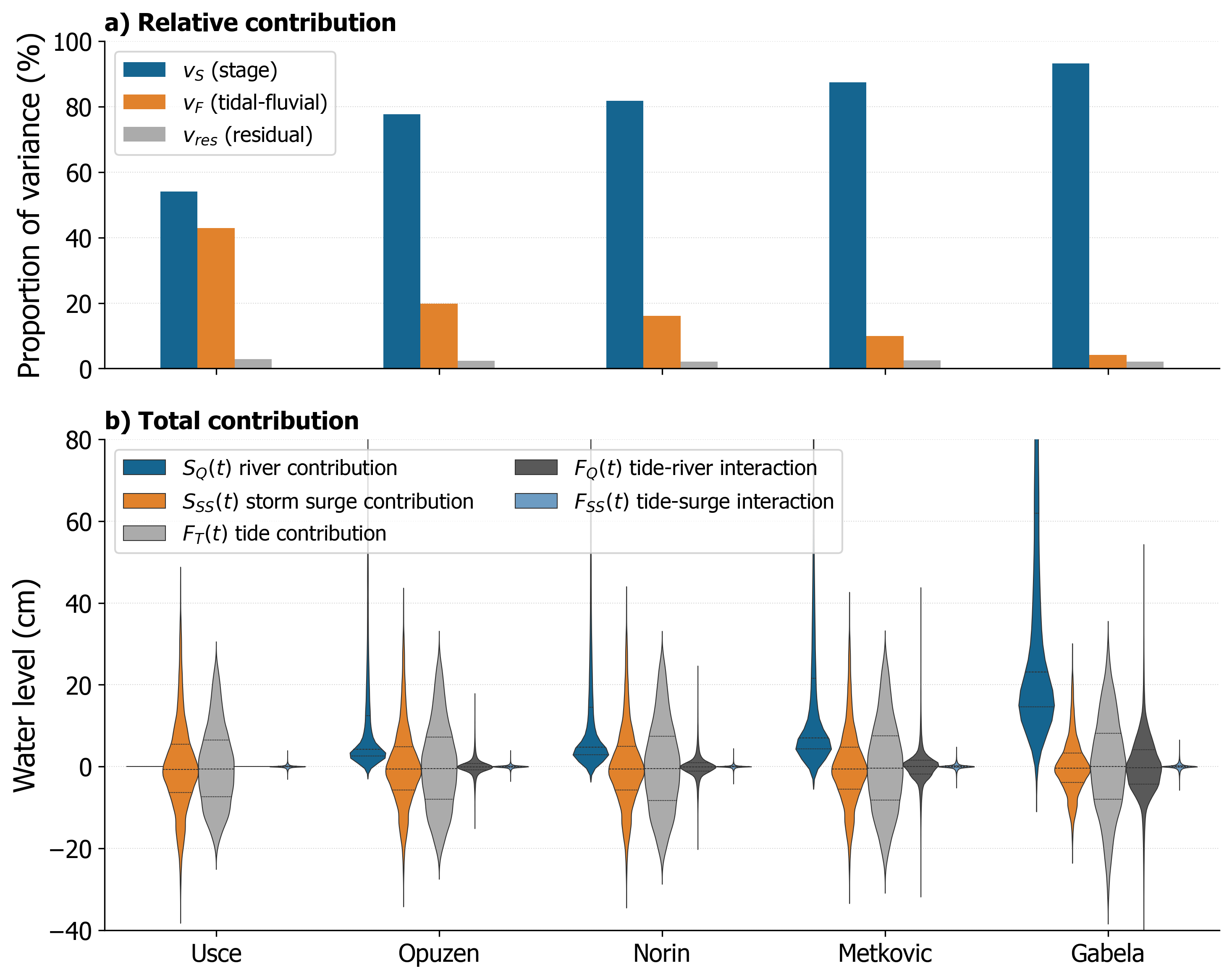}
        \label{fig:contributions}
 \caption{Relative and total contribution of partial reconstructions by $\mu$NS\_TIDE model: a) Proportion of variance explained by the stage, tidal-fluvial, and residual terms at different stations, and b) total water level contributions of river $S_{Q}$, storm surge $S_{SS}$, tides $F_{T}$, tide-river interaction $F_{Q}$ and tide-surge interaction $F_{SS}$ terms at different stations.}
 \label{fig:contributions}
\end{figure}

The tidal-fluvial term, which represents the tidal harmonics and their modulation by storm surge and river flow,
explains a smaller proportion of the variance across all stations. The highest contribution of this term is estimated at Usce ($v_F = 44\%$), where the tidal effects are more pronounced, and the lowest contribution at Gabela with only $v_F = 4\%$. These results should be interpreted alongside Fig.~\ref{fig:waterlevels}, which shows the total water levels and illustrates how the absolute value of the river stage increases strongly at the upstream stations. Therefore, the lower relative impact of the tidal-fluvial term does not necessarily reflect its absolute contribution. The residual component $v_{res}$, which represents the unexplained part of the total water level, is relatively small and accounts for about 2-3\% of the total variance at all stations. This highlights the robustness of the models in capturing the main drivers of water level variability in the Neretva River estuary.

Fig.~\ref{fig:contributions}b complements the previous figure by showing the overall contributions of the river flow, storm surge, stationary tides, tide-river interactions, and tide-surge interactions at different stations. The discharge contribution, denoted as $S_{Q}$, dominates at all stations except Usce, with the highest values observed at Gabela. In contrast, the average contribution of storm surge $S_{SS}$ remains relatively consistent across the stations, but decreases upstream. The tidal contribution $F_{T}$ is largely uniform across the stations, but its net effect is reduced upstream by the interactions, which can partially dampen the tidal amplitudes when they are out of phase with the tides. It is noteworthy that the contribution from tide-river interactions $F_{Q}$ increases upstream and reaches maximum values at Gabela. Tide-surge interaction $F_{SS}$ contributes only minimally compared to the other components. The time series of the individual contributions can be found in the Supplementary Materials in Figs.~E.1-E.5, highlighting that each of these contributions is not necessarily synchronized temporally.

\subsection{Stationary and non-stationary harmonic analysis at the Usce tidal station}

To establish the baseline for the amplitudes and phases of the main constituents, we first decomposed the observed water levels at the Usce tidal station using stationary and non-stationary HA. In the non-stationary analysis, we used $\mu$NS\_Tide, but only the storm surge term, $SS(t)$, was included as a predictor in Eqs.~\ref{eq:stage_model4} and \ref{eq:tidal_model4}, as the effects of river discharges at the coast are negligible.
The HA results, shown in Tab.~\ref{tab:usce_harmonics}, highlight the major tidal constituents. The primary identified constituents include the diurnal constituents $K_1$, $P_1$, $O_1$, and $S_1$, as well as the semi-diurnal constituents $M_2$, $S_2$, $K_2$, and $N_2$. Among them, $M_2$ has the largest amplitude of 9.05 cm, followed by $K_1$ with 7.37 cm and $S_2$ with 6.24 cm. These results confirm the microtidal conditions and the mixed tidal form of the Adriatic Sea at this location. It can be noted that $S_1$ at this coastal station is most likely caused by atmospheric effects (e.g. the diurnal sea breeze).

The non-stationary HA, which takes into account non-linear interactions between tides and storm surges, shows small temporal variations in the tidal parameters and a weak influence of the storm surge on the amplitudes and phases of the main tidal constituents during the observation period (Tab.~\ref{tab:usce_harmonics}). In particular, the temporal standard deviations $\sigma_{A_k}$ of the tidal amplitude are small compared to their temporal mean $\bar{A}_k$ and the corresponding stationary estimate $A_k$. Similarly, the circular variance $\sigma_g$ of the tidal phases over time is close to zero, indicating a small dispersion of the time-varying phases 
around their circular mean value $\bar{g}_k$.
The time series of the total water levels at the Usce tidal station, together with the reconstructed time series from the stage, tides, and tidal-surge interaction terms, are shown in Fig.~D.1 in the Supplementary Materials.

\begin{table}[htb]
    \centering
    \begin{tabular}{rl|lll|lllll}
    \toprule
    Const & $T_k$ (h) & 
    $SNR_k$ (-) &  $A_k$ (cm) & $g_k$ (\degree) & 
    $\overline{SNR}_k$ (-) & $\overline{A}_k$ (cm) & $\sigma_{A_k}$ (cm) & $\overline{g}_k$ ($\circ$) & $\sigma_{g_k}$ (-) \\
    \midrule
    $M_2$ & 12.42  & 76.29 & 9.05 & 88.5   & 79.26 & 9.04 & 0.09  & 88.23 & $7.8\cdot 10^{-6}$\\
    $K_1$ & 23.93  & 82.70 & 7.37 & 40.9   & 90.24 & 7.29 & 0.14  & 40.72 & $1.2\cdot 10^{-5}$ \\
    $S_2$ & 12.00  & 84.01 & 6.24 &  6.2   & 75.97 & 6.21 & 0.21  & 87.14 & $7.7\cdot 10^{-6}$\\
    $P_1$ & 24.07  & 49.26 & 2.56 & 31.1   & 56.21 & 2.61 & 0.05  & 31.27 & $1.6\cdot 10^{-4}$\\
    $O_1$ & 25.82  & 54.73 & 2.24 & 17.4   & 53.11 & 2.26 & 0.02  & 17.69 & $1.3\cdot 10^{-4}$\\
    $K_2$ & 11.97  & 20.53 & 1.56 & 97.4   & 19.29 & 1.56 & 0.03  & 98.04 & $1.6\cdot 10^{-4}$\\
    $N_2$ & 12.66  & 37.56 & 1.45 & 91.2   & 31.71 & 1.45 & 0.002 & 90.69 & $1.9\cdot 10^{-4}$\\
    $S_1$ & 24.00  & 8.61  & 0.92 & -127   & 10.56 & 0.96 & 0.06  & -130  & $3.1\cdot 10^{-3}$\\ 
    \bottomrule
    \end{tabular}
    \caption{Results from stationary and $\mu$NS\_Tide non-stationary analysis at the Usce tidal station, considering major constituents: stationary estimates of the signal-to-noise ratio ($SNR_k$, unitless), amplitude and amplitude standard error ($A_k$ and $\sigma_{A_k}$, cm), phase and phase circular variance ($g_k$, degrees, and $\sigma_{g_k}$, unitless), and corresponding temporal means from non-stationary analysis. Temporal means, indicated by overbars, are computed over the entire observational period.}
    \label{tab:usce_harmonics}
\end{table}

\subsection{Spatial and temporal variability of tidal constituents}

\begin{figure}[htbp]
    \centering
    \begin{subfigure}{0.24\textwidth}
        \centering
        \includegraphics[width=\textwidth]{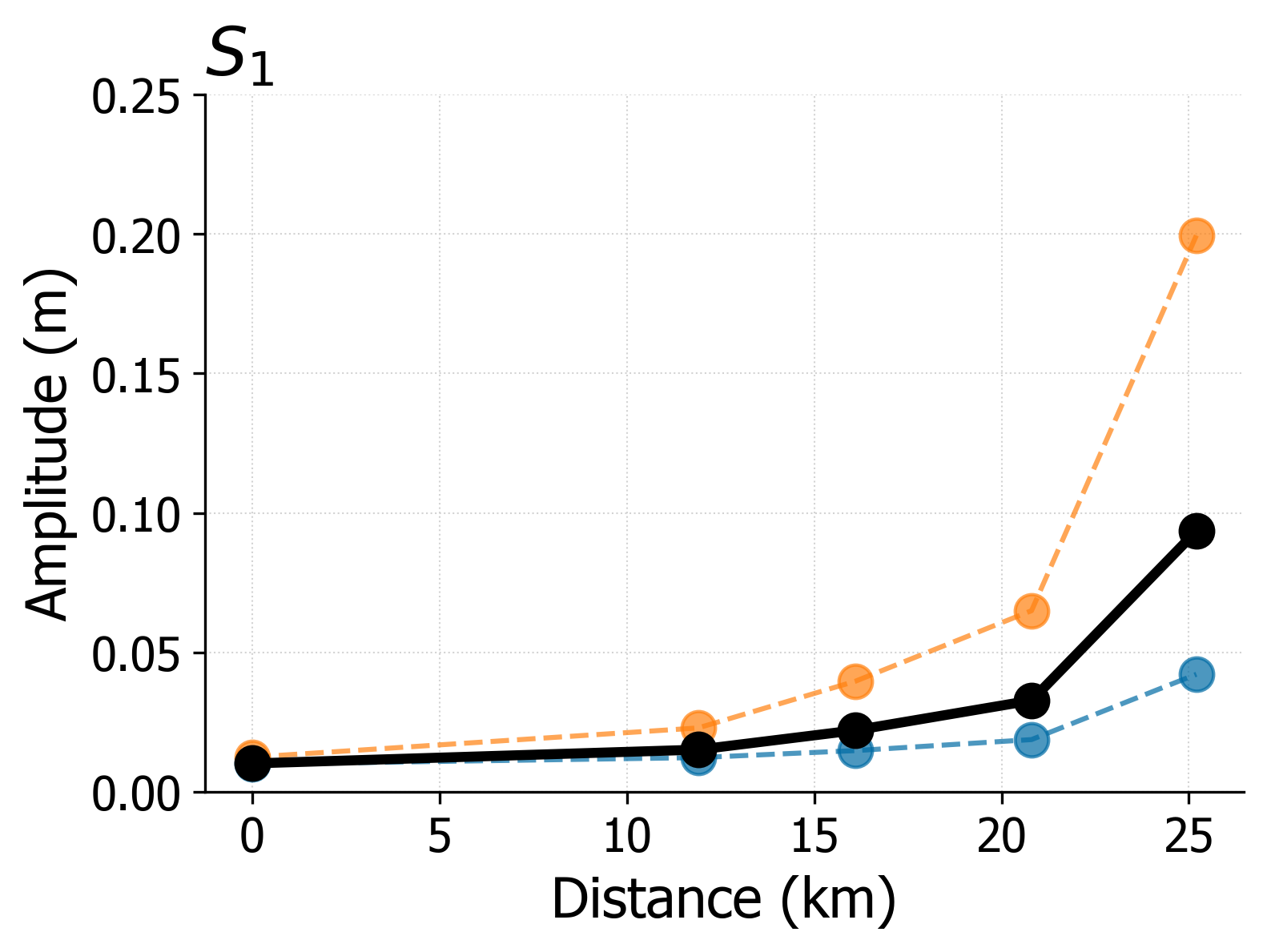}
    \end{subfigure}
    \hfill
    \begin{subfigure}{0.24\textwidth}
        \centering
        \includegraphics[width=\textwidth]{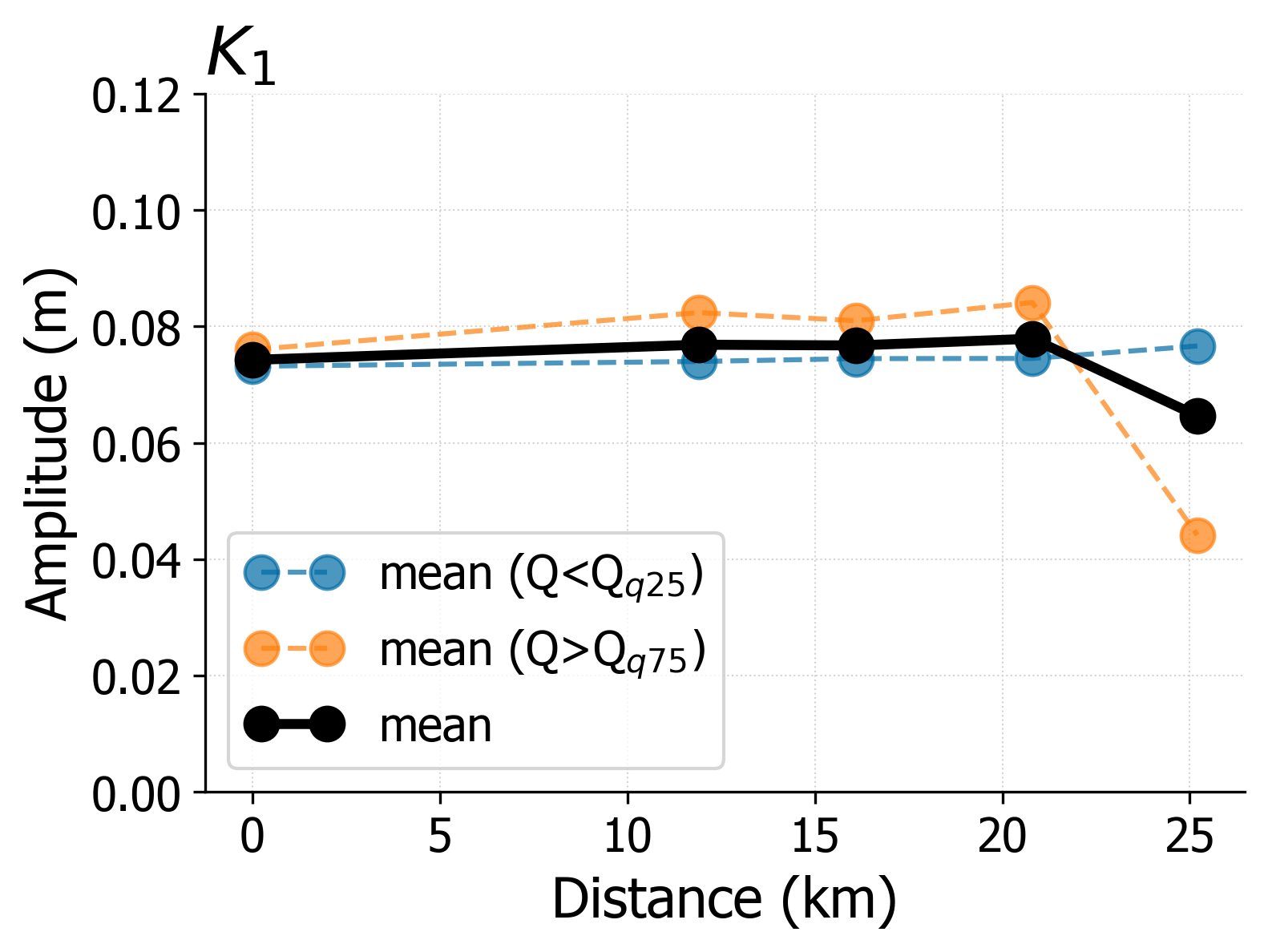}
    \end{subfigure}
    \hfill
    \begin{subfigure}{0.24\textwidth}
        \centering
        \includegraphics[width=\textwidth]{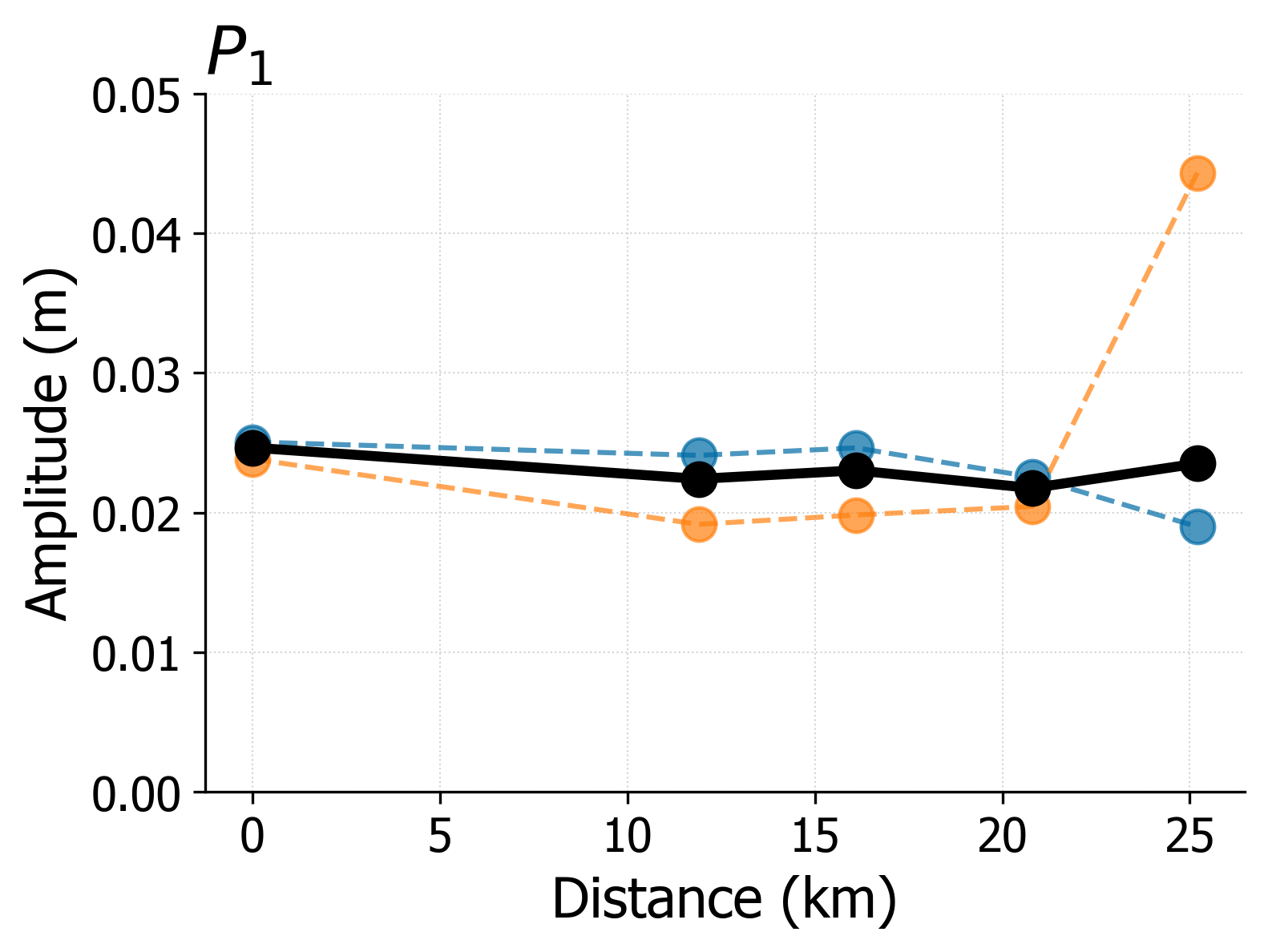}
    \end{subfigure}
    \hfill
    \begin{subfigure}{0.24\textwidth}
        \centering
        \includegraphics[width=\textwidth]{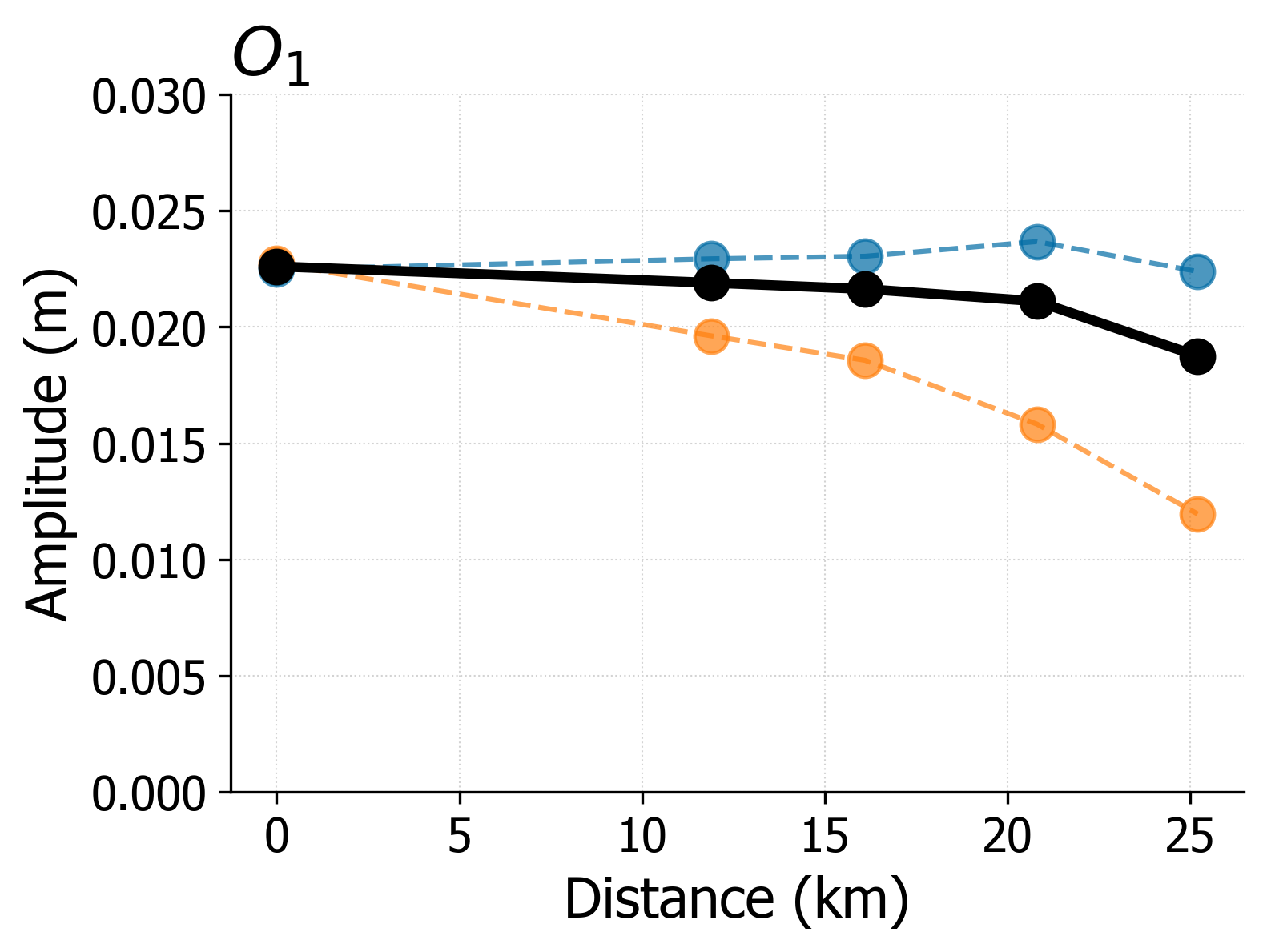}
    \end{subfigure}
    \begin{subfigure}{0.24\textwidth}
        \centering
        \includegraphics[width=\textwidth]{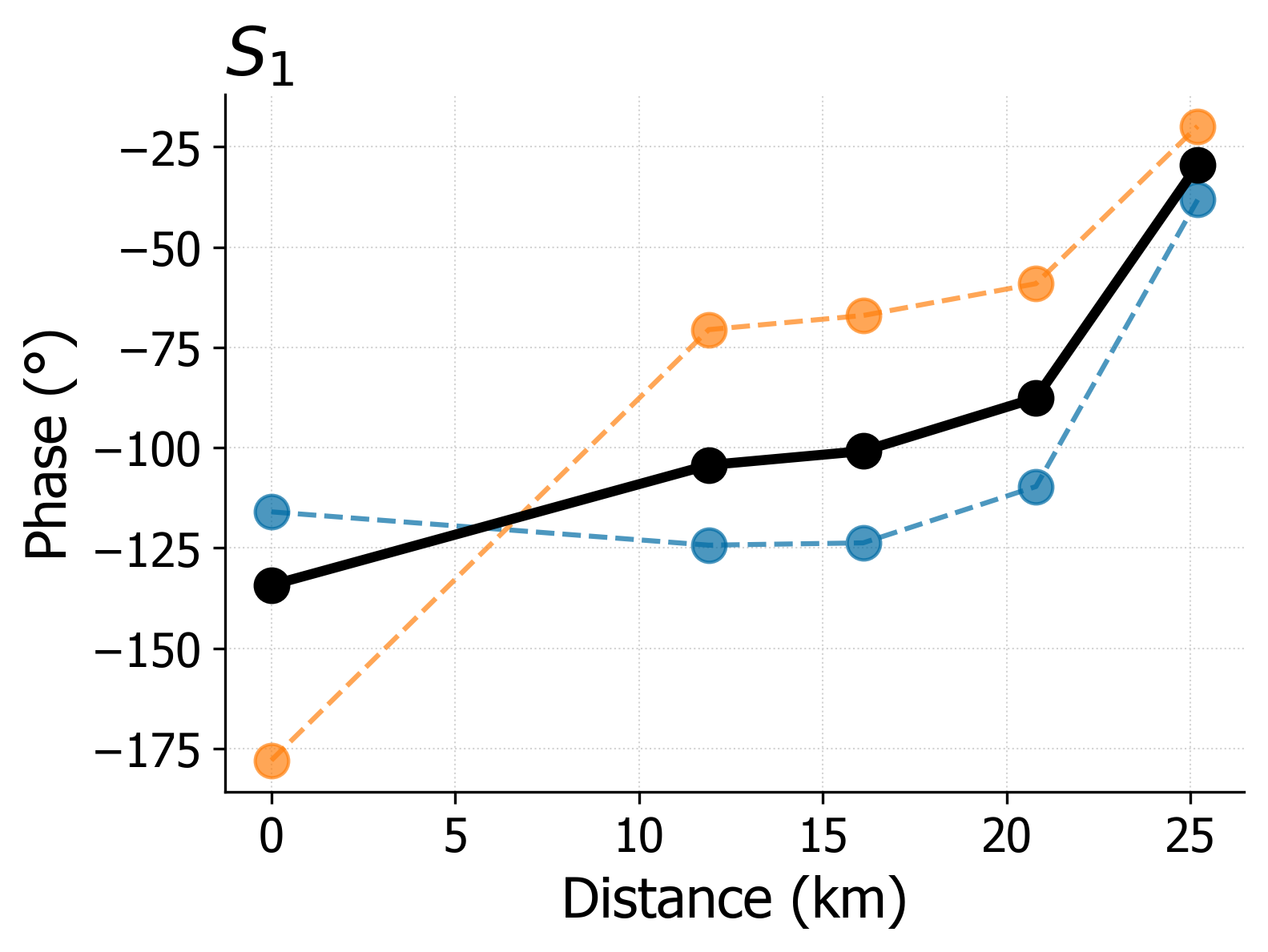}
    \end{subfigure}
    \hfill
    \begin{subfigure}{0.24\textwidth}
        \centering
        \includegraphics[width=\textwidth]{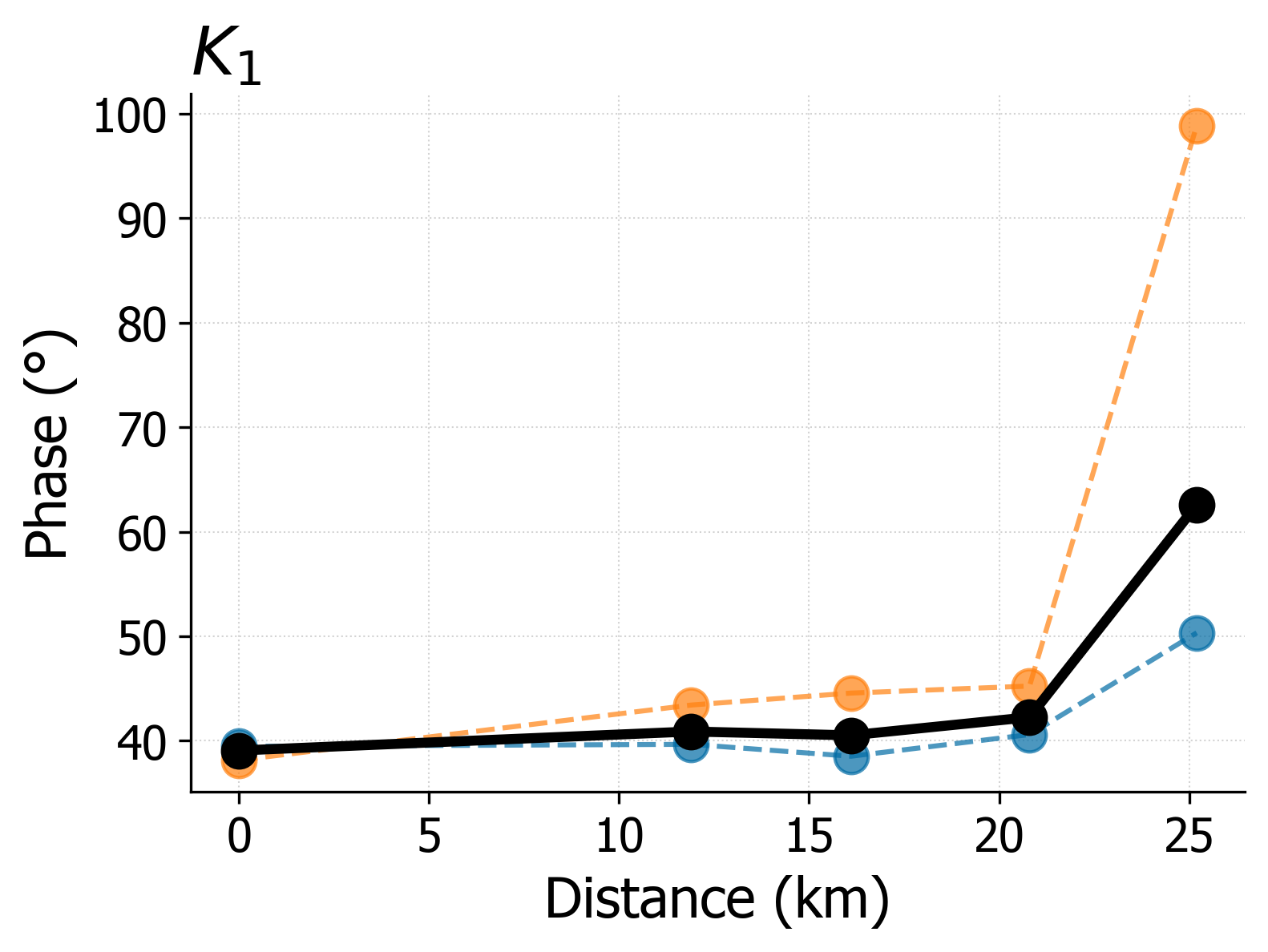}
    \end{subfigure}
    \hfill
    \begin{subfigure}{0.24\textwidth}
        \centering
        \includegraphics[width=\textwidth]{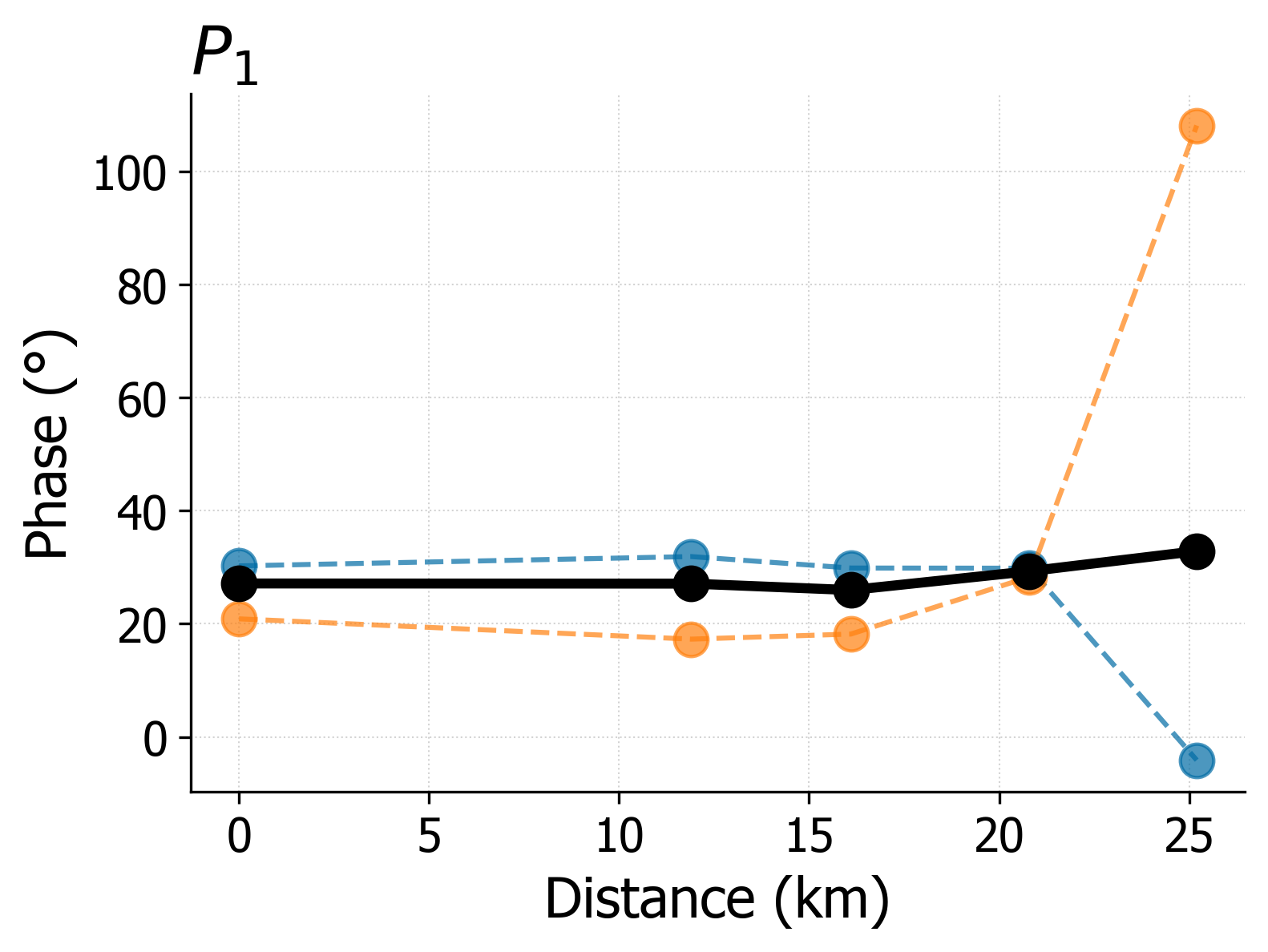}
    \end{subfigure}
    \hfill
    \begin{subfigure}{0.24\textwidth}
        \centering
        \includegraphics[width=\textwidth]{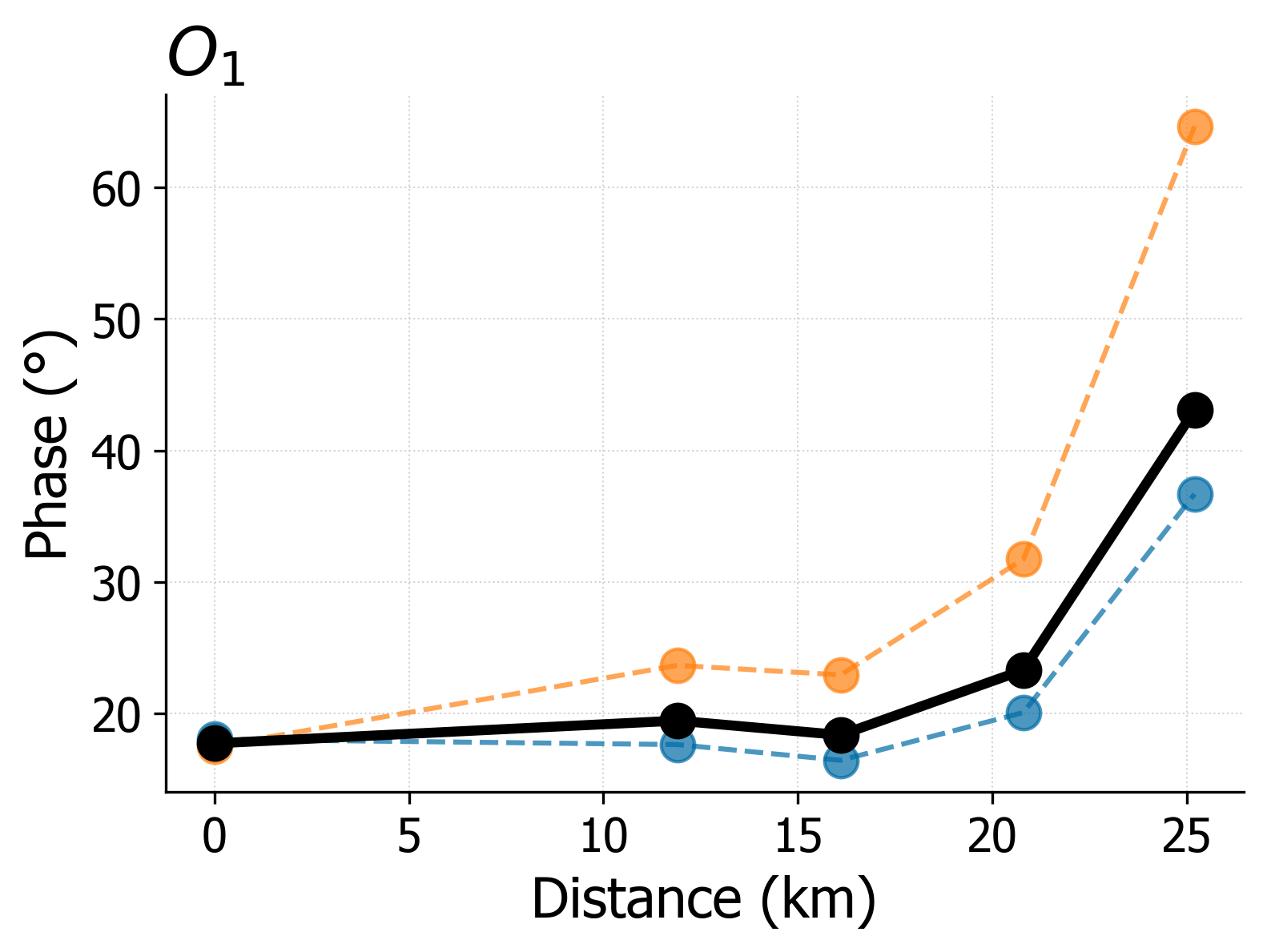}
    \end{subfigure}
   \begin{subfigure}{0.24\textwidth}
        \centering
        \includegraphics[width=\textwidth]{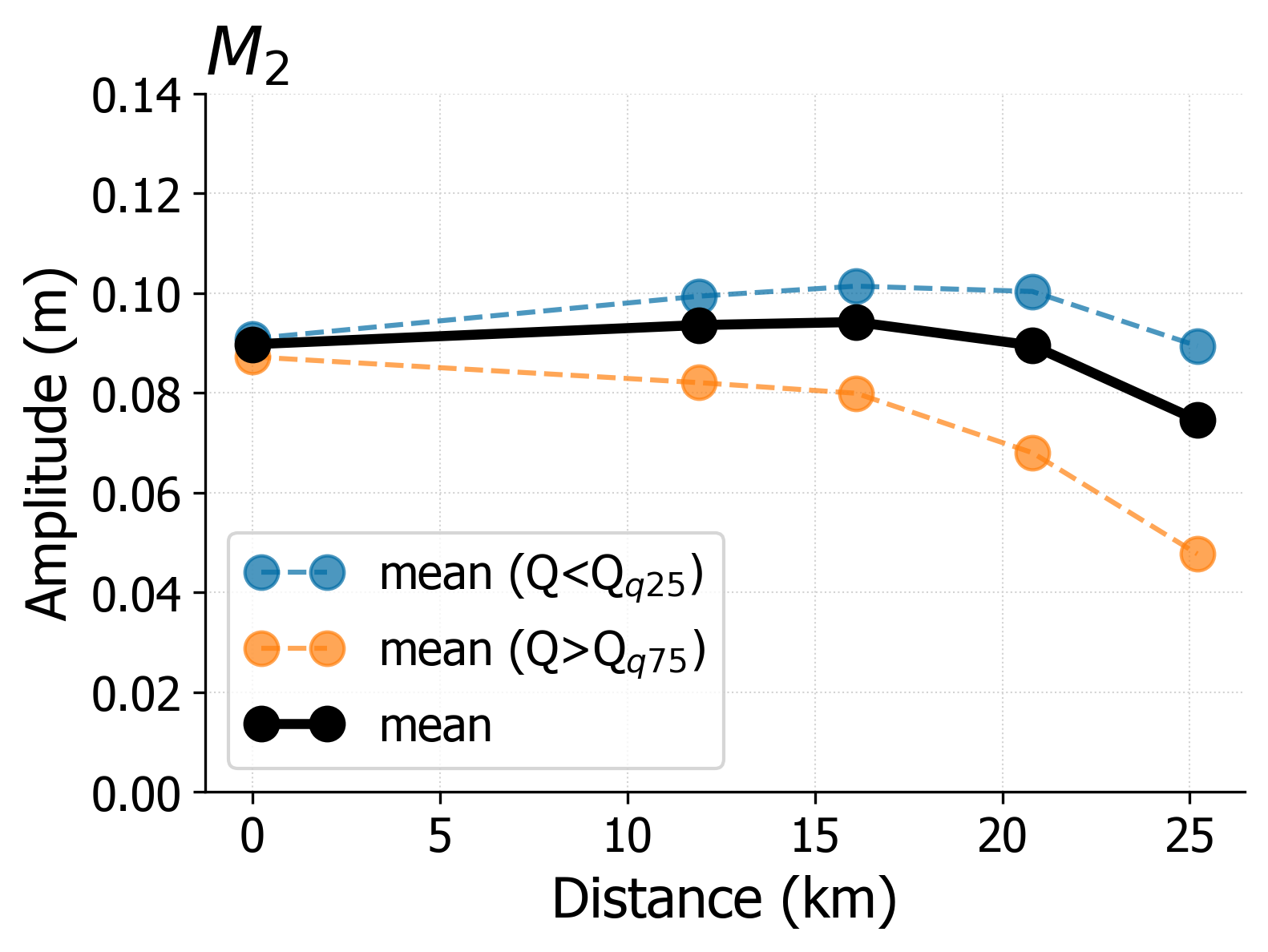}
    \end{subfigure}
    \hfill
    \begin{subfigure}{0.24\textwidth}
        \centering
        \includegraphics[width=\textwidth]{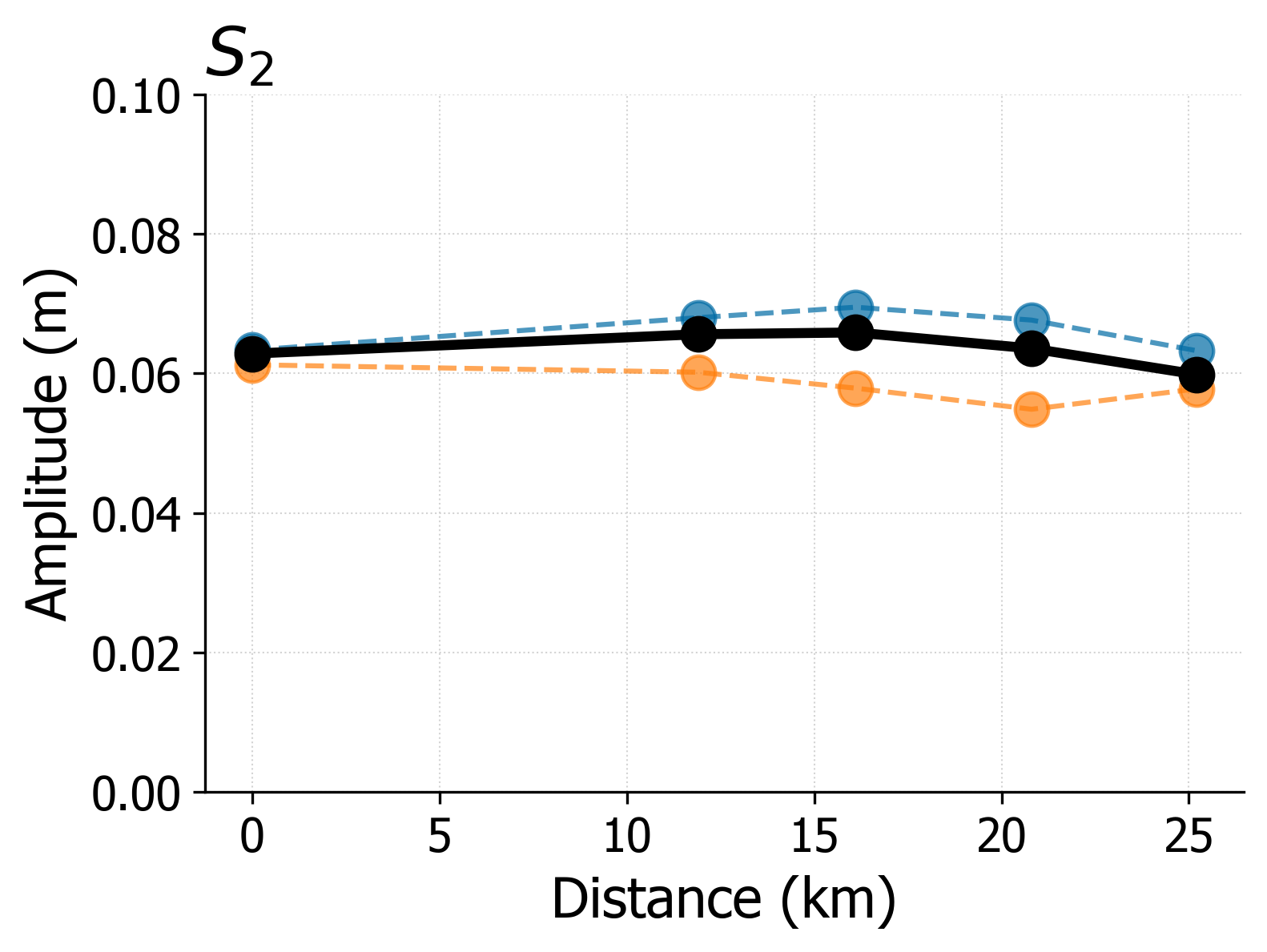}
    \end{subfigure}
    \hfill
    \begin{subfigure}{0.24\textwidth}
        \centering
        \includegraphics[width=\textwidth]{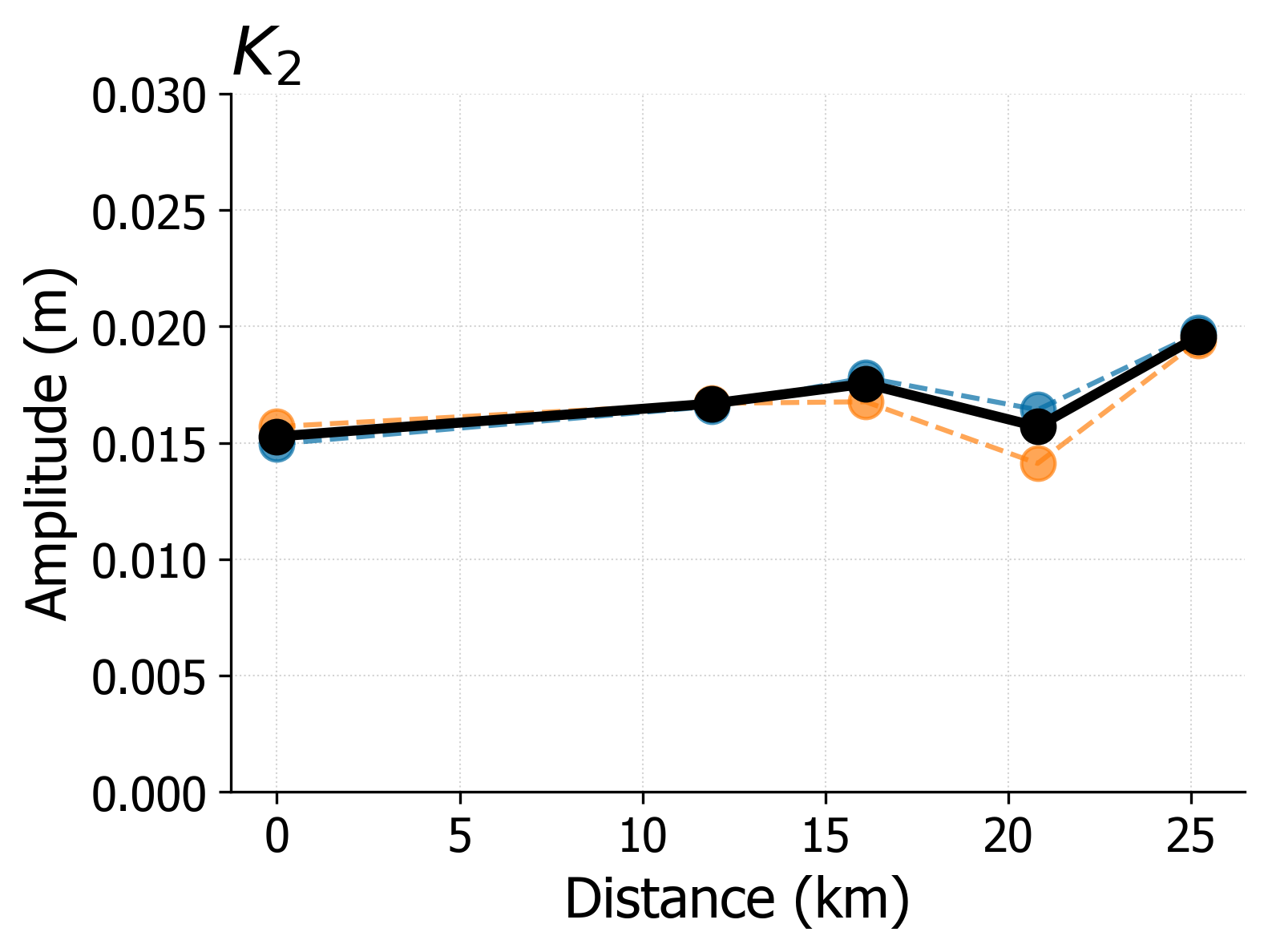}
    \end{subfigure}
        \hfill
    \begin{subfigure}{0.24\textwidth}
        \centering
        \includegraphics[width=\textwidth]{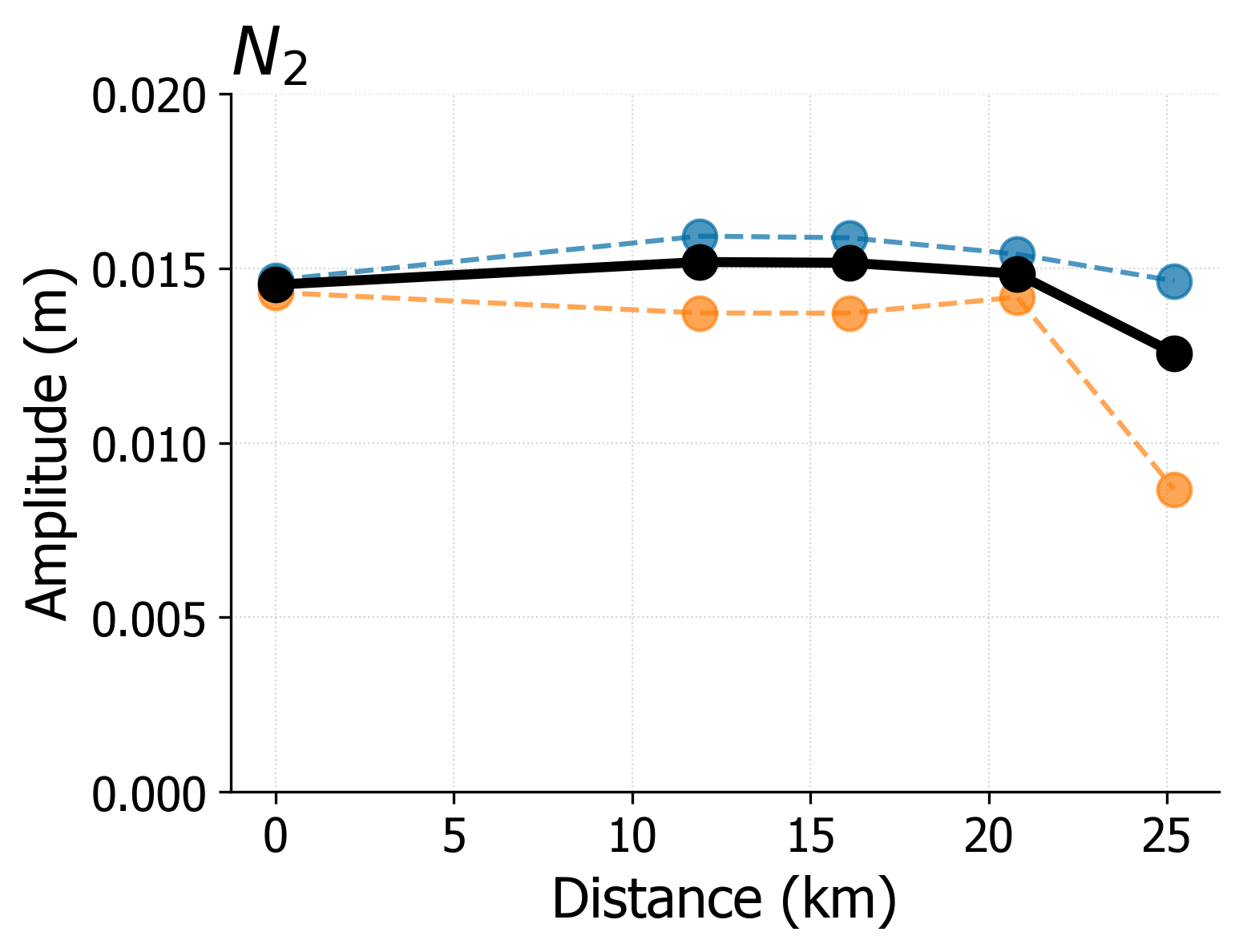}
    \end{subfigure}
    \begin{subfigure}{0.24\textwidth}
        \centering
        \includegraphics[width=\textwidth]{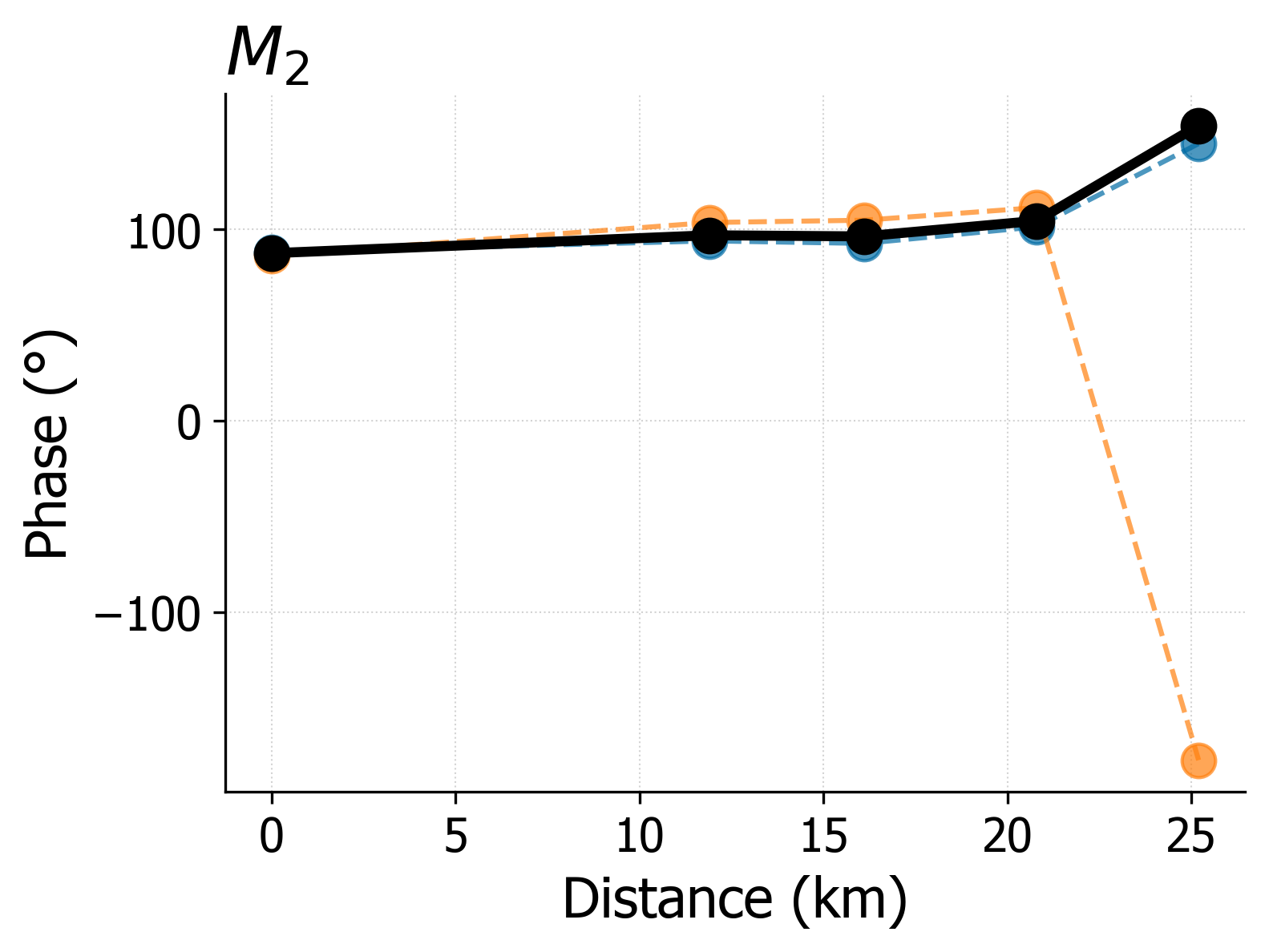}
    \end{subfigure}
    \hfill
    \begin{subfigure}{0.24\textwidth}
        \centering
        \includegraphics[width=\textwidth]{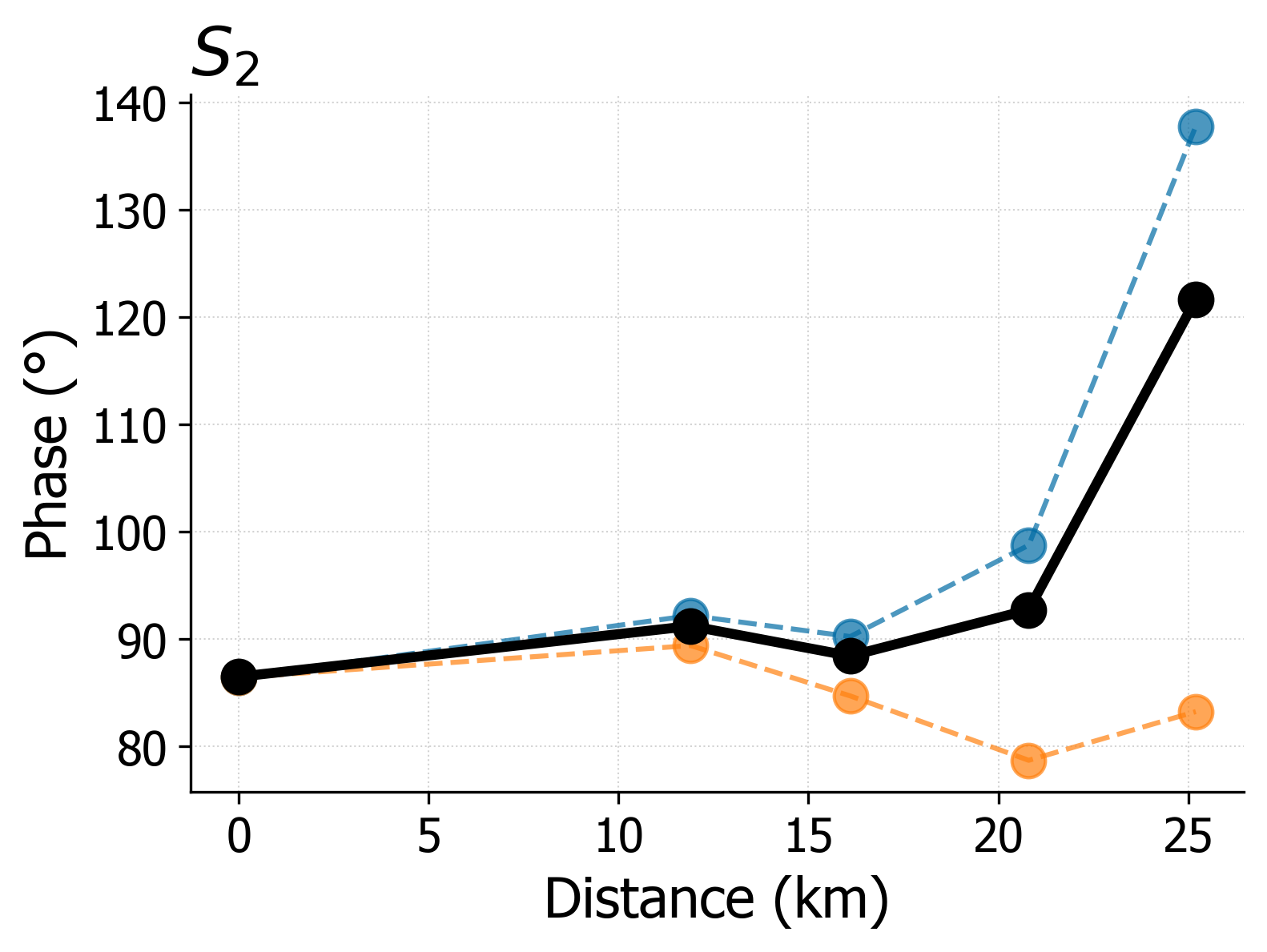}
    \end{subfigure}
    \hfill
    \begin{subfigure}{0.24\textwidth}
        \centering
        \includegraphics[width=\textwidth]{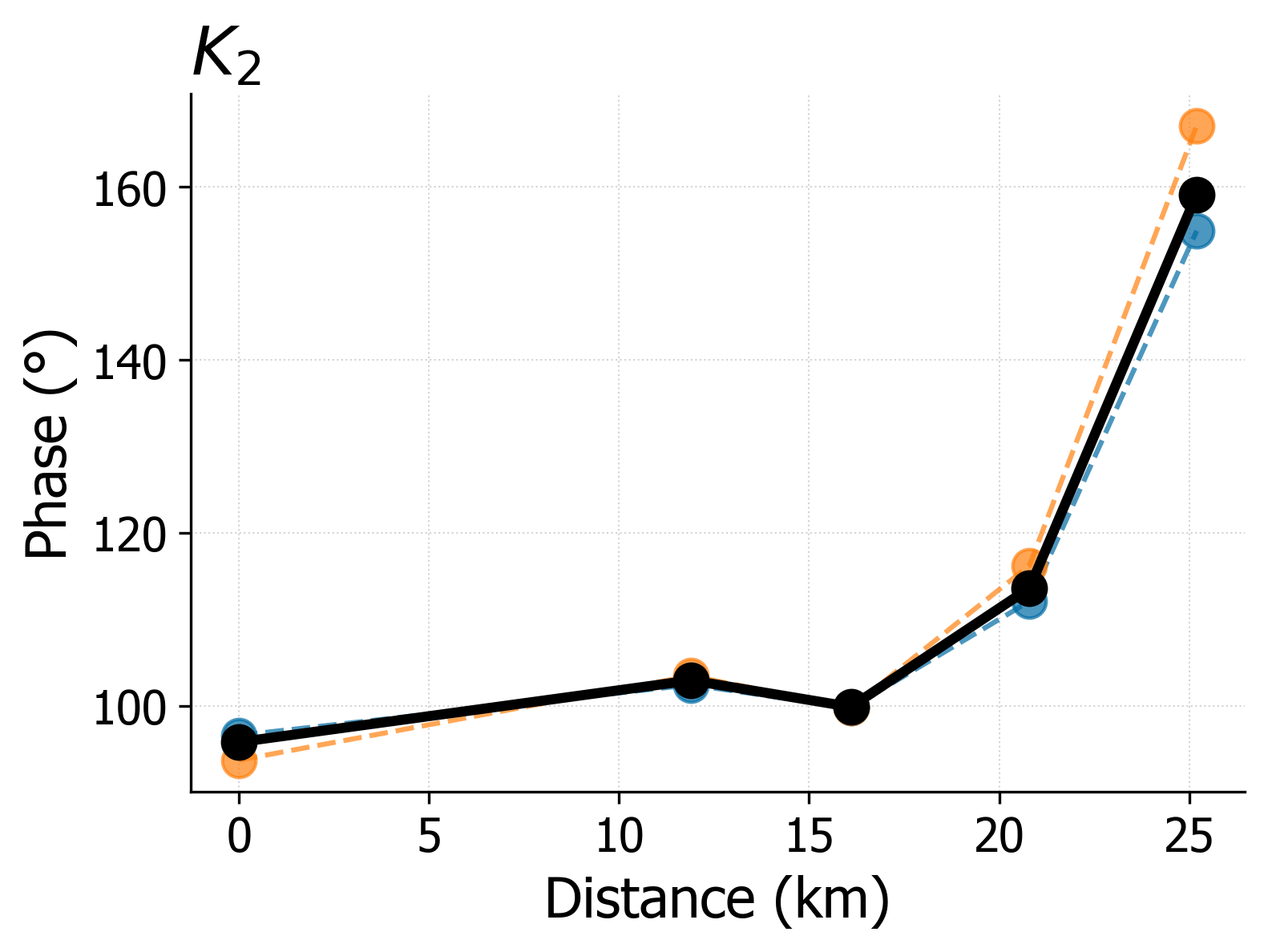}
    \end{subfigure}
        \hfill
    \begin{subfigure}{0.24\textwidth}
        \centering
        \includegraphics[width=\textwidth]{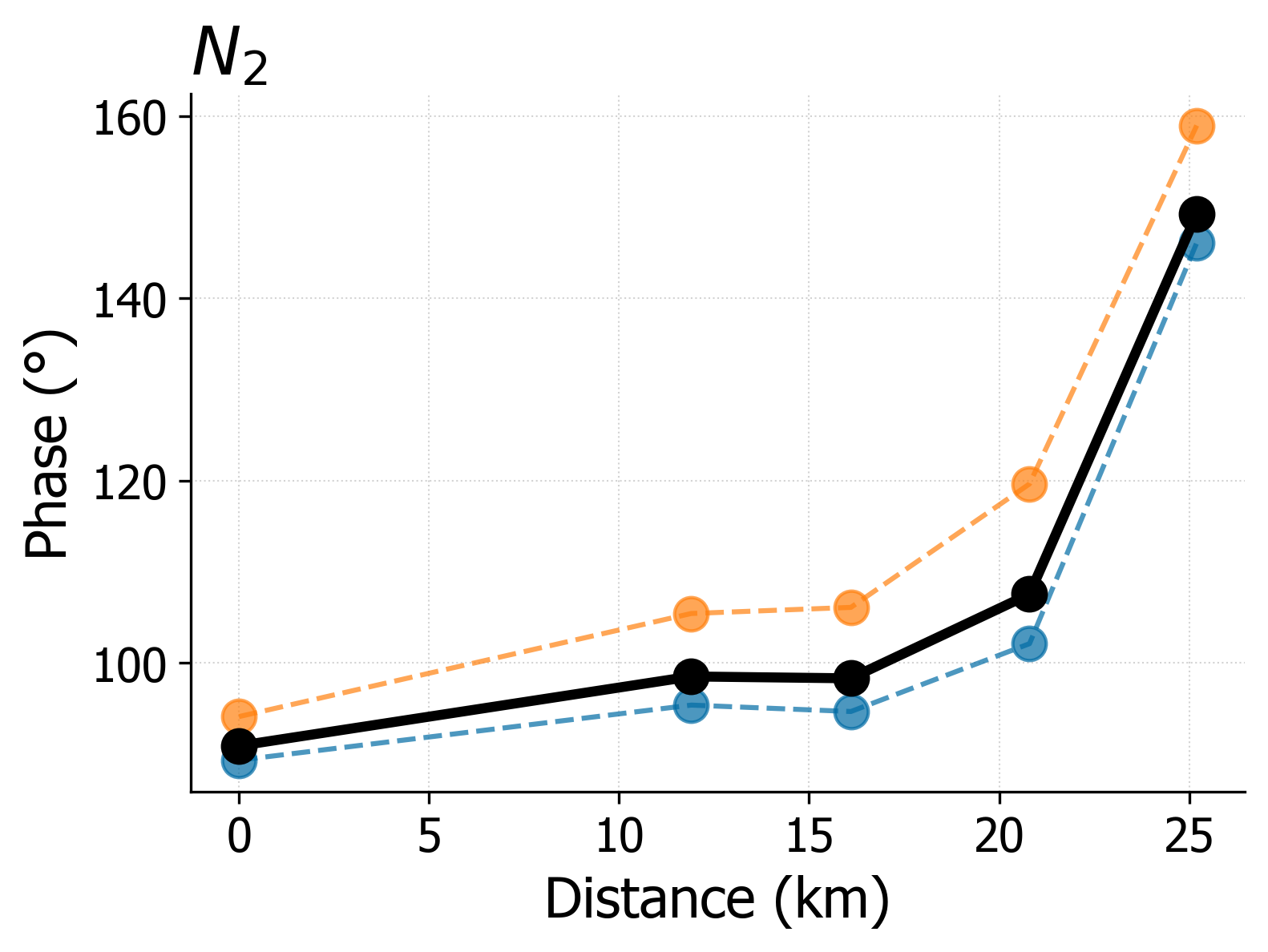}
    \end{subfigure}
    \caption{Amplitude and phase variations across the station locations for main diurnal and semi-diurnal constituents conditioned to the river flow regime: average flow conditions, high-flow conditions (river discharge above 75th-percentile), and low-flow conditions (river discharge below 25th-percentile). Station locations: Usce (0 rkm), Opuzen (11.9 rkm), Norin (16.1 rkm), Metković (20.8 rkm), and Gabela (25.2 rkm).}
    \label{fig:spatial_variability}
\end{figure}

Fig.~\ref{fig:spatial_variability} presents the amplitudes and phases for the main diurnal ($K_1$, $P_1$, $O_1$, $S_1$) and semi-diurnal ($M_2$, $S_2$, $K_2$, $N_2$) tidal constituents as a function of the distance from the river mouth. Each figure compares the mean tidal parameters for three characteristic regimes: average flow, high flow (river discharge above the 75th percentile; blue), and low flow  (river discharge below the 25th percentile; orange). 
The temporal mean and variance of the tidal parameters estimated for these constituents are reported in Tab.~E.1 of the Supplementary Materials.

The amplitudes of the $S_1$ constituent show the largest changes along the river, with mean amplitudes close to 1 cm at Usce and higher than 9 cm at the most upstream station, Gabela.  
The distance between the blue and orange curves also shows that the upstream increase of $S_1$ is more pronounced for high flow than for average and low $Q$ values. The phase of $S_1$ remains stable for most stations within Opuzen and Metkovic, with a sharp increase at Gabela.

Interestingly, the $K_1$ constituent behaves differently from $S_1$. The amplitudes remain relatively constant along the downstream and middle reaches and decrease slightly at the upstream station Gabela. At low flow, $K_1$ maintains a constant amplitude, similar to average flow. At high flow, the amplitudes of $K_1$ remain stable but decrease at the Gabela station. The phase of $K_1$ remains relatively constant along the river but exhibits significant shifts at the Gabela station depending on the flow conditions.

For the $P_1$ constituent, the amplitude remains relatively low at all stations under all flow conditions, with only a moderate increase at the Gabela station at high flow. The phase of $P_1$ remains constant under all flow conditions along the river, but shows a strong dependence on the flow regime of the river at Gabela.

The $O_1$ constituent displays the smallest amplitudes among the diurnal constituents. Its amplitude remains low and stable along the river under all flow conditions, with only a slight decrease in the upstream direction. In contrast to $S_1$ and $K_1$, the amplitudes of $O_1$ decrease at high flow and increase at low flow compared to the average values. The phase of $O_1$ remains relatively constant along most of the river, but shows a shift at the Gabela station.

All semi-diurnal constituents show relatively similar patterns. The $M_2$, $S_2$, and $N_2$ constituents show relatively uniform amplitudes, with a slight increase at the midstream stations, followed by a decrease further upstream at the Metkovic and Gabela stations. However, the $K_2$ constituent shows a consistent increase in amplitudes in an upstream direction. For all semi-diurnal constituents, the amplitude increases at high flow and decreases at low flow. A phase shift in the upstream direction is observed for all components, with the Gabela station responding strongly to the flow.

\section{Discussion} \label{sec:discussion}

\subsection{Interaction of power peaking and tides}

To explain the spatial variation of tidal constituents and the specific influence of power peaking on signal variability, we performed a numerical experiment based on two STREAM model simulations: one with measured river flow $Q_{mes}$ (simulation A) and the other with low-passed river flow $Q_{filt}$ (simulation B) as upstream boundary conditions. Since frequencies above 0.03 cph are removed from $Q_{filt}$, simulation B represents a baseline scenario without the influence of peaking from the hydropower plants, which is characterized by a 24-hour period oscillation. The non-stationary $\mu$NS\_Tide model is then applied to the water level results from two simulations to investigate the tidal parameters of constituents potentially affected by the power peaking ($K_1$, $K_2$, $S_1$, $S_2$ and $P_1$).

\begin{figure}[htbp]
    \centering
    \begin{subfigure}{0.3\textwidth}
   (a) Measured
        \centering
        \includegraphics[width=\textwidth]{figs/mNS_TIDE_K1_amp_long.png}
    \end{subfigure}
    \hfill
    \begin{subfigure}{0.3\textwidth}
    (b) Simulation A ($Q_{mes}$)
        \centering
        \includegraphics[width=\textwidth]{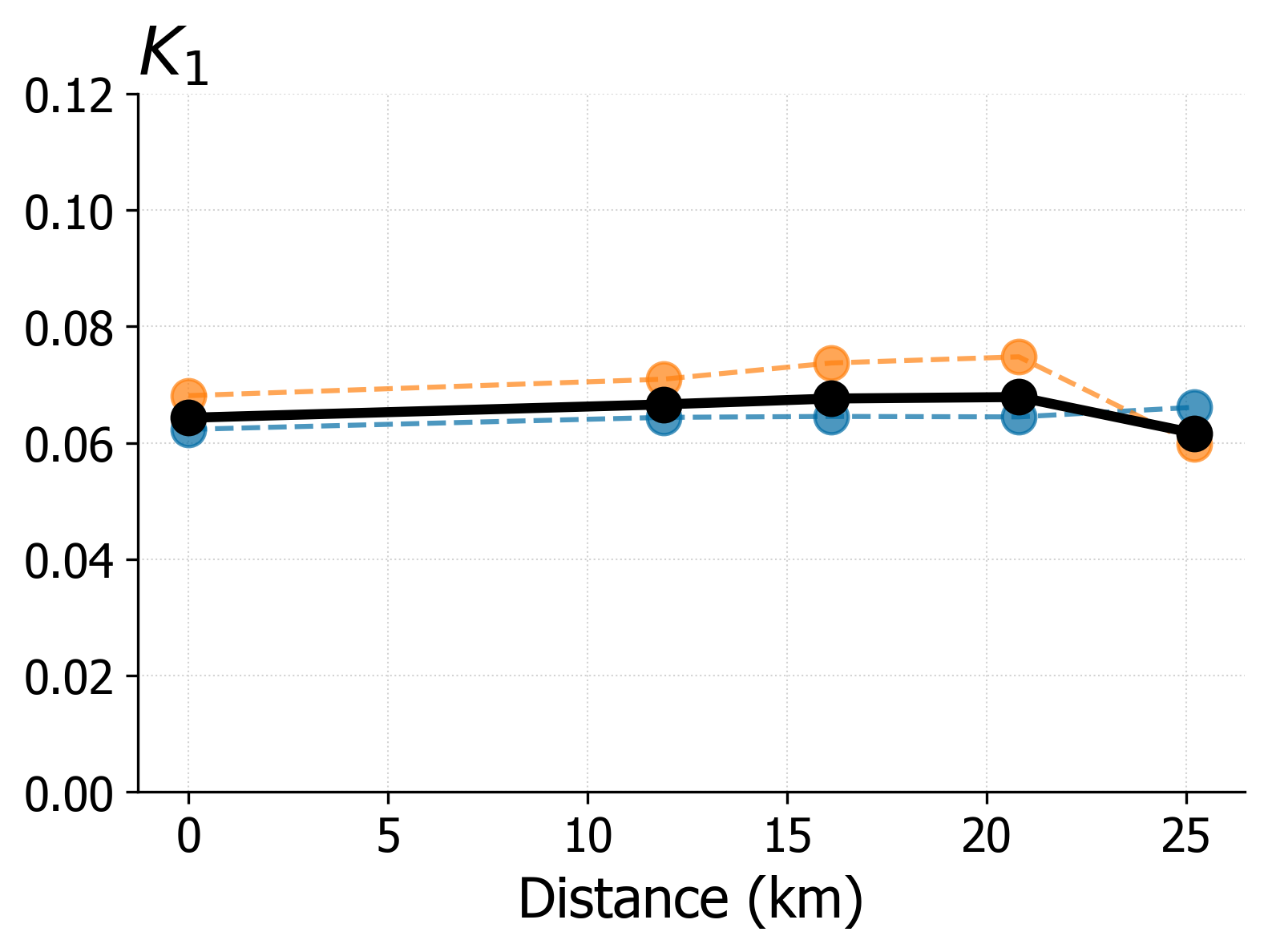}
    \end{subfigure}
    \hfill
    \begin{subfigure}{0.3\textwidth}
    (c) Simulation B ($Q_{filt}$)
        \centering
        \includegraphics[width=\textwidth]{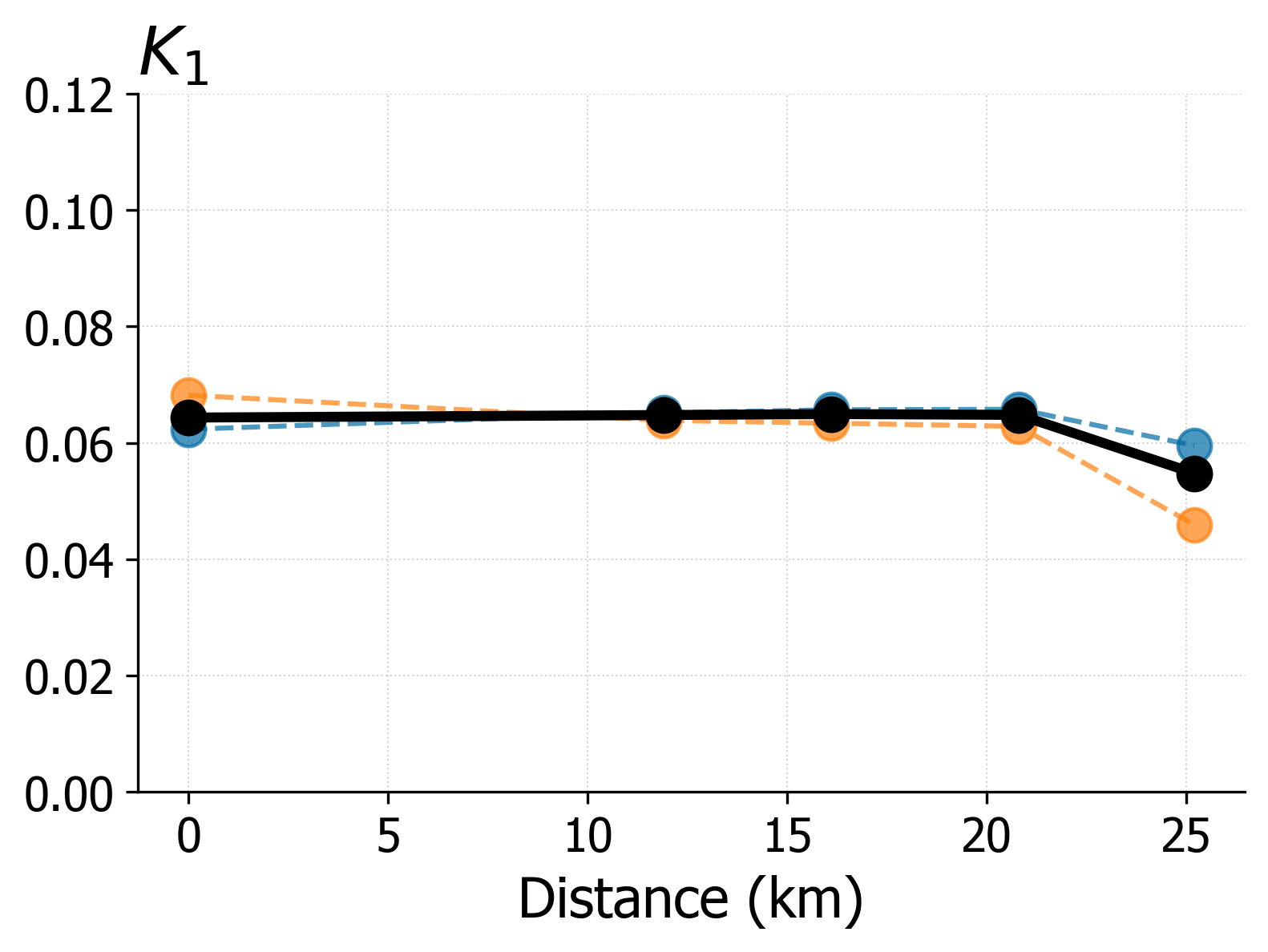}
    \end{subfigure}
    \begin{subfigure}{0.3\textwidth}
        \centering
        \includegraphics[width=\textwidth]{figs/mNS_TIDE_K2_amp_long.png}
    \end{subfigure}
    \hfill
    \begin{subfigure}{0.3\textwidth}
        \centering
        \includegraphics[width=\textwidth]{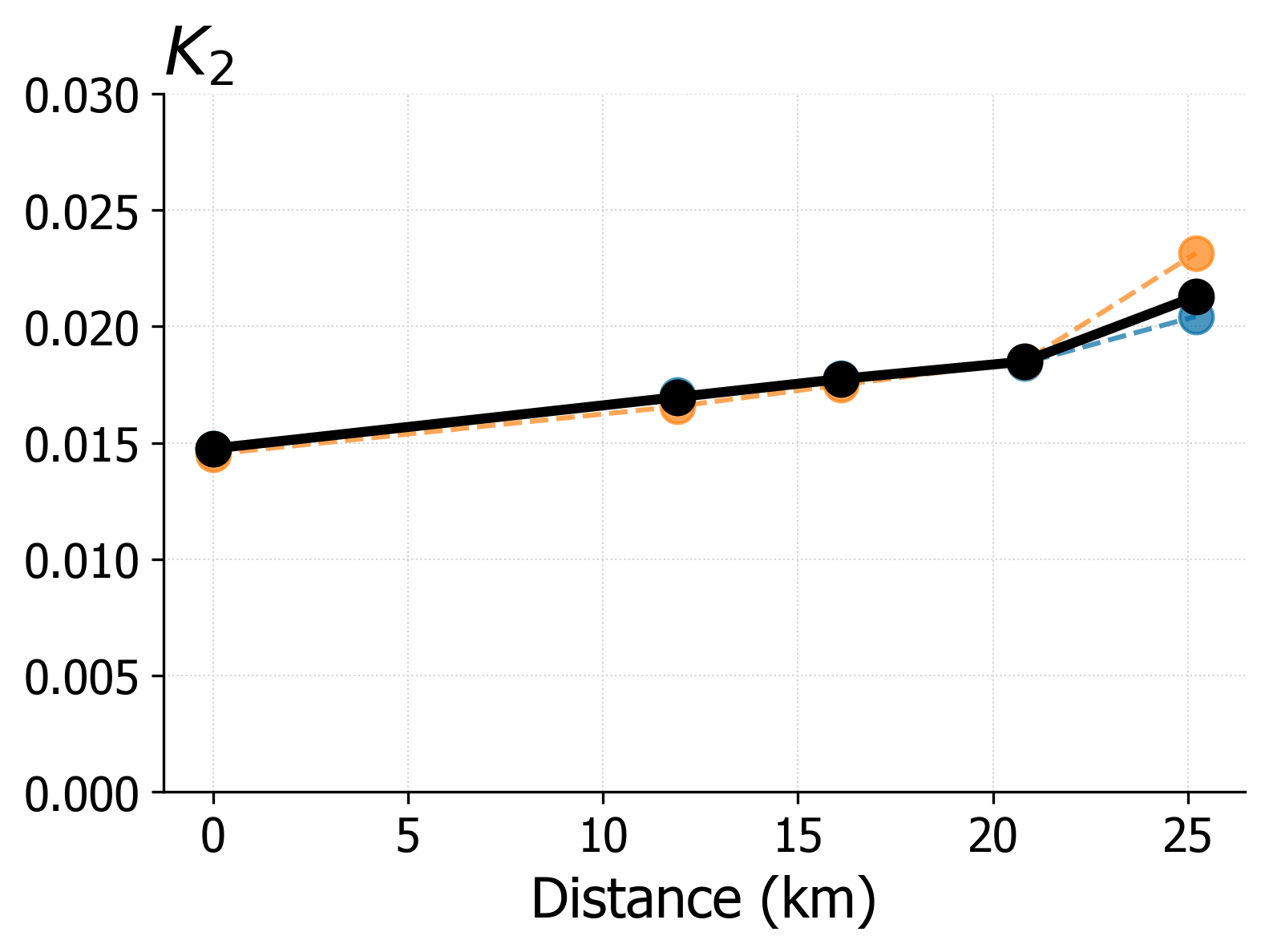}
    \end{subfigure}
    \hfill
    \begin{subfigure}{0.3\textwidth}
        \centering
        \includegraphics[width=\textwidth]{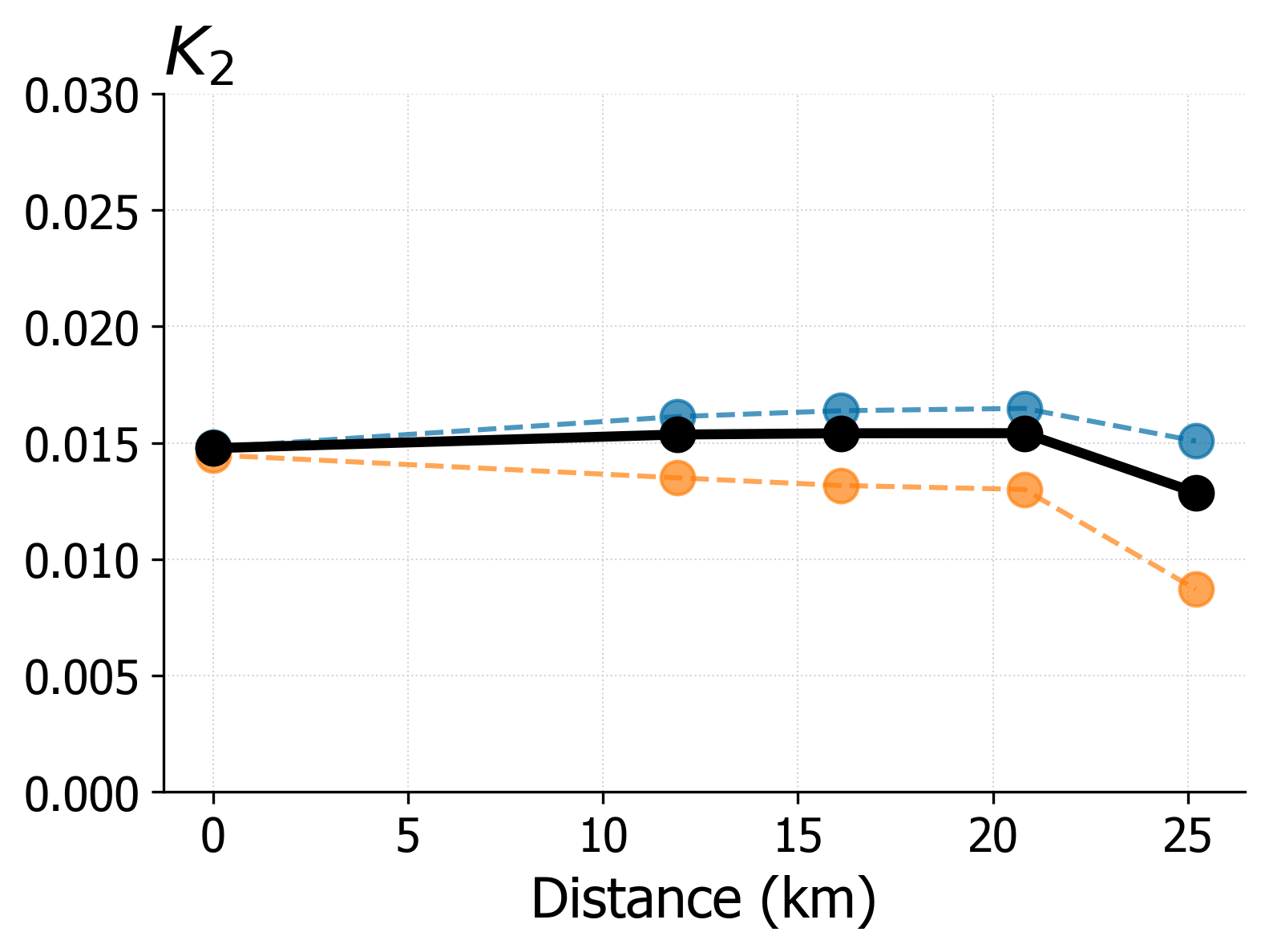}
    \end{subfigure}
   \begin{subfigure}{0.3\textwidth}
        \centering
        \includegraphics[width=\textwidth]{figs/mNS_TIDE_P1_amp_long.png}
    \end{subfigure}
    \hfill
    \begin{subfigure}{0.3\textwidth}
        \centering
        \includegraphics[width=\textwidth]{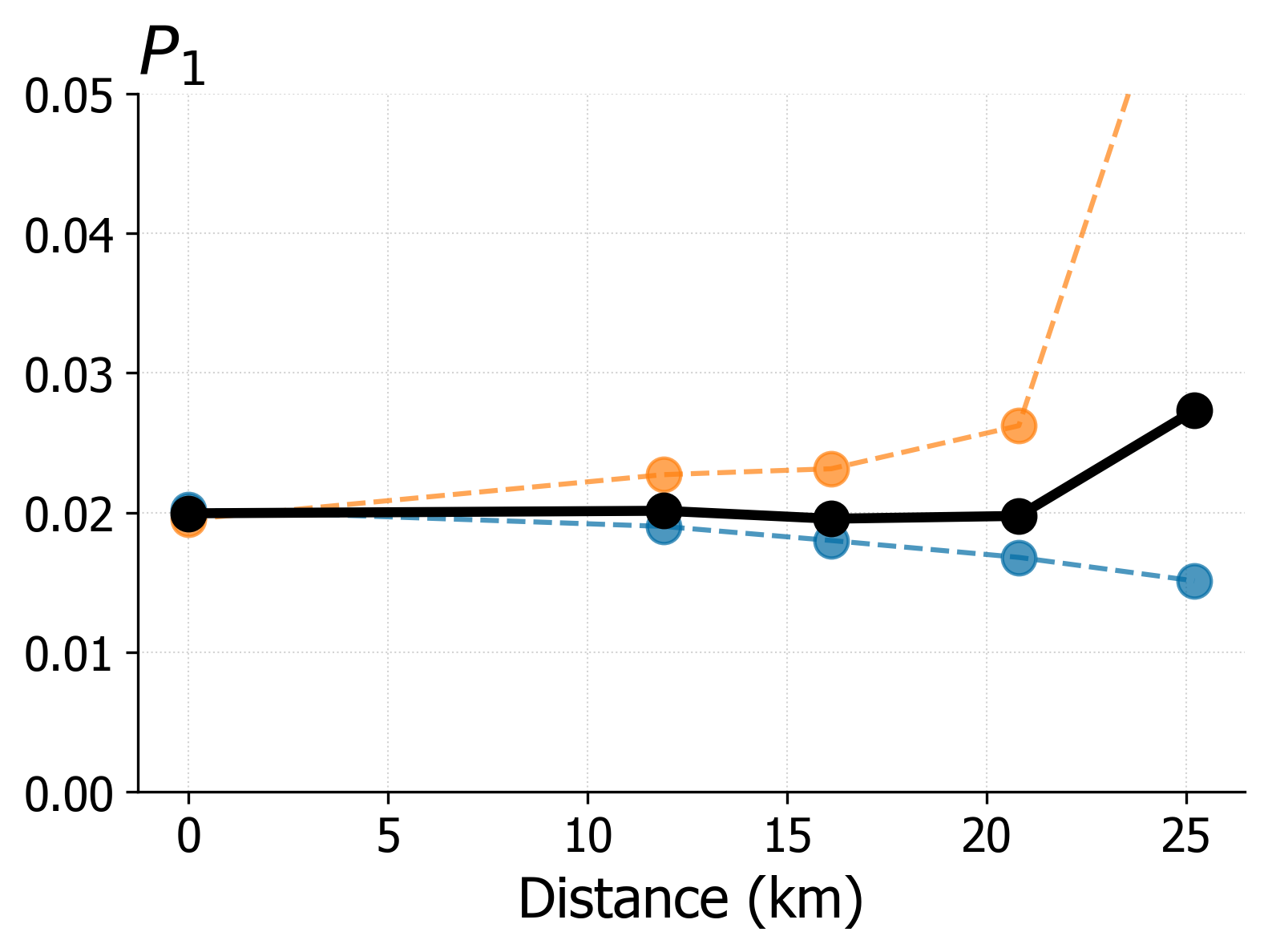}
    \end{subfigure}
    \hfill
    \begin{subfigure}{0.3\textwidth}
        \centering
        \includegraphics[width=\textwidth]{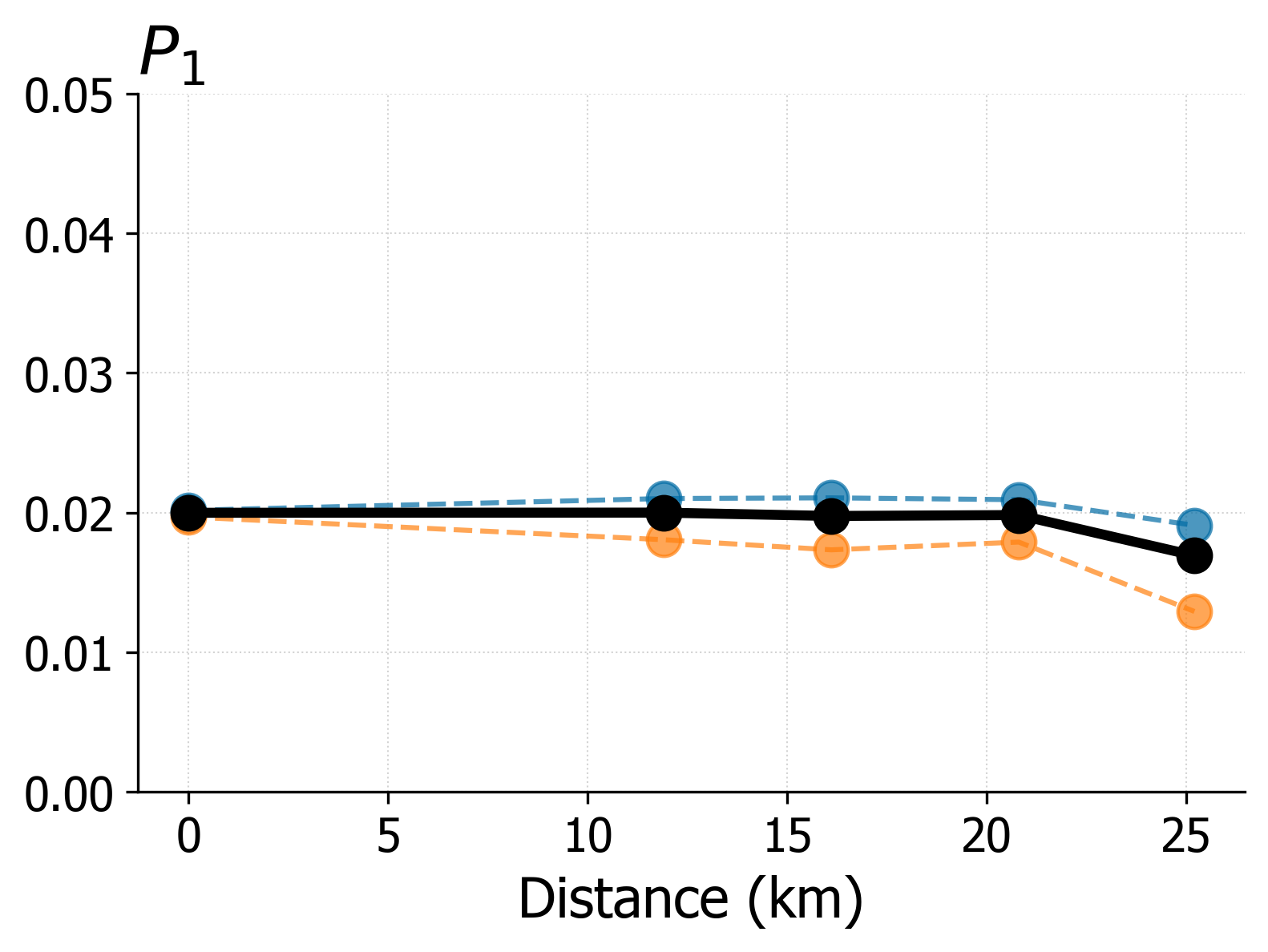}
    \end{subfigure}
    \begin{subfigure}{0.3\textwidth}
        \centering
        \includegraphics[width=\textwidth]{figs/mNS_TIDE_S1_amp_long.png}
    \end{subfigure}
    \hfill
    \begin{subfigure}{0.3\textwidth}
        \centering
        \includegraphics[width=\textwidth]{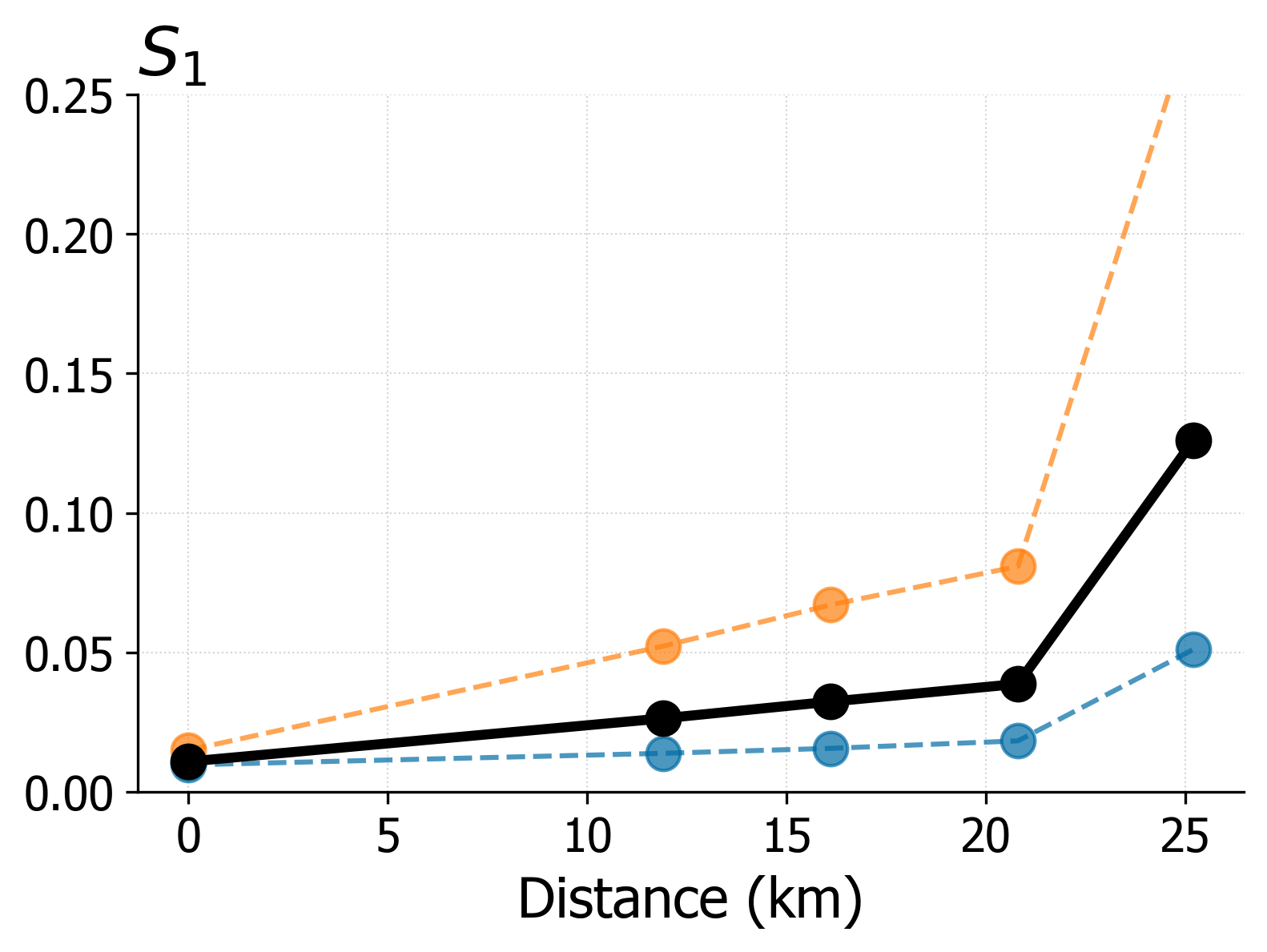}
    \end{subfigure}
    \hfill
    \begin{subfigure}{0.3\textwidth}
        \centering
        \includegraphics[width=\textwidth]{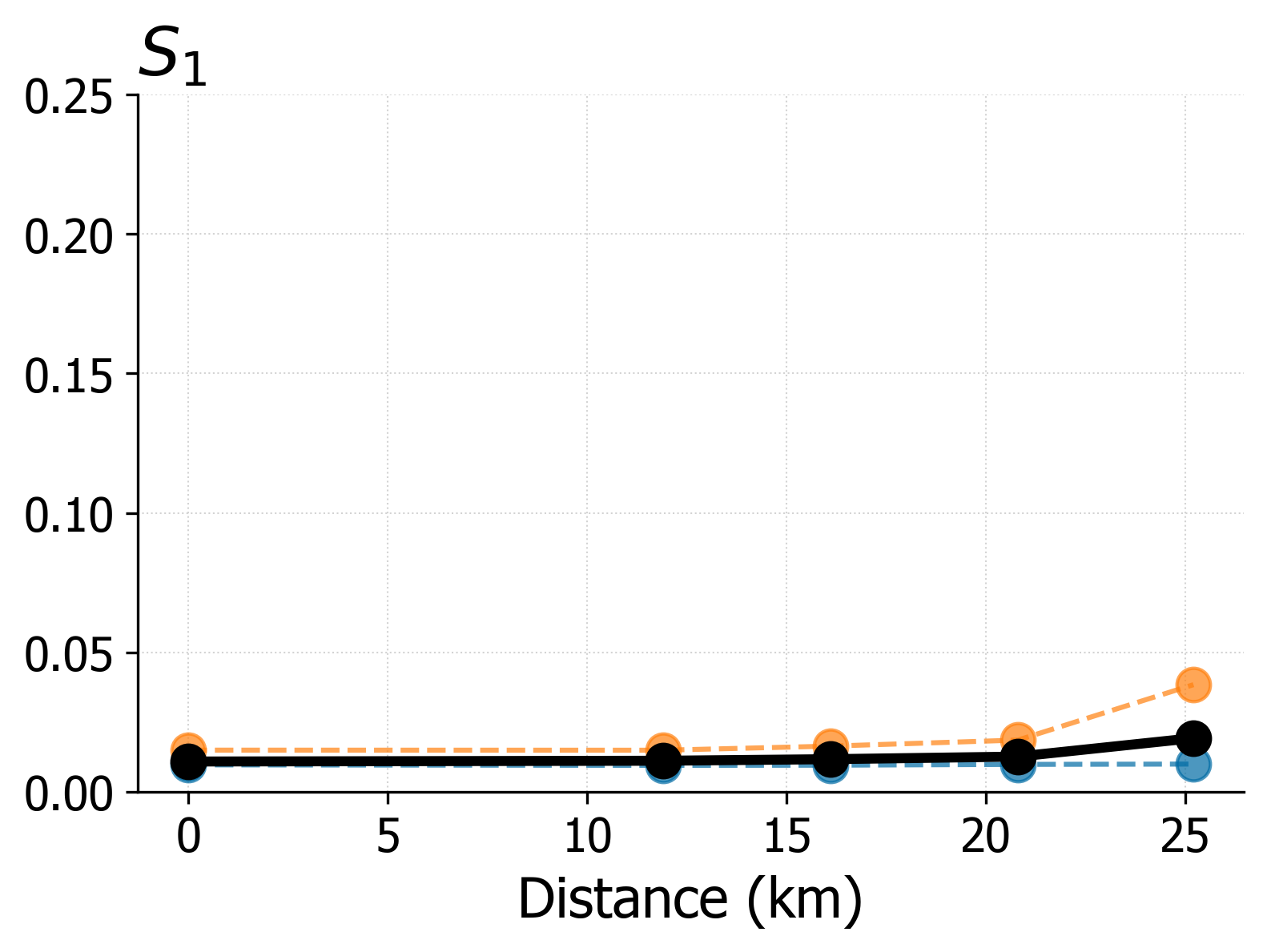}
    \end{subfigure}
    \begin{subfigure}{0.3\textwidth}
        \centering
        \includegraphics[width=\textwidth]{figs/mNS_TIDE_S2_amp_long.png}
    \end{subfigure}
    \hfill
    \begin{subfigure}{0.3\textwidth}
        \centering
        \includegraphics[width=\textwidth]{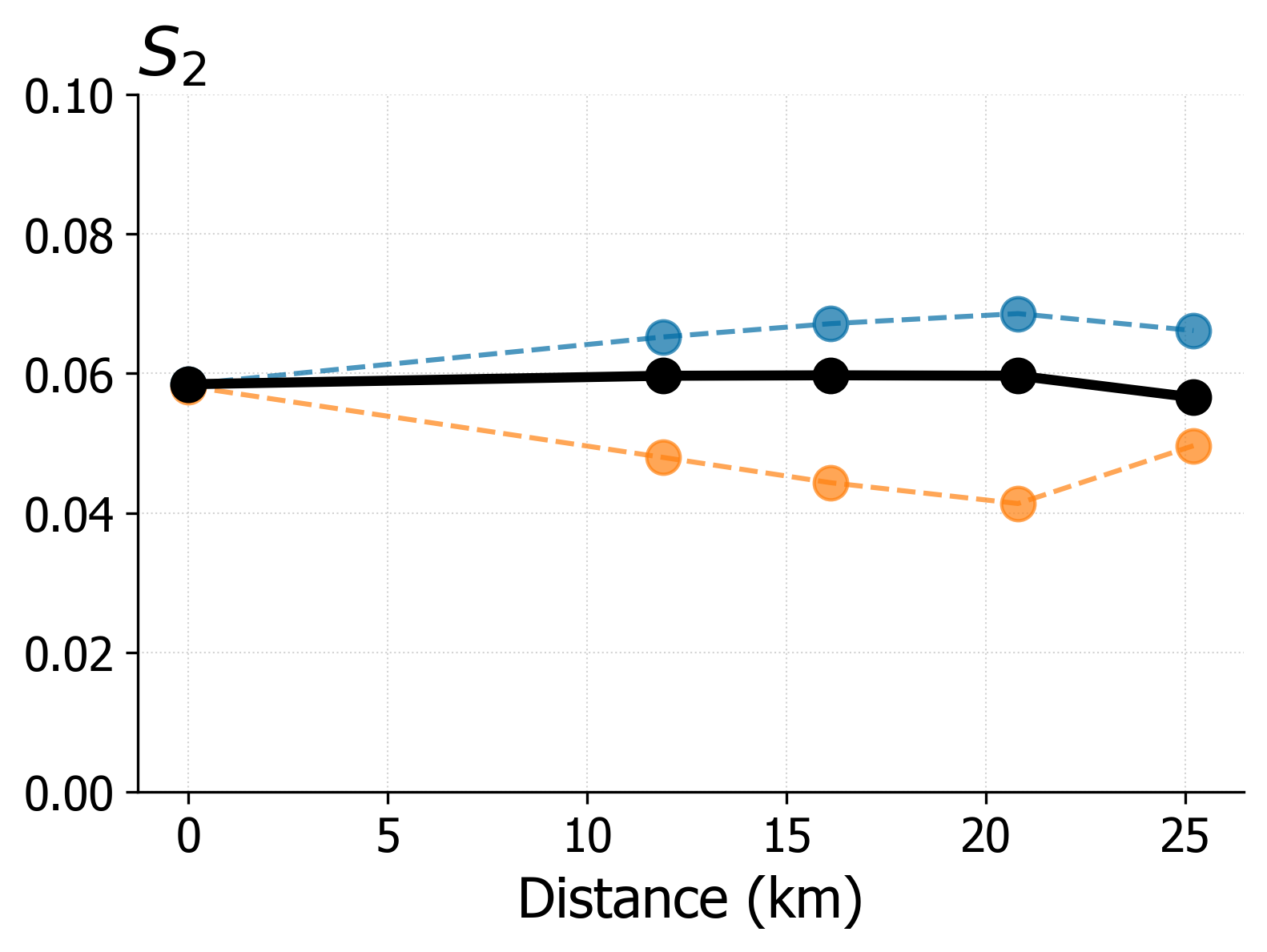}
    \end{subfigure}
    \hfill
    \begin{subfigure}{0.3\textwidth}
        \centering
        \includegraphics[width=\textwidth]{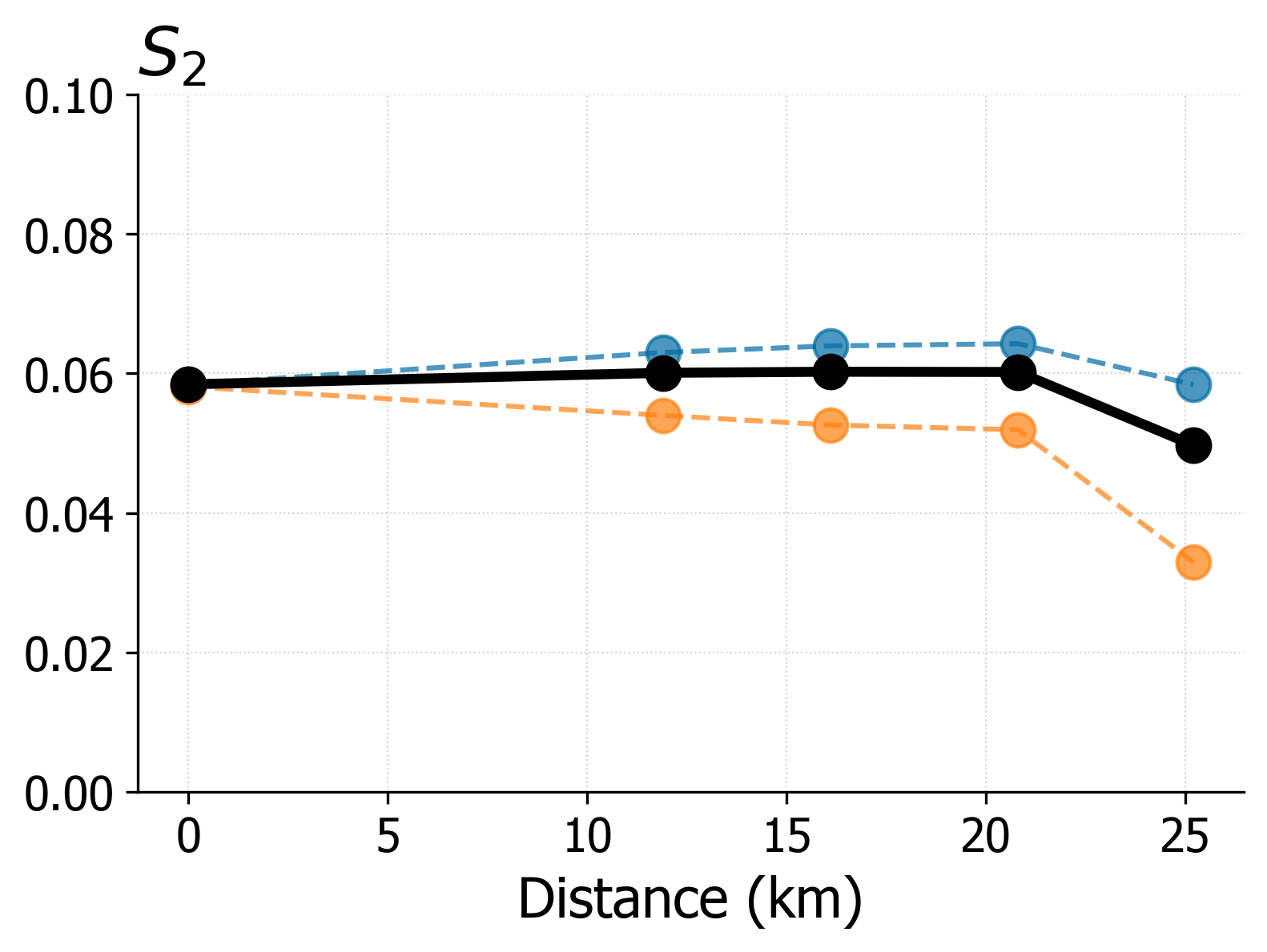}
    \end{subfigure}
    \caption{Changes across the stations of the $\mu$NS\_Tide amplitudes for main diurnal and semi-diurnal constituents, during average flow conditions, high-flow conditions (river discharge above 75th-percentile), and low-flow conditions (river discharge below 25th-percentile) obtained for: a) measured water levels, b) water levels simulated with observed river flow (Simulation A), and c) water levels simulated with filtered river flow (Simulation B). Station locations: Usce (0 rkm), Opuzen (11.9 rkm), Norin (16.1 rkm), Metković (20.8 rkm), and Gabela (25.2 rkm).}
    \label{fig:discussion}
\end{figure}

Fig.~\ref{fig:discussion} compares the three estimates of tidal amplitudes obtained from measured and STREAM-simulated water levels. Other diurnal and semi-diurnal components are not shown as there are no significant differences between the three estimates.
The comparison for some relevant amplitude ratios ($K_1/K_2$, $S_1/S_2$, $K_1/S_1$, and $P_1/S_1$) is shown in Fig.~F.1 in the Supplementary Materials.

The agreement between Fig.~\ref{fig:discussion}a and \ref{fig:discussion}b confirms that the STREAM model reliably reproduces 
the tidal-fluvial dynamics along the Neretva River under different flow conditions. In particular, the observed variations of the considered tidal amplitudes across the stations are well reproduced in simulation A. For instance, the noticeable increase in $S_1$ amplitude at Gabela for high flow  (orange curve) is clearly visible in both Fig.~\ref{fig:discussion}a and \ref{fig:discussion}b.
Some tidal parameters estimated in Simulation A are more sensitive to flow regimes than the corresponding parameters estimated on the observations, e.g. $P_1$, $S_1$ and $S_2$, especially at upstream stations. On the other hand, the amplitudes of $K_1$ at Gabela are less sensitive to the flow condition in simulation A than on the observed water level. These minor differences could be attributed to local variations in channel geometry or additional factors not fully accounted for by the model.

In simulation B, filtering the high-frequency flow components leads to smaller amplitude changes at the stations.
Specifically, the upstream increase of the $S_1$ constituent visible in Figs.~\ref{fig:discussion}a and \ref{fig:discussion}b is almost absent in simulation B. This confirms that the operation of hydropower plants adds a considerable amount of energy at the $S_1$ frequency through high-frequency discharge fluctuations, especially in the upstream section of the river.
Power peaking also slightly increases the amplitudes of the $K_2$ constituent, which behaves like an overtide of $K_1$, and that would otherwise decrease further upstream. Similarly, without the influence of the dam operations, the amplitudes of $S_2$ would decrease earlier upstream. The $K_1$ constituent appears to be the least affected by the high-frequency fluctuations, but shows a more uniform, less variable behavior without the power peaking, possibly because some of its energy is transferred nonlinearly to $K_2$. A similar behavior can be expected between $S_1$ and $S_2$, with part of the total amplitude of $S_2$ being stimulated by $S_1$ in the presence of power peaking.
On the other hand, $P_1$ shows a clear increase at upstream locations with power peaking, which could be attributed in this case to the annual modulation of $S_1$ (or, equivalently, of the hydropower operations), as $P_1$ and $S_1$ frequencies are separated by 1 cycle/year. Similar increases in $S_1$ and $P_1$ amplitudes have been observed in other regulated estuaries, such as the Columbia River \citep{Jay2015a,Lobo2024} and Qiantang River \citep{Zhou2024}.

\subsection{Decomposition of total water levels and tide-surge-river interactions}

Previous studies mostly used numerical simulation scenarios to quantify the nonlinear interactions between tides, storm surge, and river signals \citep[e.g.][]{Hu2023,Kumbier2018,Qiu2022}. For instance, \citet{Dinapoli2021} isolated the sources of the interaction term by sequentially turning off the advective, bottom friction, and shallow water effect terms in the dynamical equations.
In contrast, using harmonic analysis, wavelet transforms, and numerical filtering for a specific storm event,
\citet{Spicer2019} suggested a decomposition of the total observed water levels into three components: a) a stationary tide, b) a low-frequency surge including influences from river discharge, wind, and pressure-driven storm surge, and c) a high-frequency tide-surge-river interaction. This last term represents how tides modify the surge and river signals or, conversely, how these external forcings modulate the tides. Using a similar signal decomposition in hydrodynamic simulations, \citet{Xiao2021} further quantified the relative importance of tide-river and tide-surge interactions to the overall tide-surge-river interaction during compound floods.

In our approach, the explicit use of non-stationary predictors in the NS\_Tide basis functions allows a finer decomposition of the water level variability and the definitive attribution of low- and high-frequency components to different sources.
As such, the low-frequency non-tidal water levels (stage term) are described as a function of both the river discharge and the storm surge.
The high-frequency non-stationary tides are separated into a stationary purely tidal component, as well as tide-river and tide-surge interaction terms.
Additionally, compared to previous studies, a higher frequency resolution is achieved by resolving multiple tidal constituents within each tidal band rather than just species.
To the best of our knowledge, this is the first attempt to explicitly quantify the nonlinear interactions between tides, storm surge and river flow on a constituent-by-constituent basis.

Moreover, while not detailed here for brevity, new adjusted t-student tests were introduced in the NS\_Tide package to test the significance of the non-stationary predictors for each model term and constituent. These tests effectively confirmed the significance of the tide-surge interaction term for up to 10 tidal constituents depending on the station, despite its small contribution in the Neretva River estuary compared to other signal components.

While the tide-surge interaction was relatively small in our application, previous studies have shown that these interactions in other estuaries can dominate the variability during intense storm conditions.
For instance, in the macrotidal Penobscot estuary, the high-frequency tide-surge-river interaction can be amplified to more than double the low-frequency storm surge levels \citep{Spicer2019,Spicer2021}. 
In the Delaware Bay Estuary, the tide-surge-river interaction mostly affects the total water levels during compound flood-hurricane events. During these events, the tide-river interactions dampen the water levels in the upper estuary, 
whereas tide-surge interactions enhance water levels in the downstream surge-dominated region; in the transition zone, a tipping point is reached where these two effects cancel each other \citep{Xiao2021}. 
Such systems at risk of experiencing strong tide-surge-river interactions and associated flooding may be ideal test cases for future applications of our model.

\subsection{A new versatile formulation for NS\_Tide}

This study demonstrates an application of a new Python package for non-stationary HA originating from the NS\_Tide MATLAB code presented in \citet{matte2013}. Compared to the original version, the new package allows for a more versatile definition of the non-stationary model, with predictors specified by the user independently for the stage and tidal-fluvial model terms for each specific application. Preliminary analyses based on GAM can guide the selection of covariates and the specific functional form of each term.
The software also incorporates the recent advances in uncertainty estimation and circular statistics presented in \citet{Innocenti2022}, as well as additional objective criteria for constituent and predictor selection. 
Among these criteria, the package includes a t-test adjusted to account for tidal parameter dependence that assesses the significance, for example, of the tide-river and tide-surge interactions for each constituent. 

The different model formulations tested in this study confirmed that including the tidal range in Eqs.~\ref{eq:stage_model} and \ref{eq:tidal_model} produces suboptimal results and may introduce nonphysical water level oscillations when multiple constituents per tidal bands are included. Removing the range term or replacing it with other regressors, such as storm surge, improves the model predictability and prevents overfitting, as also shown by \citet{Wu2022a} and \citet{Cai2023}. 

A practical solution to improve the model efficiency, interpretability, and predictive performance was to replace the original non-linear discharge functions with quadratic and linear expressions, respectively, for the stage and tidal-fluvial NS\_Tide terms. These formulations were derived from both theoretical considerations on microtidal estuary dynamics and preliminary explorations based on semi-parametric GAM. 
However, different functional forms may better adjust the tidal-fluvial model in other contexts; for example, a quadratic form could help capture the nonlinear tide-discharge relationship in mesotidal or macrotidal estuaries.
Equally important, other systems and applications may require the use of interaction terms such as the discharge-surge product, $Q \cdot SS$.
In the Neretva River and possibly other microtidal rivers, these additional nonlinear terms did not lead to significant improvements in the reconstructed water levels due to the small tidal ranges characterizing this environment, which were of similar or smaller magnitude than the response to other forcing variables.

\section{Conclusion}

This study investigated the interactions between tides, storm surge, river flow, and power peaking in the Neretva River microtidal estuary, Croatia, using a new non-stationary Harmonic Analysis (HA) formulation built from the existing NS\_Tide model. 
Key contributions include the presentation of the improved NS\_Tide, available as a Python and a MATLAB package, the validation of a new non-stationary HA formulation for microtidal environments (referred to as $\mu$NS\_Tide), and the corresponding insights from the analysis of the microtidal estuarine response to storm surges and power peaking. 

The results showed that a non-stationary HA that uses a linear storm surge together with a linear and quadratic river discharge successfully reproduce the water level dynamics in the entire Neretva estuary, with good agreement between measured and reconstructed water levels from tide-dominated to flow-driven locations.
With the proposed model it was also possible to disentangle the interaction of tides, storm surge and fluvial processes in a complex microtidal environment. Specifically, the non-stationary HA identified the following contributions to the tidal-fluvial dynamics:
\begin{itemize}
    \item The stage component is significantly stronger than the tidal-fluvial term, as expected for a microtidal estuary.  
    \item The river variability dominates the stage at all stations except at the coastal station Usce.
    \item Storm surge effects are consistent along the estuarine stations and decreases upstream along the tidal river section.
    \item The stationary tidal component of the tidal-fluvial term has a similar amplitude at 
    all locations, with a smaller relative net effect upstream due to tide-river interactions that dampen tidal amplitudes when they are out of phase.
    \item The interactions between tides and river flow, although weaker than stationary tides, increased upstream and peaked at Gabela, while interactions between tides and storm surges were minimal compared to other components.
\end{itemize}

Overall, the tidal parameters of both the diurnal and semi-diurnal constituents show moderate and consistent changes along the river, with a small amplitude decrease in the upstream direction. As notable exceptions, the $S_1$ constituent shows a strong amplitude increase at the upstream stations, and the $K_2$ constituent presents 
a significant phase shift. The sensitivity to high and low flow conditions is also noticeable, with the constituent amplitudes generally decreasing at high flow, except for $S_1$ and $P_1$. 
However, some noticeable differences were observed at the Gabela station for the $K_1$ and $P_1$ diurnal constituents.

These results emphasize the importance of high-frequency discharge fluctuations in modulating the tidal dynamics in the Neretva River. Power peaking and similar high-frequency signals increase the tidal amplitudes and phase shifts, especially in the transitional tidal river zone.
This is critical for understanding the dynamics of tides and water levels in microtidal estuaries, where even small changes in river discharge can significantly affect tidal propagation.
Considering the valuable insights obtained with the presented approach for the Neretva River estuary, future research should investigate the adaptability and effectiveness of this method in estuaries with more pronounced tide-surge-river dynamics.

\section*{Acknowledgment}
This research was supported by the Croatian Science Foundation under the project "Compound Flooding in Coastal Croatia under Present and Future Climate" (4SeaFlood, project number: IP-2022-10-7598), and the University of Rijeka projects uniri-iskusni-tehnic-23-83, and uniri-iskusni-tehnic-23-74.
The authors thank Croatian Waters and Agency for watershed of the Adriatic Sea (BiH) for their support in providing measurement data.

\section*{Data and code availability}
New NS\_Tide code is available upon request to Pascal Matte (Pascal.Matte@ec.gc.ca) and Silvia Innocenti (silvia.innocenti@ec.gc.ca).
Data used for this study is available upon request from Croatian Waters and Agency for watershed of the Adriatic sea (BiH).

\printcredits


\bibliographystyle{cas-model2-names}

\bibliography{cas-refs}

\end{document}


\maketitle

\newpage

\section{Measured data}

\begin{figure}[h!]
 \centering
 \includegraphics[width=0.85\textwidth]{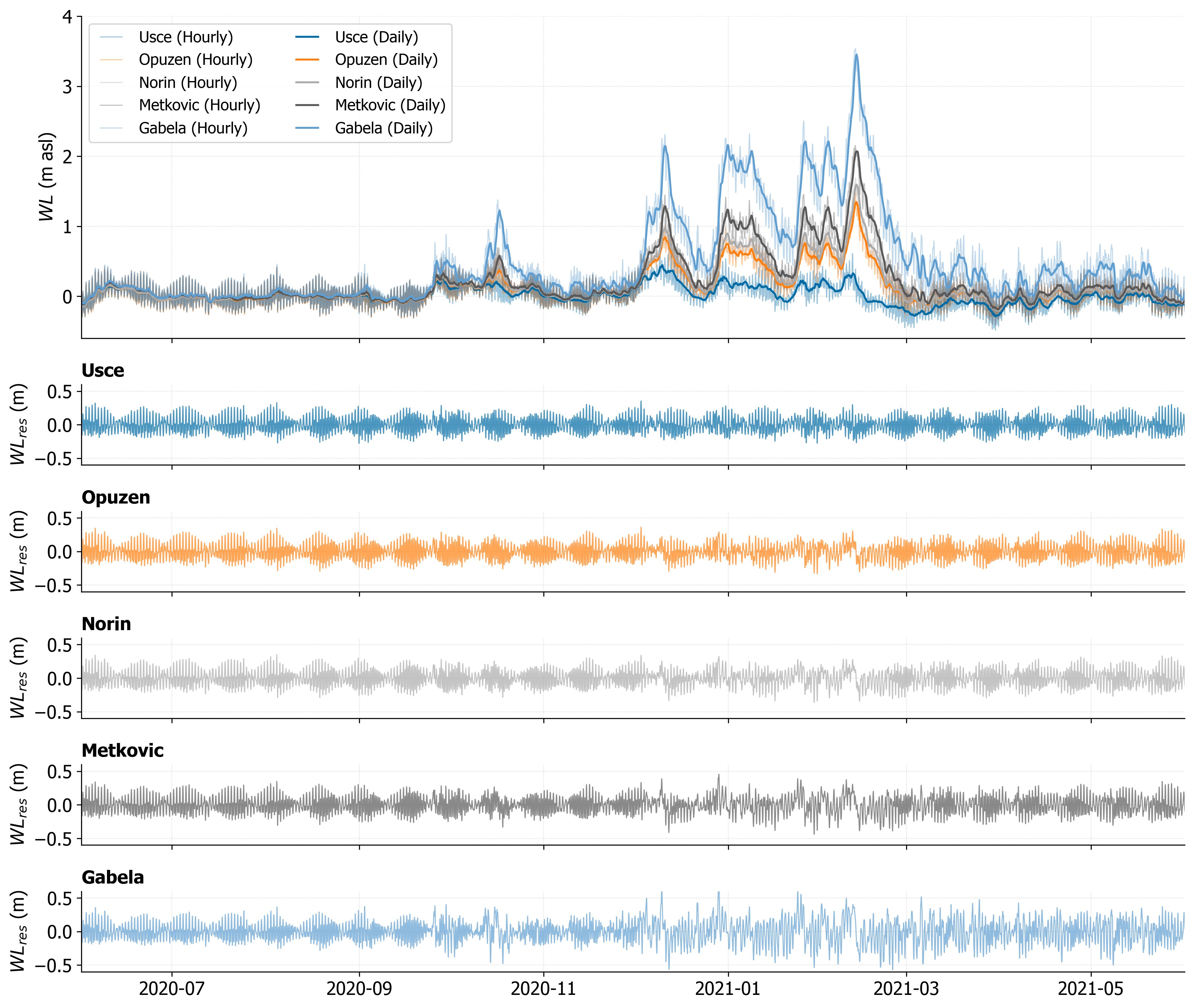}
 \caption{Detail of the measured water level time series and sub-daily residuals (difference between total water levels and 24-hour moving average) at different stations (June 2020 to June 2021).}
 \label{fig:waterlevels1year}
\end{figure}

\begin{figure}[h!]
 \centering
 \includegraphics[width=0.85\textwidth]{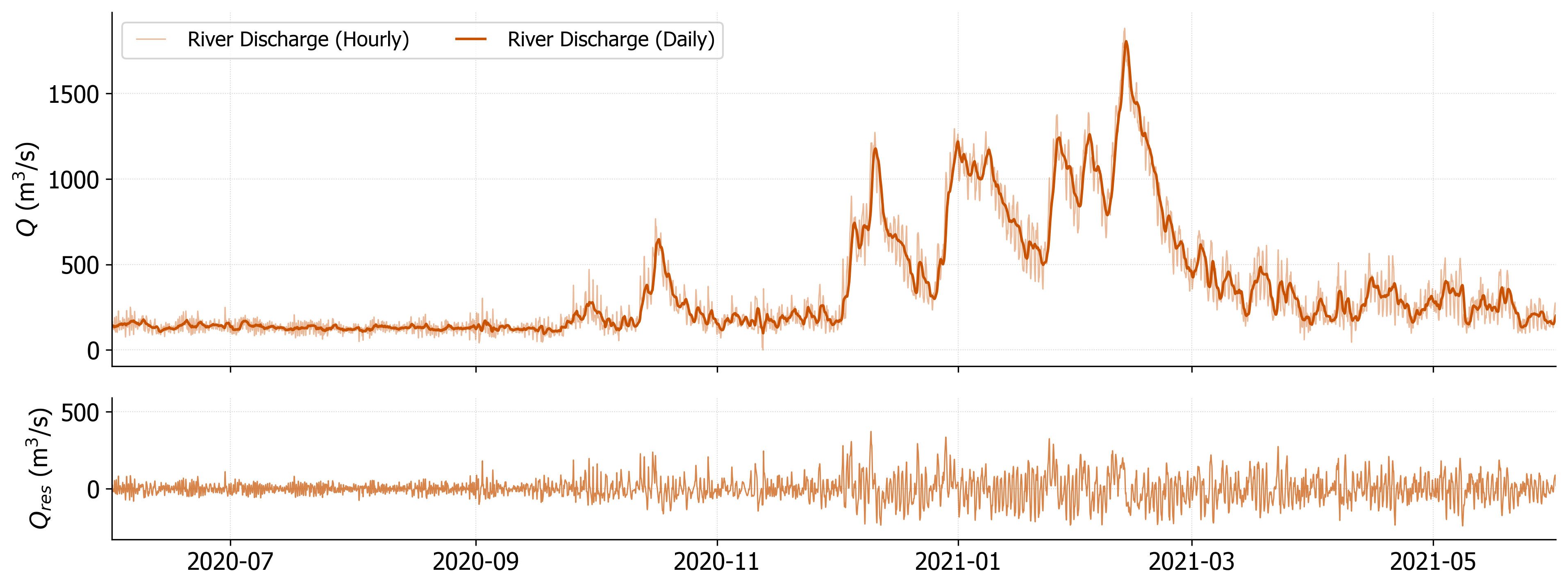}
 \caption{Detail of the measured discharge time series and sub-daily residuals (difference between total discharge and 24-hour moving average) at the Metkovic station (June 2020 to June 2021).}
 \label{fig:flowrate1year}
\end{figure}

\begin{figure}[h!]
 \centering
 \includegraphics[width=0.85\textwidth]{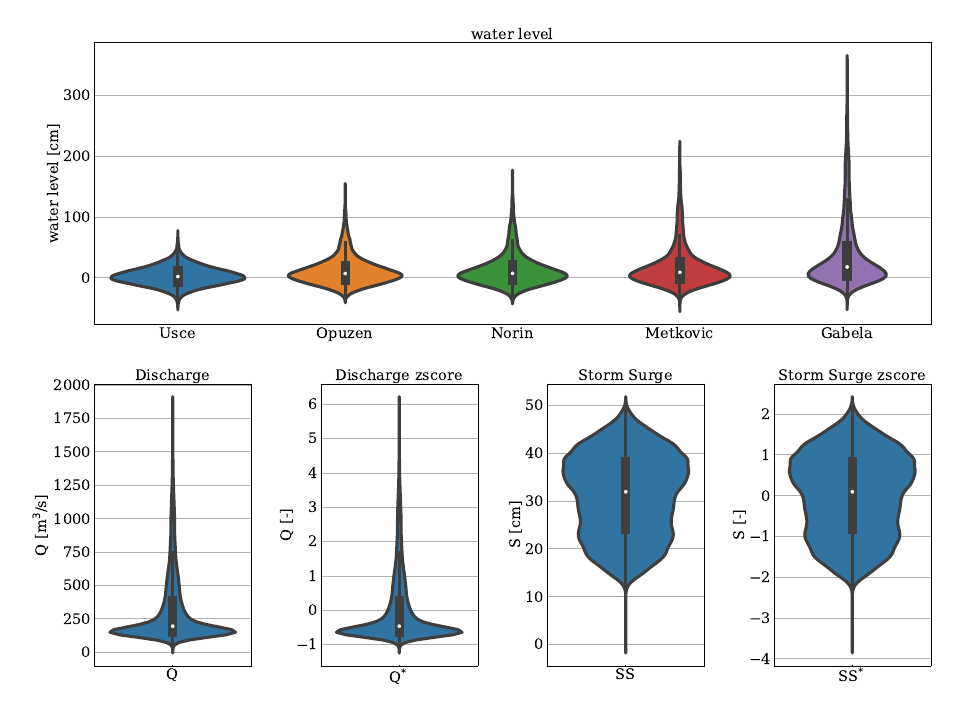}
 \caption{Probability distribution of response variables and covariates: water levels at the five tidal stations (upper panel), river discharge and storm surge and corresponding z-scores, i.e., centred scaled variables (bottom panels).}
 \label{fig:dataviolins}
\end{figure}

\section{Generalized additive model results}

\begin{figure}[ht]
 \centering
 \includegraphics[width=0.95\textwidth]{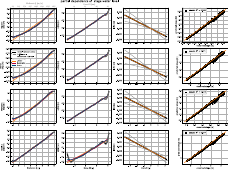}
 \caption{Marginal effect of river discharge $Q$, storm surge $SS$, and coastal tidal range $R$ on the stage component of the water level as estimated by a Generalized Additive Model (GAM), with $s(.)$ indicating a 5-term smoothing spline of order 3. The x-axes are unit-less since they report the reduced and scaled $Q$, $SS$, and $R$ variables.} 
 \label{fig:gam_stage}
\end{figure}

\begin{figure}[ht]
 \centering
 \includegraphics[width=0.95\textwidth]{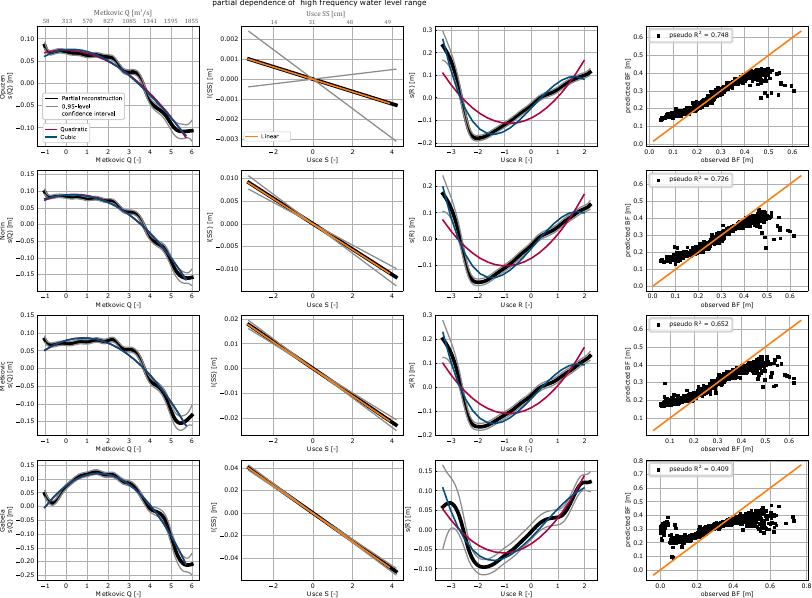}
 \caption{Marginal effect of river discharge $Q$, storm surge $SS$, and coastal tidal range $R$ on the high-frequency water level range as estimated by a Generalized Additive Model (GAM), with $s(.)$ indicating a 5-term 3-order smoothing splines and $l(.)$ a linear term. } 
 \label{fig:gam_hf}
\end{figure}

\section{Spectral analysis of river discharge}

\begin{figure}[ht]
 \centering
 \includegraphics[width=0.95\textwidth]{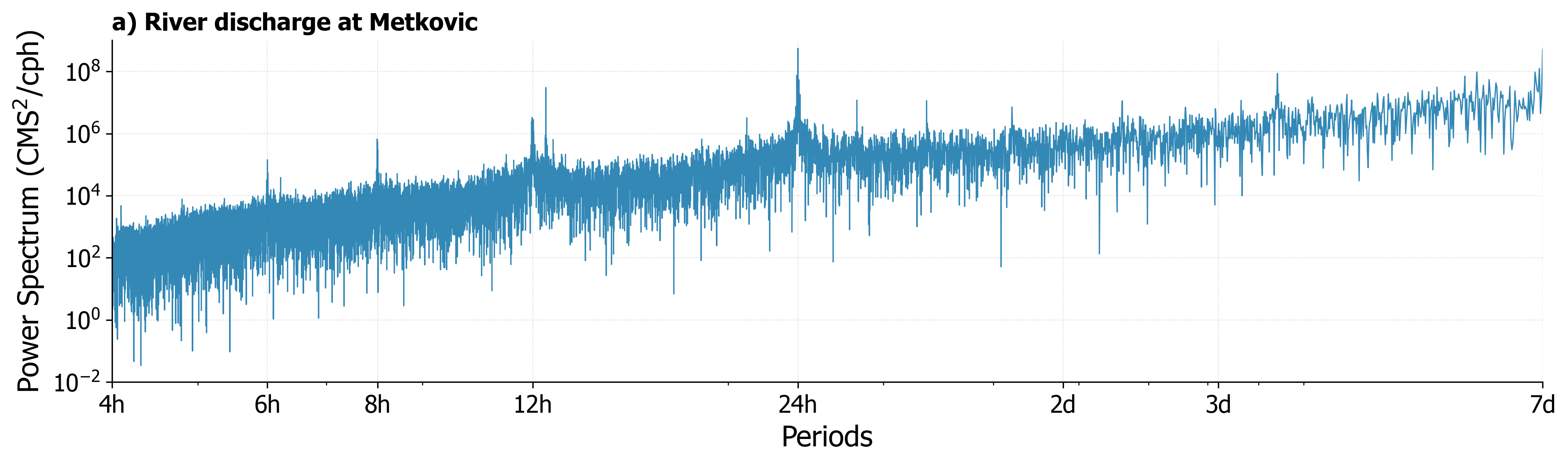}
 \includegraphics[width=0.95\textwidth]{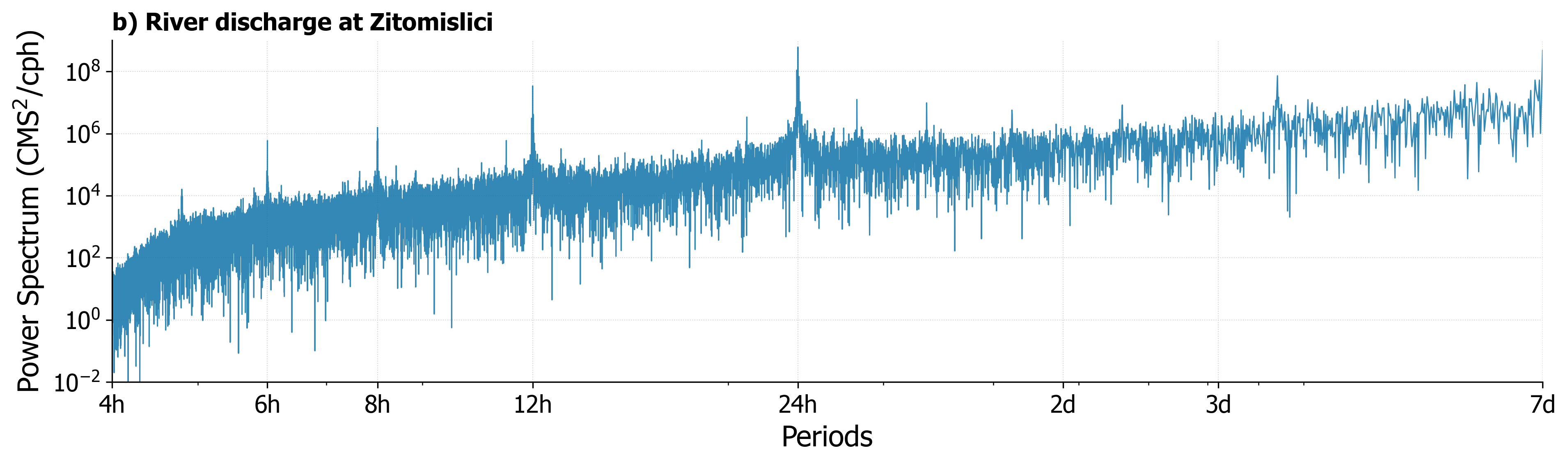} \caption{Power spectrum of measured river discharge at a) Metkovic and b) Zitomislici.}
 \label{fig:Q_spectrum}
\end{figure}

\section{Reconstruction of water levels at the different stations}


\begin{figure}[ht]
 \centering
 \includegraphics[width=0.95\textwidth]{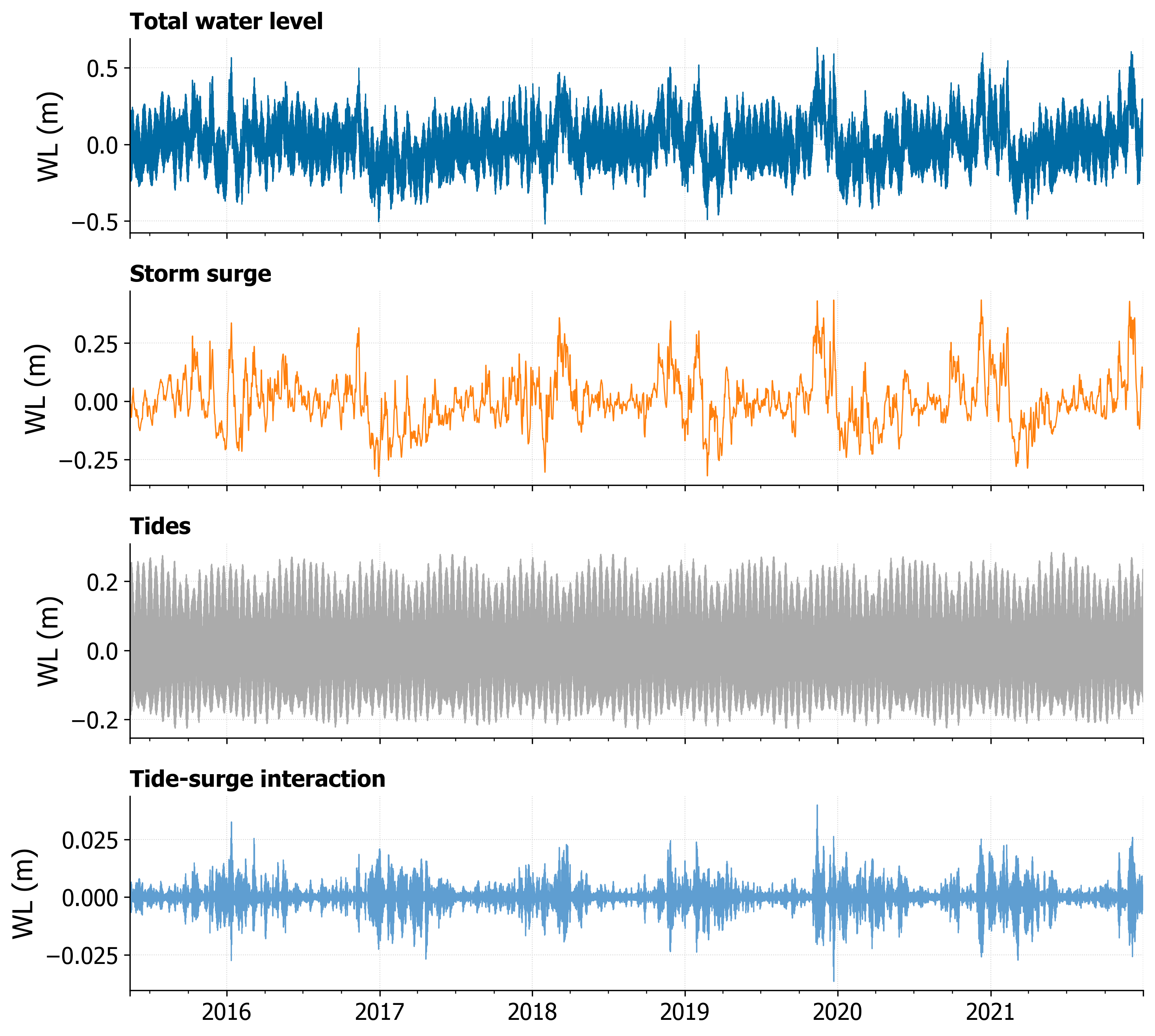}
 \caption{Reconstructed time series of total water levels, using $\mu$NS\_Tide with only linear $SS$ in stage and tidal terms, at the coastal tidal Usce and individual contributions of storm surge, tides, and tide-surge interaction.}
 \label{fig:contributions_usce}
\end{figure}

\begin{figure}[ht]
 \centering
 \includegraphics[width=0.95\textwidth]{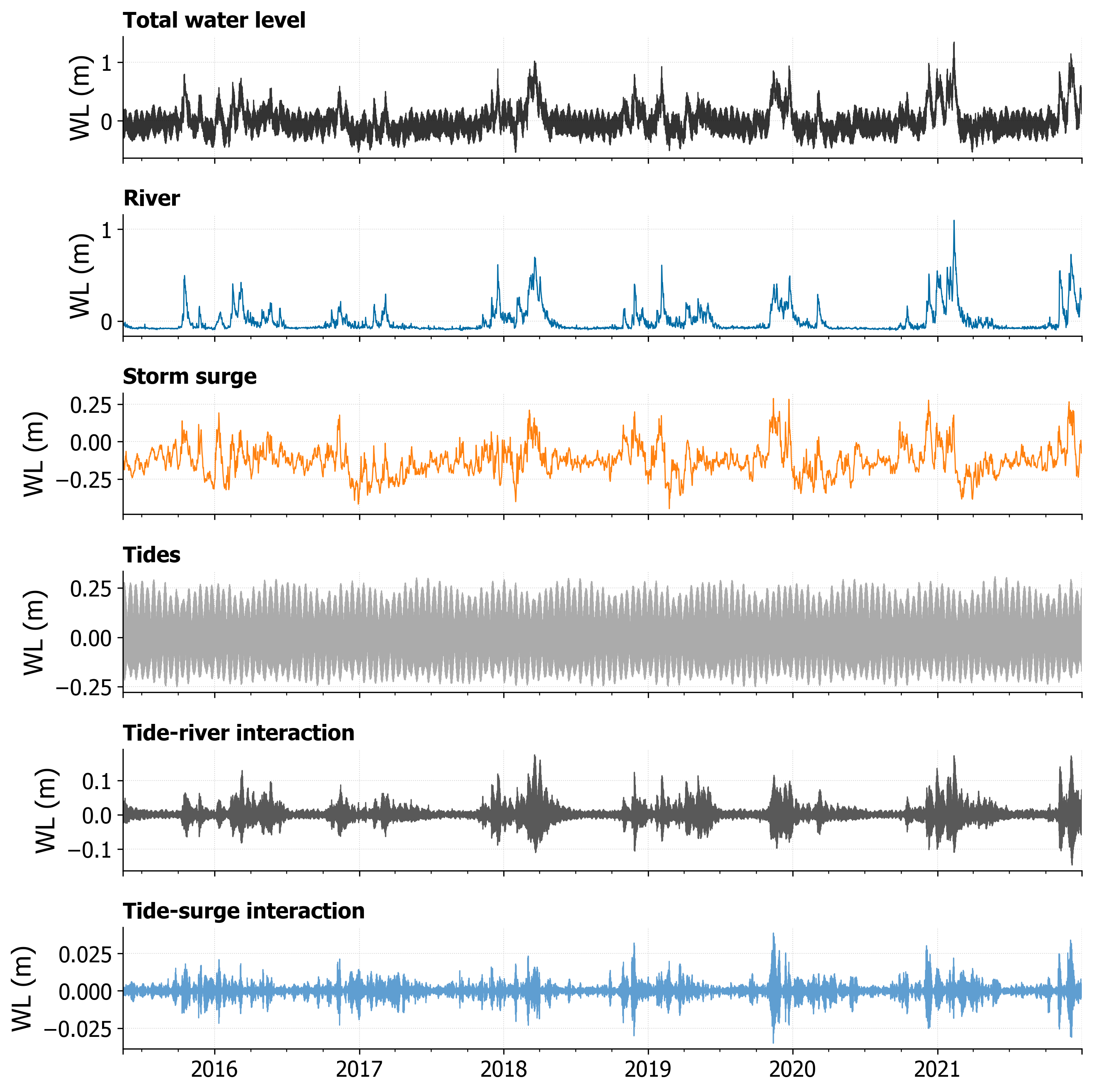}
 \caption{Reconstructed time series of total water levels, using $\mu$NS\_Tide, at Opuzen and individual contributions of discharge, storm surge, tides, tide-river interaction and tide-surge interaction.}
 \label{fig:opuzen_cont_ts}
\end{figure}

\begin{figure}[ht]
 \centering
 \includegraphics[width=0.95\textwidth]{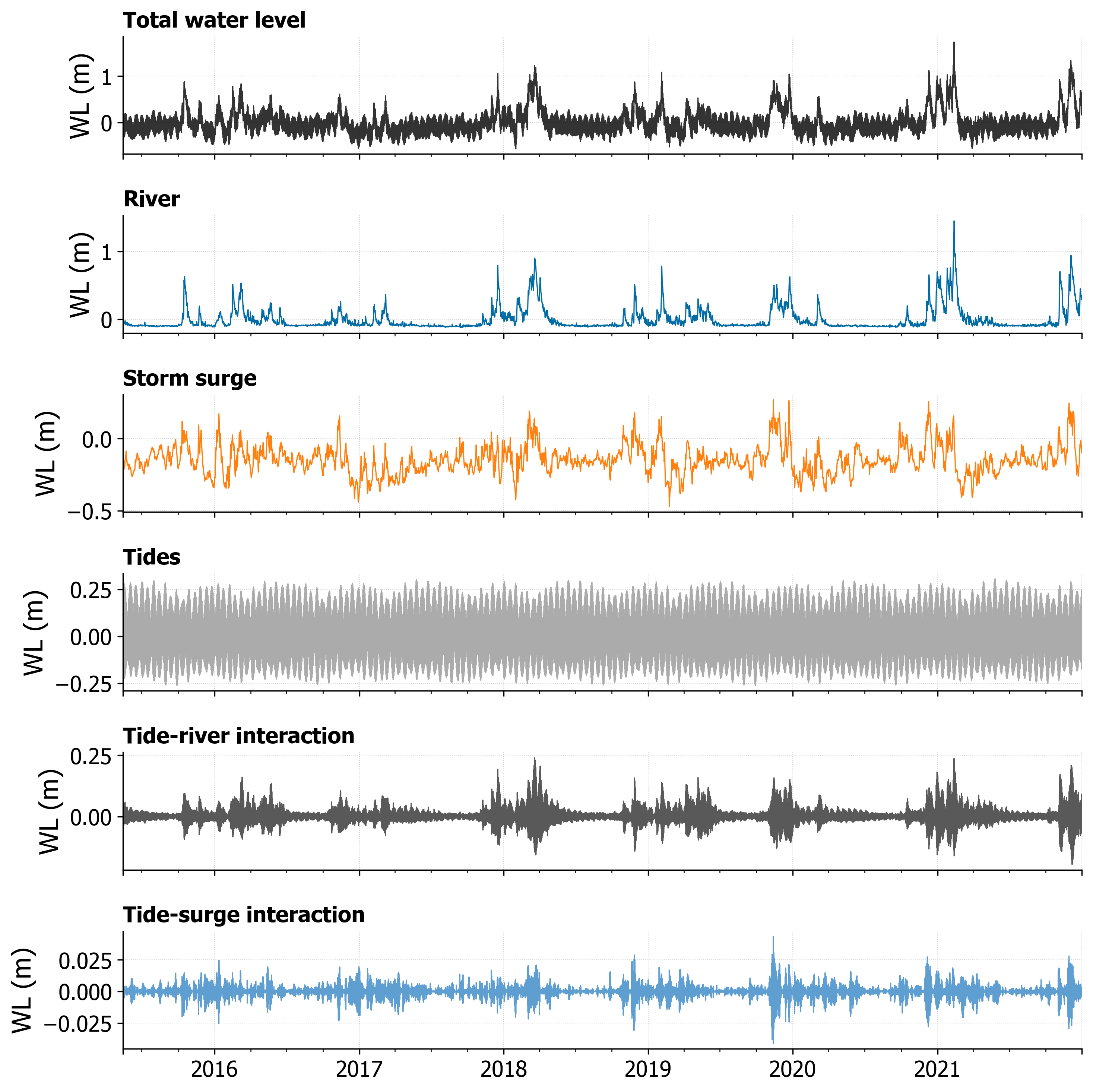}
 \caption{Reconstructed time series of total water levels, using $\mu$NS\_Tide, at Norin and individual contributions of discharge, storm surge, tides, tide-river interaction and tide-surge interaction.}
 \label{fig:norin_cont_ts}
\end{figure}

\begin{figure}[ht]
 \centering
 \includegraphics[width=0.95\textwidth]{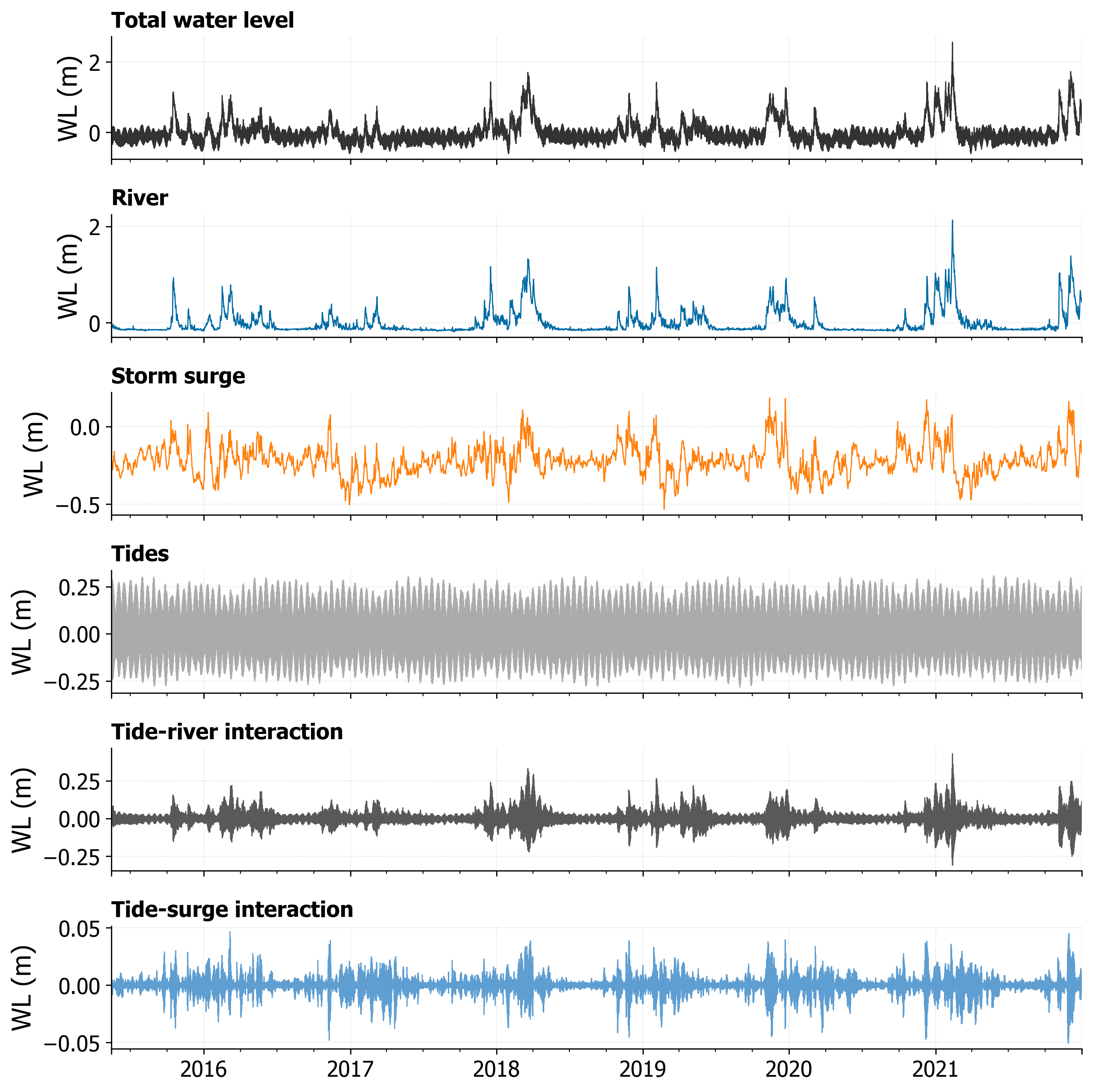}
 \caption{Reconstructed time series of total water levels, using $\mu$NS\_Tide, at Metkovic and individual contributions of discharge, storm surge, tides, tide-river interaction and tide-surge interaction.}
 \label{fig:metkovic_cont_ts}
\end{figure}

\begin{figure}[ht]
 \centering
 \includegraphics[width=0.95\textwidth]{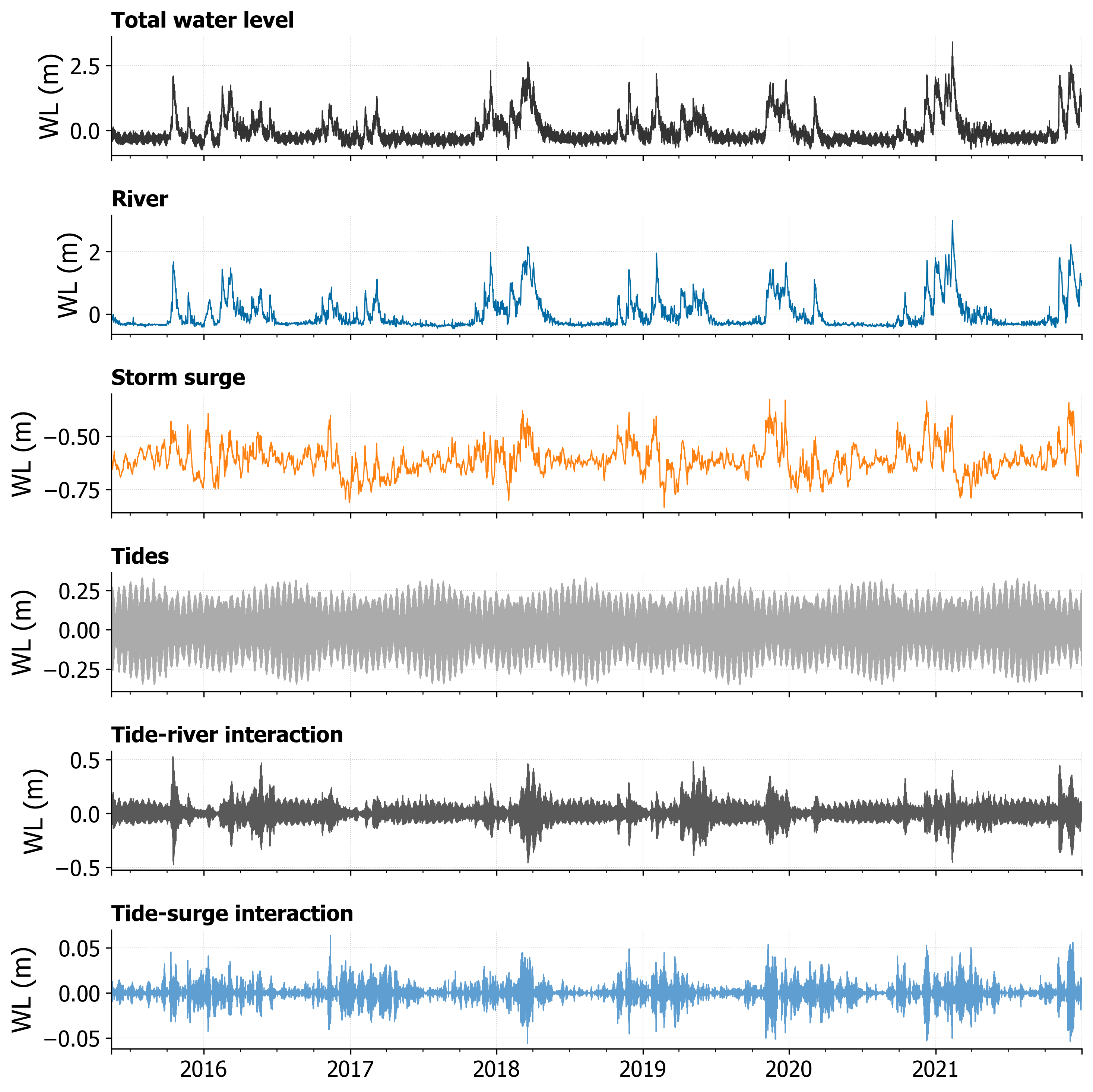}
 \caption{Reconstructed time series of total water levels, using $\mu$NS\_Tide, at Gabela and individual contributions of discharge, storm surge, tides, tide-river interaction and tide-surge interaction.}
 \label{fig:gabela_cont_ts}
\end{figure}

\section{Results of non-stationary harmonic tidal analysis}

\begin{table}[h!]
    \centering
    
     \begin{tabular}{r|l|ll|rr}
     \toprule
     Const & $\overline{SNR}_k$ (-) & $\overline{A}_k$ (cm) & $\sigma_{A_k}$ (cm) & $\overline{g}_k$ ($\circ$) & $\sigma_{g_k}$ (-) \\
    \midrule
     \multicolumn{6}{c}{\textbf{Opuzen, 53 constituents with median SNR$\geq$2 }} \\
     \midrule
 $M_2$   & 81.51 & 9.36 & 0.78 & 96.83 & 3.67 $\cdot10^{-3}$  \\
 $K_1$   & 58.54 & 7.67 & 0.40 & 40.80 & 0.47 $\cdot10^{-3}$  \\
 $S_2$   & 63.25 & 6.56 & 0.41 & 91.20 & 0.35 $\cdot10^{-3}$  \\
 $P_1$   & 23.37 & 2.23 & 0.22 & 27.38 & 8.81 $\cdot10^{-3}$  \\
 $O_1$   & 37.85 & 2.19 & 0.16 & 19.42 & 1.48 $\cdot10^{-3}$  \\
 $K_2$   & 18.90 & 1.67 & 0.06 & 102.94 & 0.38 $\cdot10^{-3}$  \\
 $N_2$   & 27.85 & 1.52 & 0.10 & 98.54 & 3.75 $\cdot10^{-3}$  \\
 $S_1$   & 11.28 & 1.62 & 0.57 & -105.86 & 67.93 $\cdot10^{-3}$  \\
    
     \midrule
     \multicolumn{6}{c}{\textbf{Norin, 54 constituents with median SNR$\geq$2}} \\
     \midrule
 $M_2$   & 132.08 & 9.42 & 0.96 & 96.24 & 6.05 $\cdot10^{-3}$  \\
 $K_1$   & 54.67 & 7.65 & 0.33 & 40.43 & 1.16 $\cdot10^{-3}$  \\
 $S_2$   & 102.16 & 6.59 & 0.57 & 88.41 & 1.22 $\cdot10^{-3}$  \\
 $P_1$   & 20.99 & 2.28 & 0.22 & 26.26 & 5.70 $\cdot10^{-3}$  \\
 $O_1$   & 34.54 & 2.16 & 0.22 & 18.33 & 1.90 $\cdot10^{-3}$  \\
 $K_2$   & 21.59 & 1.75 & 0.08 & 99.88 & 0.11 $\cdot10^{-3}$  \\
 $N_2$   & 39.71 & 1.52 & 0.10 & 98.33 & 4.73 $\cdot10^{-3}$  \\
 $S_1$   & 14.33 & 2.31 & 1.25 & -102.06 & 74.99 $\cdot10^{-3}$  \\

\midrule
\multicolumn{6}{c}{\textbf{Metkovic, 47 constituents with median SNR$\geq$2}} \\
\midrule

 $M_2$ & 29.54 & 8.95 & 1.47 & 104.36 & 5.86 $\cdot10^{-3}$  \\
 $K_1$ & 33.73 & 7.77 & 0.48 & 42.16 & 0.70 $\cdot10^{-3}$  \\
 $S_2$ & 24.27 & 6.36 & 0.64 & 92.70 & 15.73 $\cdot10^{-3}$  \\
 $P_1$ & 13.79 & 2.16 & 0.12 & 29.64 & 0.06 $\cdot10^{-3}$  \\
 $O_1$ & 24.81 & 2.11 & 0.36 & 23.27 & 8.48 $\cdot10^{-3}$  \\
 $K_2$ & 10.39 & 1.57 & 0.11 & 113.60 & 1.08 $\cdot10^{-3}$  \\
 $N_2$ & 16.95 & 1.48 & 0.10 & 107.56 & 10.74 $\cdot10^{-3}$  \\
 $S_1$ & 12.69 & 3.36 & 2.24 & -88.99 & 62.94 $\cdot10^{-3}$  \\

\midrule
\multicolumn{6}{c}{\textbf{Gabela, 45 constituents with median SNR$\geq$2}} \\\midrule
 $M_2$ & 51.70 & 7.45 & 1.74 & 154.15 & 7.27 $\cdot10^{-3}$\\	
 $K_1$ & 19.34 & 6.44 & 1.37 & 62.56 & 9.42 $\cdot10^{-3}$\\	
 $S_2$ & 38.78 & 5.98 & 0.77 & 121.62 & 8.72 $\cdot10^{-3}$\\	
 $P_1$ & 6.31 & 2.34 & 1.72 & 33.02 & 29.41 $\cdot10^{-3}$\\	
 $O_1$ & 14.60 & 1.87 & 0.45 & 42.99 & 5.91 $\cdot10^{-3}$\\	
 $K_2$ & 12.98 & 1.95 & 0.01 & 159.09 & 0.52 $\cdot10^{-3}$\\	
 $N_2$ & 15.51 & 1.25 & 0.27 & 149.25 & 1.74 $\cdot10^{-3}$\\	
 $S_1$ & 20.18 & 9.33 & 7.32 & -30.24 & 0.88 $\cdot10^{-3}$\\	

     \bottomrule
     \end{tabular}
    
     \caption{$\mu$NS\_Tide results for constituents with mean amplitude larger than 1 cm: mean signal-to-noise ratio ($\overline{SNR}_k$, unitless), mean amplitude and amplitude standard error ($\overline{A}_k$ and $\sigma_{A_k}$, cm), phase and phase circular variance  ($\overline{g}_k$, degrees,  and $\sigma_{g_k}$, unitless). Temporal means, indicated by overbars, are computed over the entire observational period.}
     \label{tab:constituents_all}
 \end{table}

\section{Spatial and temporal variations of tidal constituents}
\begin{figure}[h!]
    \centering
    \begin{subfigure}{0.3\textwidth}
    (a) Measured
        \centering
        \includegraphics[width=\textwidth]{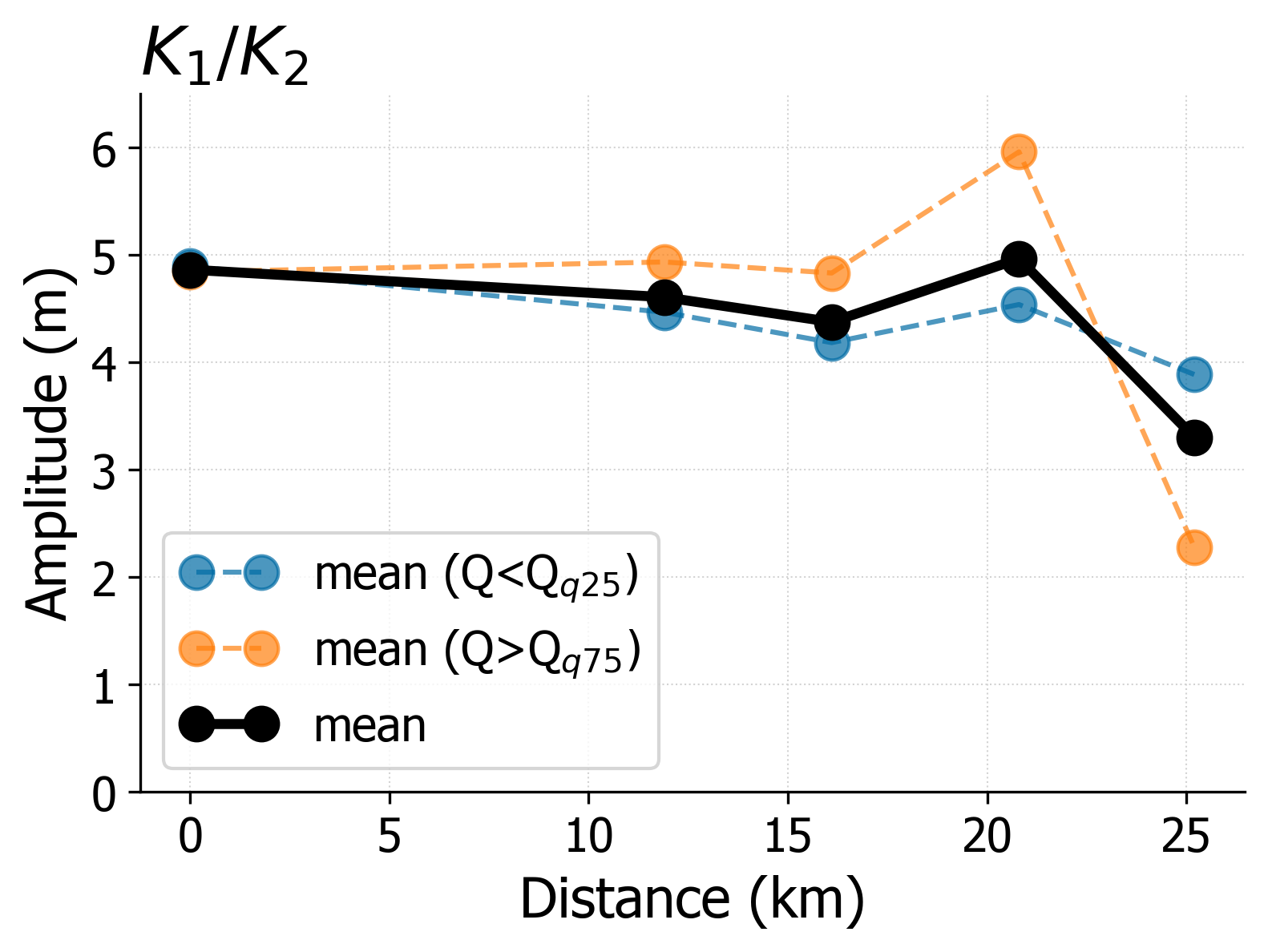}
    \end{subfigure}
    \hfill
    \begin{subfigure}{0.3\textwidth}
     (b) Simulation A ($Q_{mes}$)
       \centering
        \includegraphics[width=\textwidth]{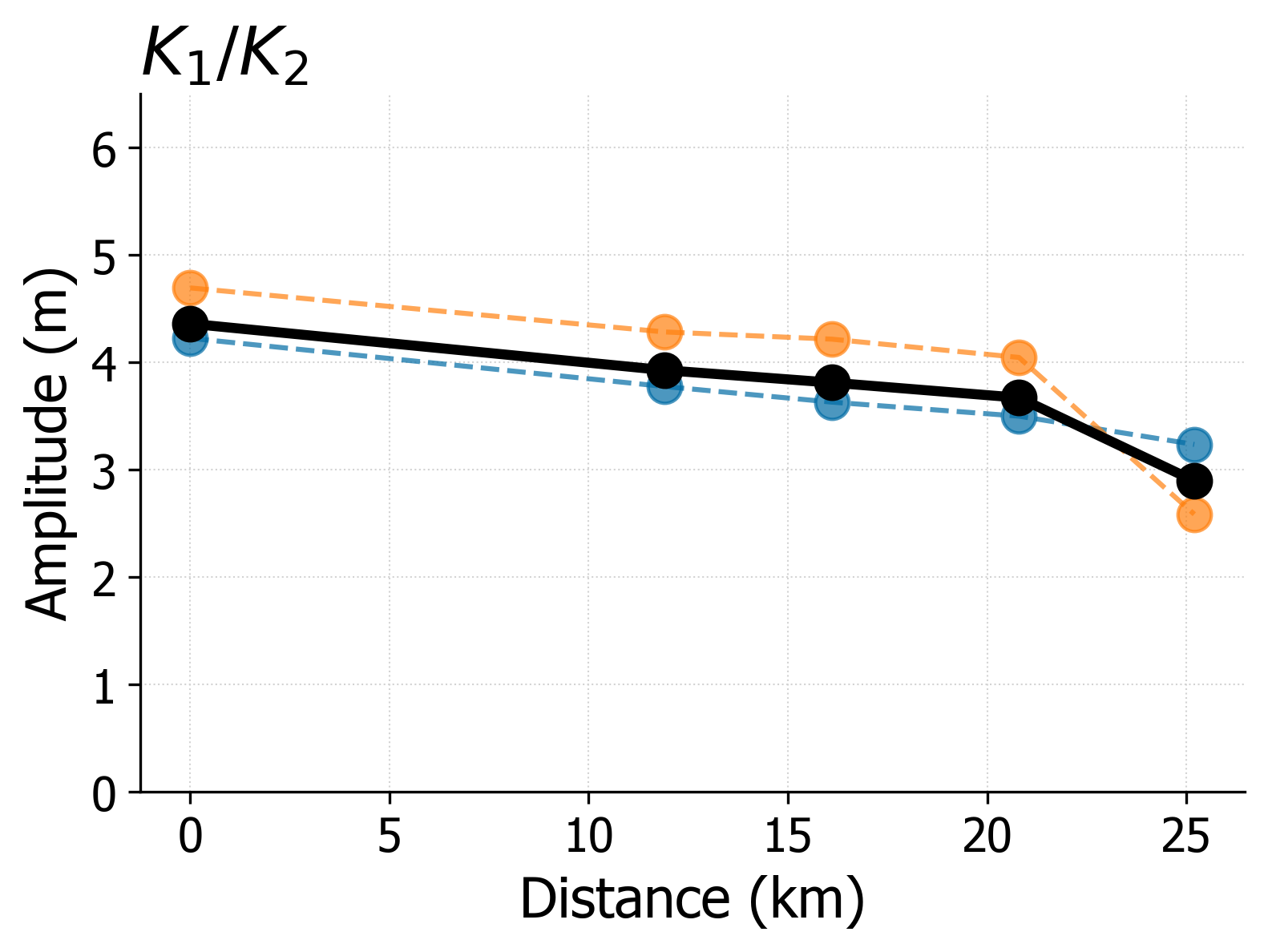}
    \end{subfigure}
    \hfill
    \begin{subfigure}{0.3\textwidth}
    (c) Simulation B ($Q_{filt}$)
        \centering
        \includegraphics[width=\textwidth]{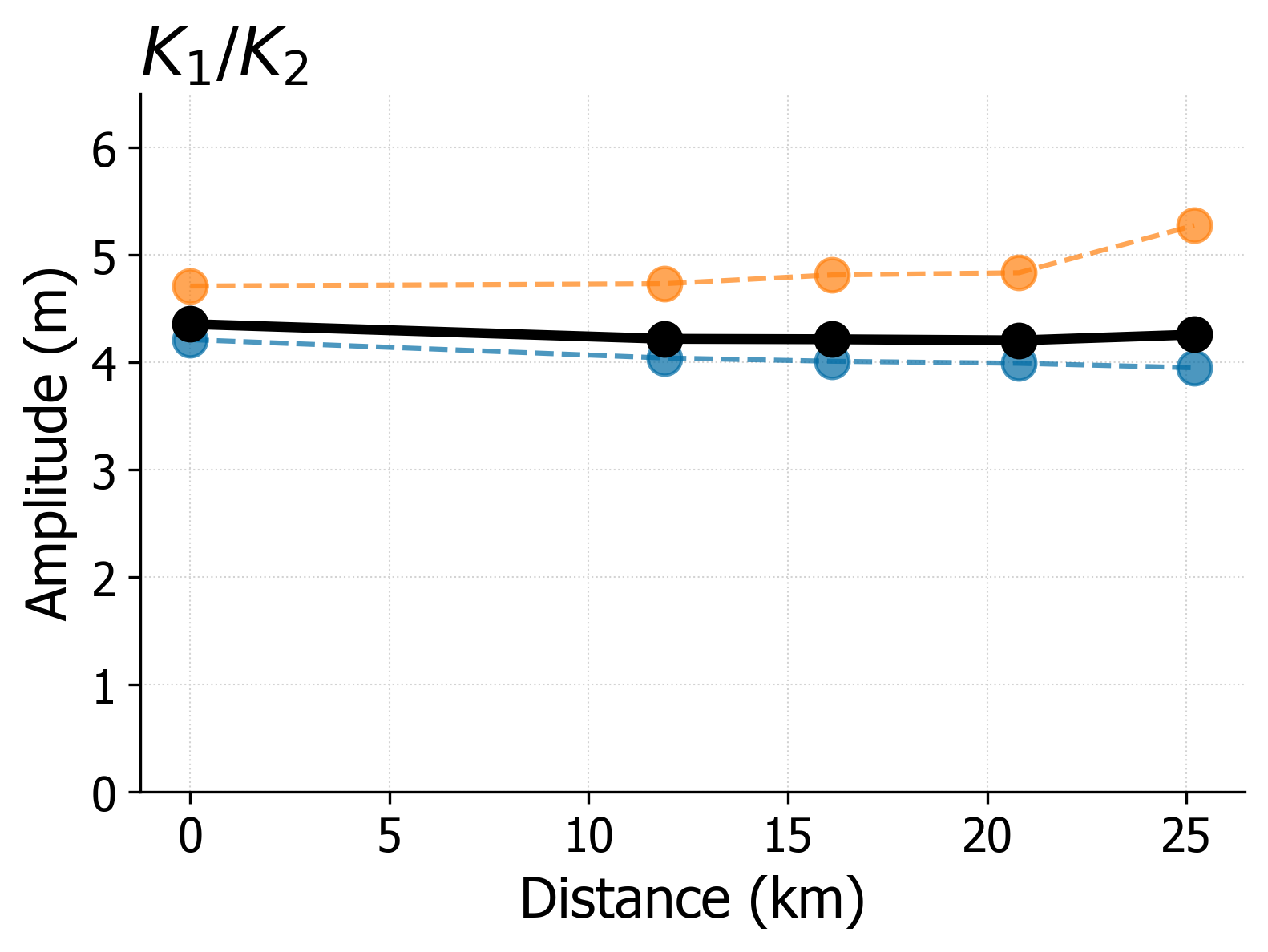}
    \end{subfigure}
    \begin{subfigure}{0.3\textwidth}
        \centering
        \includegraphics[width=\textwidth]{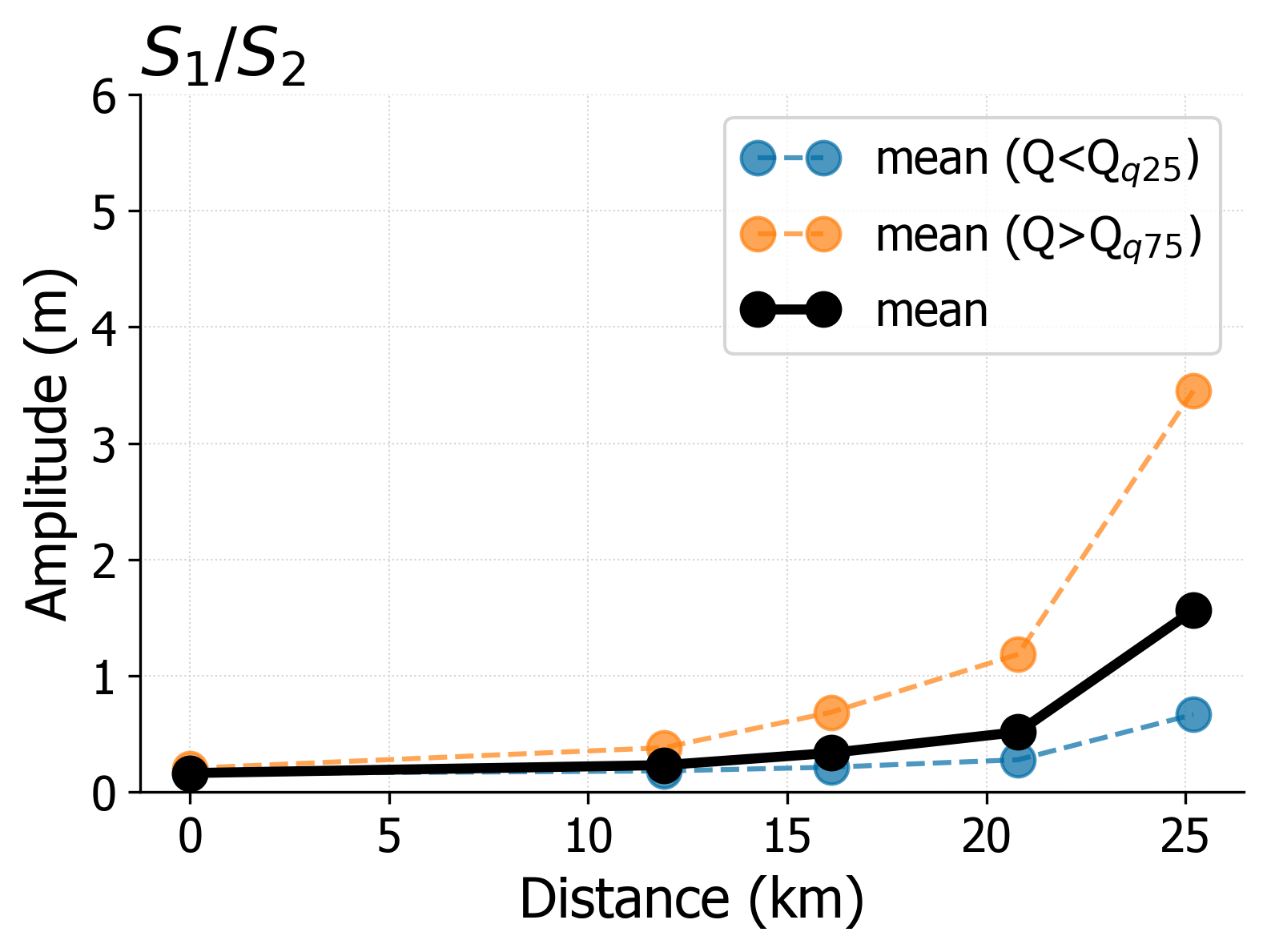}
    \end{subfigure}
    \hfill
    \begin{subfigure}{0.3\textwidth}
        \centering
        \includegraphics[width=\textwidth]{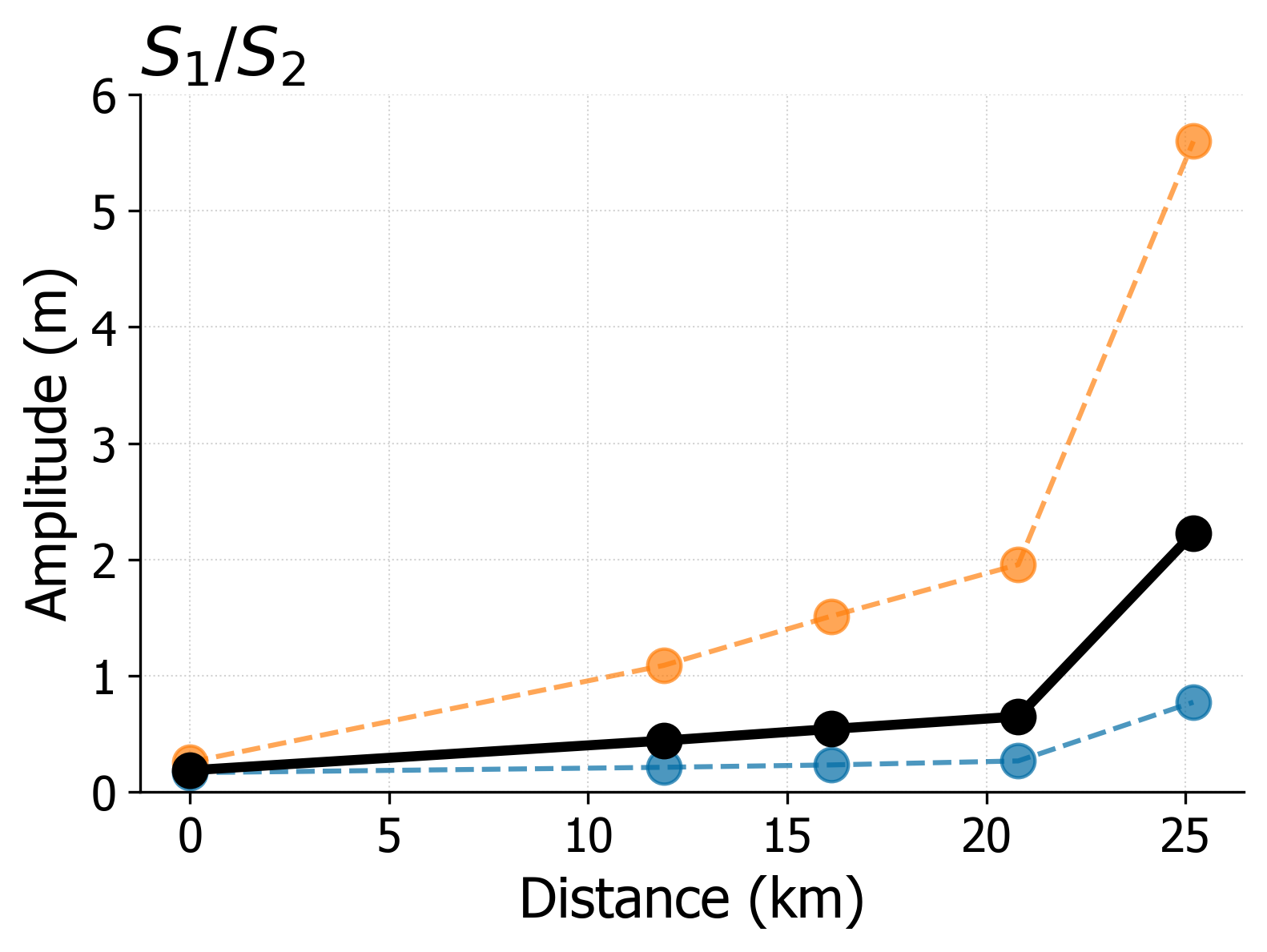}
    \end{subfigure}
    \hfill
    \begin{subfigure}{0.3\textwidth}
        \centering
        \includegraphics[width=\textwidth]{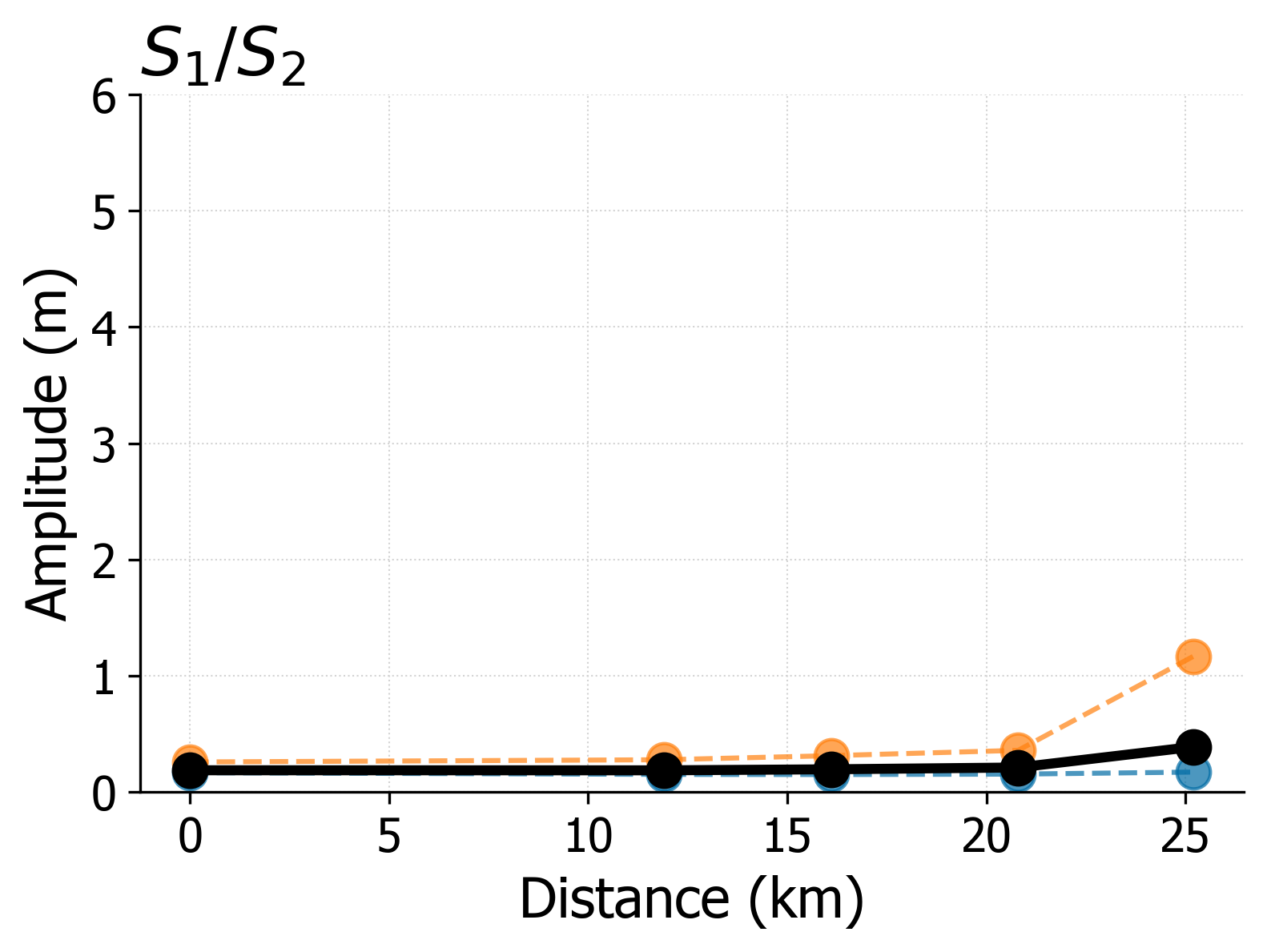}
    \end{subfigure}
    \begin{subfigure}{0.3\textwidth}
        \centering
        \includegraphics[width=\textwidth]{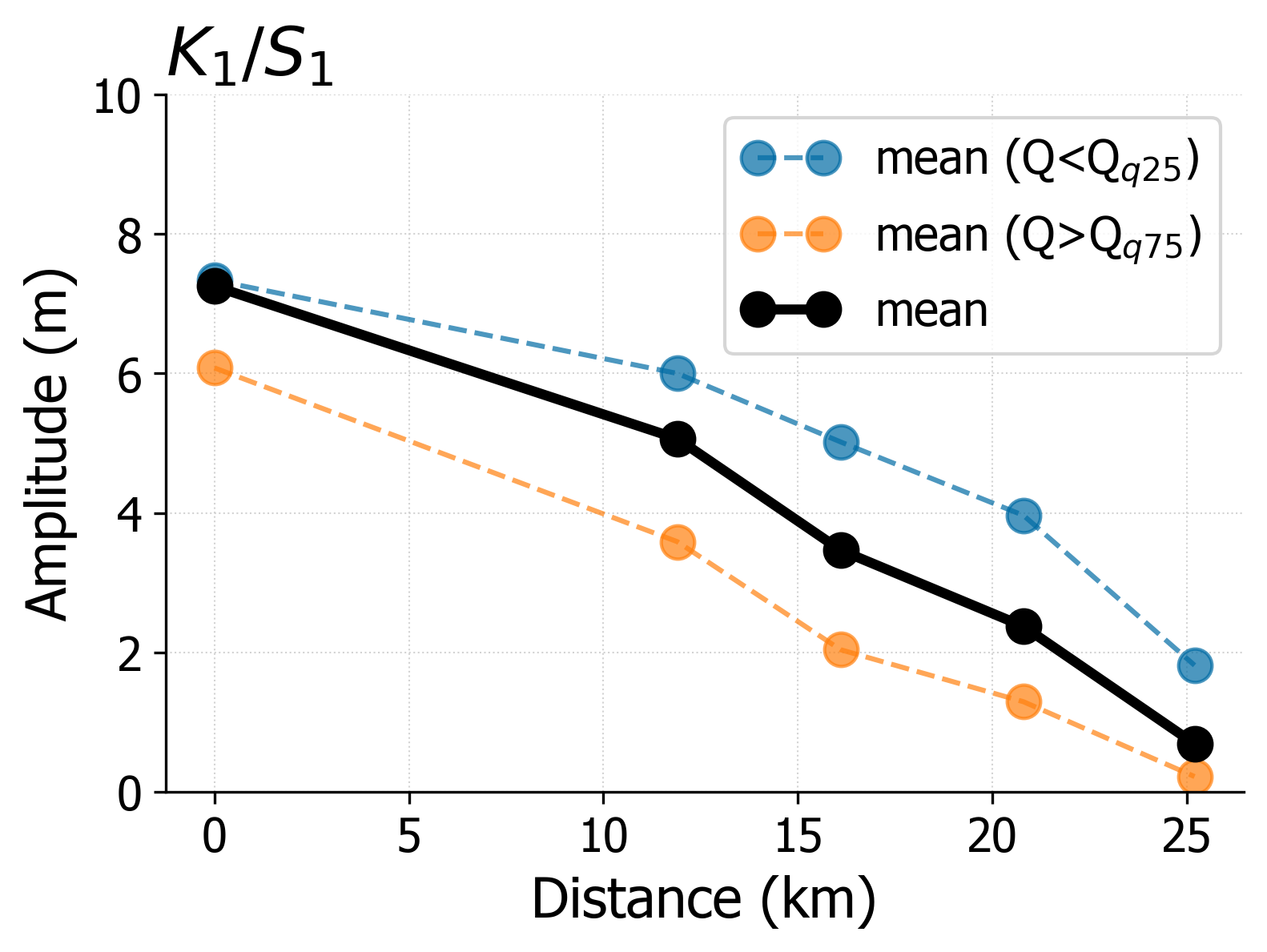}
    \end{subfigure}
    \hfill
    \begin{subfigure}{0.3\textwidth}
        \centering
        \includegraphics[width=\textwidth]{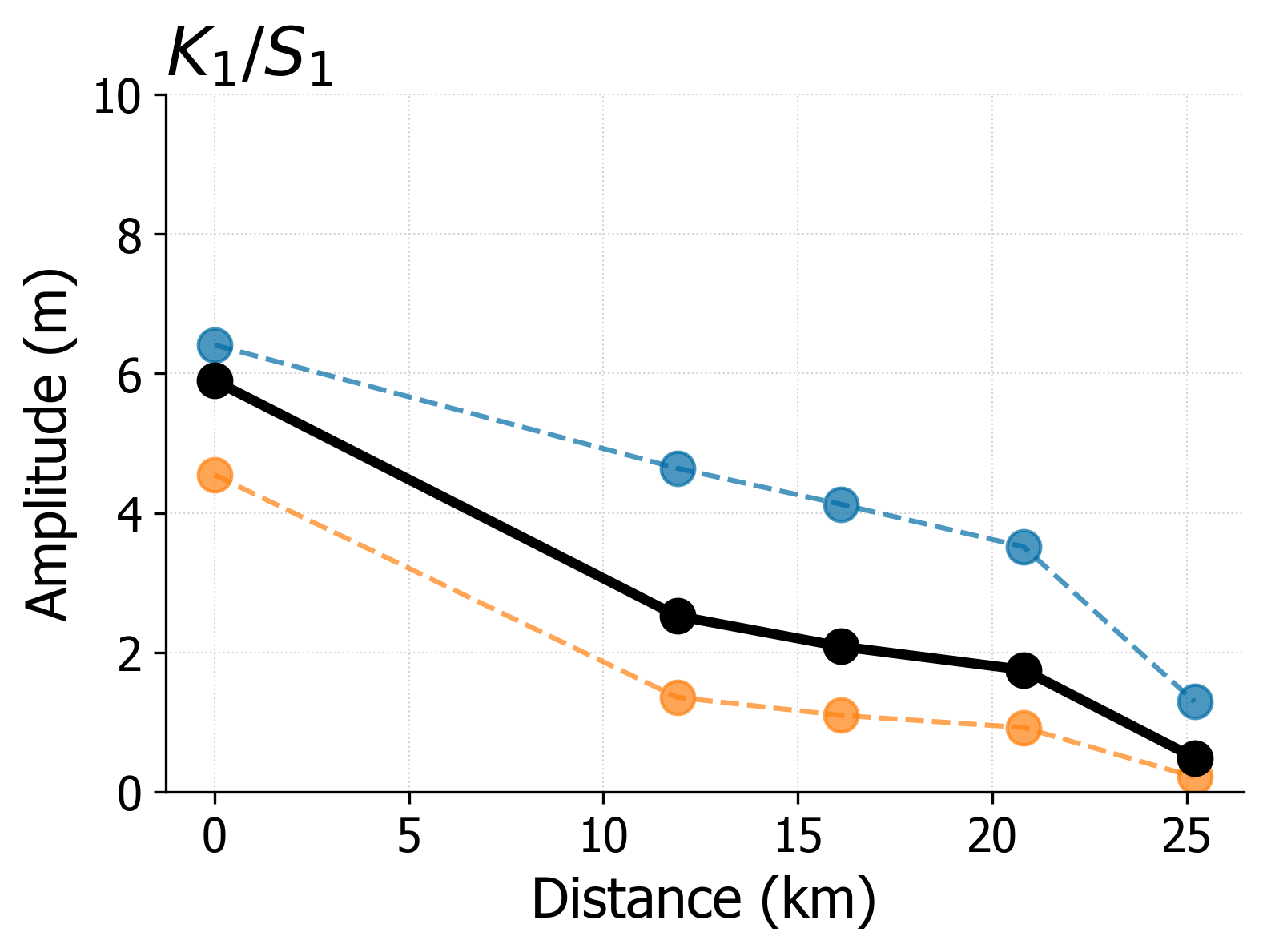}
    \end{subfigure}
    \hfill
    \begin{subfigure}{0.3\textwidth}
        \centering
        \includegraphics[width=\textwidth]{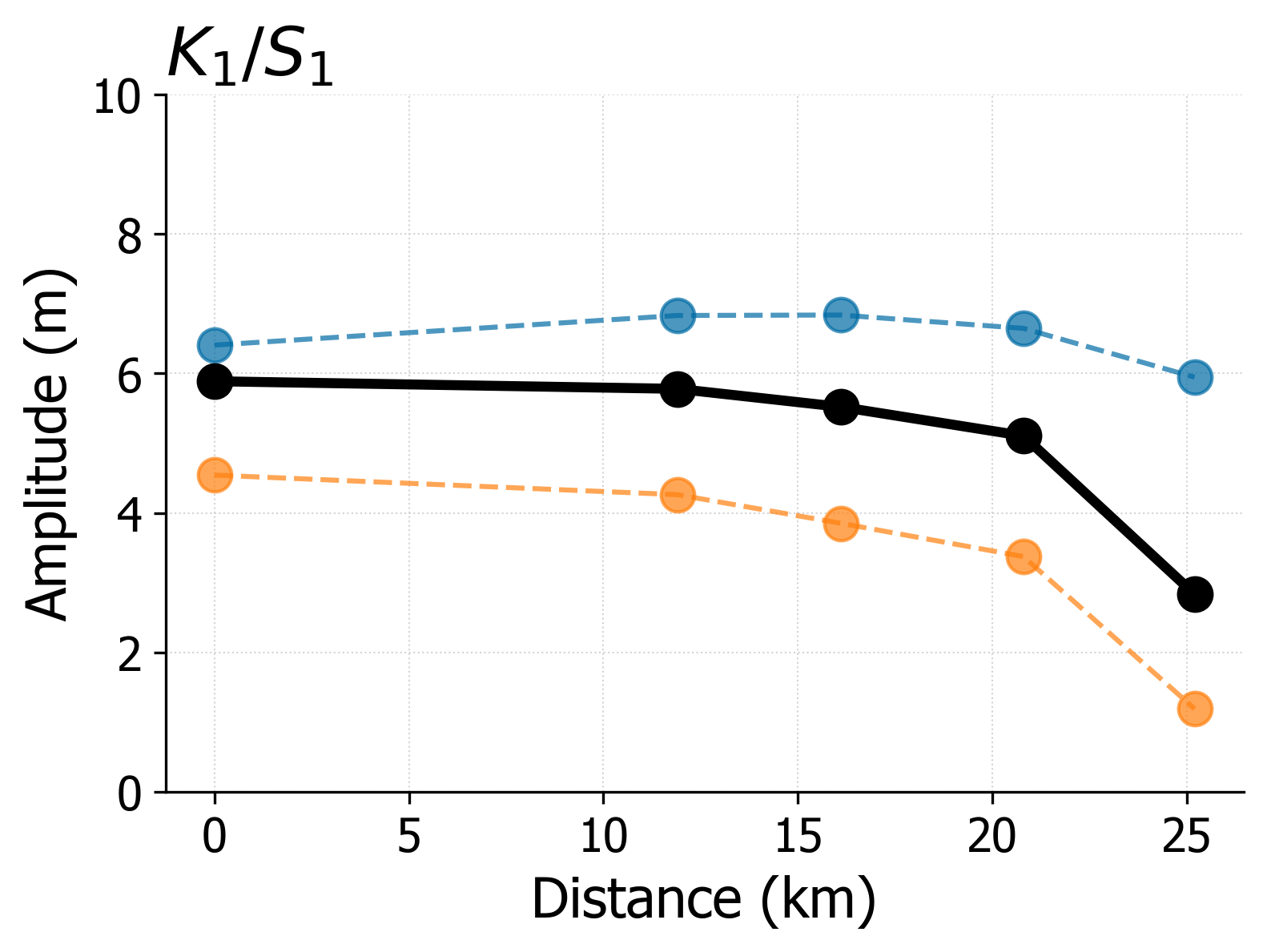}
    \end{subfigure}
    \begin{subfigure}{0.3\textwidth}
        \centering
        \includegraphics[width=\textwidth]{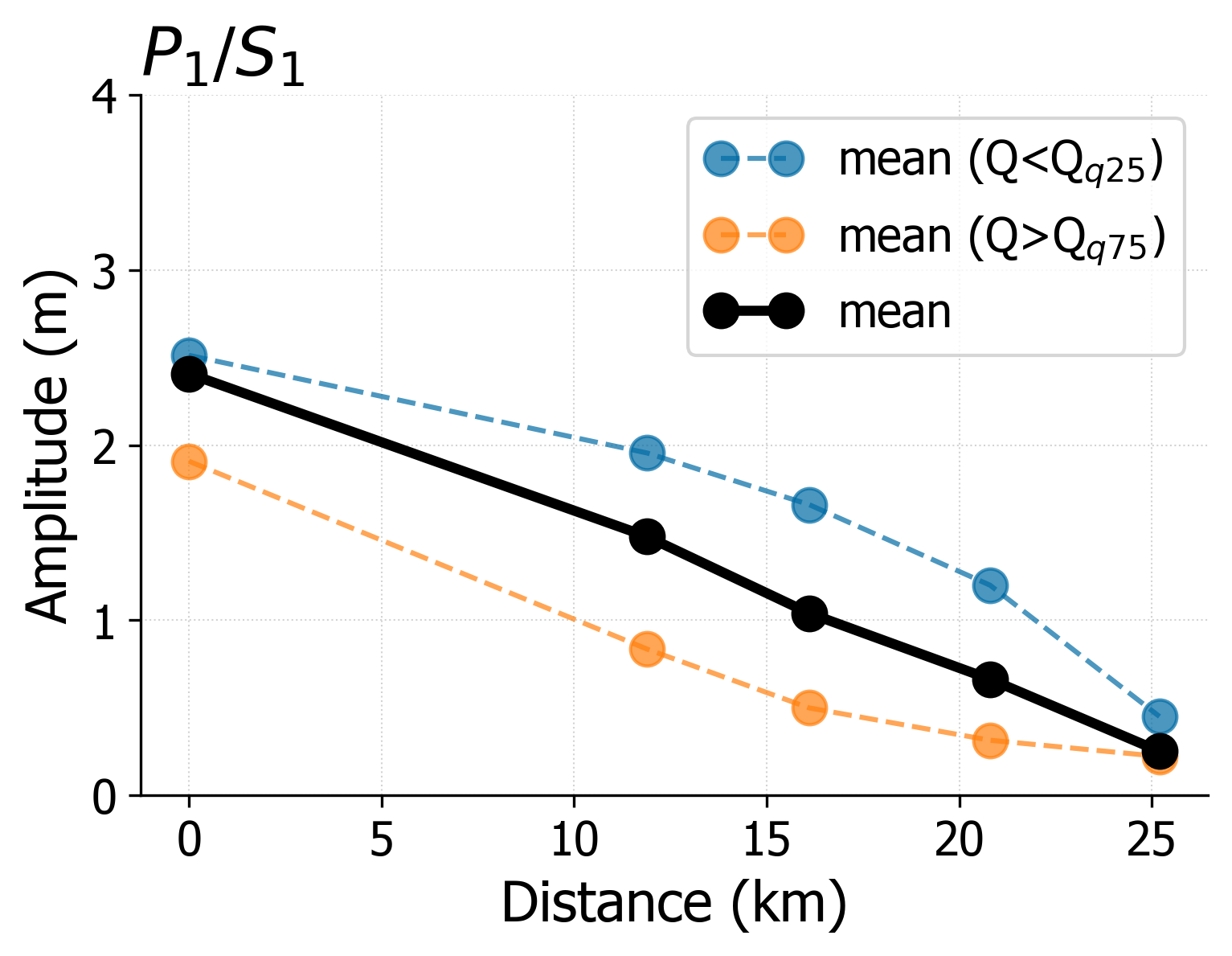}
    \end{subfigure}
    \hfill
    \begin{subfigure}{0.3\textwidth}
        \centering
        \includegraphics[width=\textwidth]{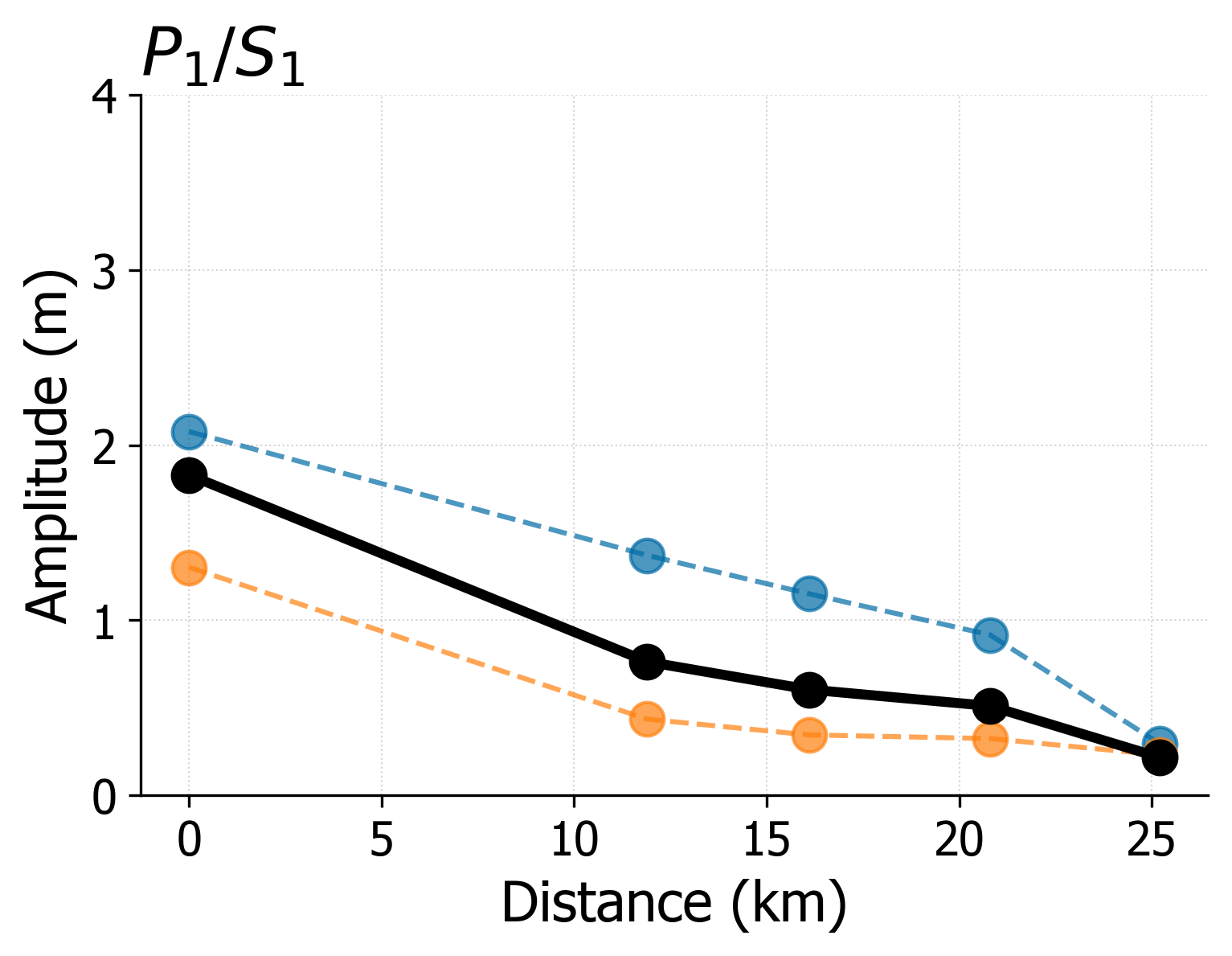}
    \end{subfigure}
    \hfill
    \begin{subfigure}{0.3\textwidth}
        \centering
        \includegraphics[width=\textwidth]{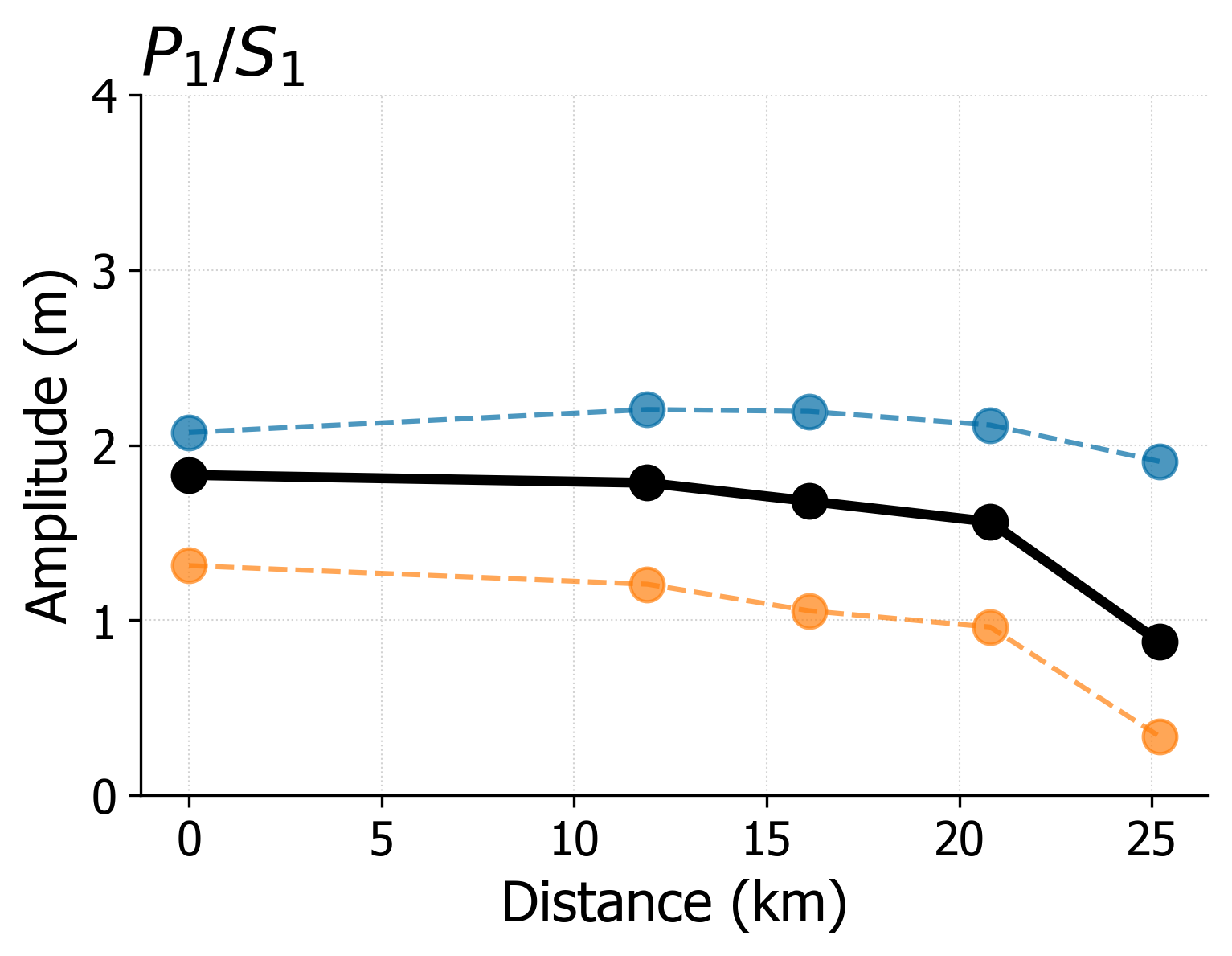}
    \end{subfigure}
  \caption{Reconstructed amplitude and phase of the ratios of diurnal and semi-diurnal constituents, during average flow conditions, high-flow conditions (river discharge above 75th-percentile), and low-flow conditions (river discharge below 25th-percentile) for: a) measured water levels, b) water levels simulated with observed river flow (Simulation A), and c) water levels simulated with filtered river flow (Simulation B).}
    \label{fig:discussion_ratios}
\end{figure}